\documentclass[pdflatex,sn-mathphys-num]{sn-jnl}

\usepackage{graphicx}    
\usepackage{subcaption}  
\usepackage{multirow}%
\usepackage{amsmath,amssymb,amsfonts}%
\usepackage{amsthm}%
\usepackage{mathrsfs}%
\usepackage[title]{appendix}%
\usepackage{xcolor}%
\usepackage{textcomp}%
\usepackage{manyfoot}%
\usepackage{booktabs}%
\usepackage{algorithm}%
\usepackage{algorithmicx}%
\usepackage{algpseudocode}%
\usepackage{listings}%

\graphicspath{{./}{figs/}}

\newcommand{\prospector}{\texttt{Prospector}}

\newcommand{\dynesty}{\texttt{dynesty}}
\newcommand{\phoenix}{\texttt{Phoenix}}
\newcommand{\tlusty}{\texttt{tlusty}}
\newcommand{\prism}{\texttt{Prism}}

\newcommand{\zsol}{{\rm Z}_{\odot}}

\newcommand{\zspec}{z_{\rm spec}}
\newcommand{\Av}{A_{\rm V}}
\newcommand{\rwater}{r_{\rm H_{2}O}}

\newcommand{\ha}{H$\alpha$}
\newcommand{\hb}{H$\beta$}
\newcommand{\pab}{Pa-$\beta$}
\newcommand{\pag}{Pa-$\gamma$}

\newcommand{\nii}{[N\,{\sc ii}]}
\newcommand{\oiii}{[O\,{\sc iii}]}

\newcommand{\oi}{O\,{\sc i}\,$\lambda$8446}

\newcommand{\niia}{[N\,{\sc ii}]\,$\lambda$6549}
\newcommand{\niib}{[N\,{\sc ii}]\,$\lambda$6585}

\newcommand{\oiiia}{[O\,{\sc iii}]\,$\lambda$4960}
\newcommand{\oiiib}{[O\,{\sc iii}]\,$\lambda$5008}

\newcommand{\heib}{He\,$\textsc{i}\,\lambda$1.28$\mu$m}

\newcommand{\kms}{{\rm km\,s^{-1}}}
\newcommand{\cms}{{\rm cm\,s^{-2}}}

\newcommand{\cgs}{{\rm erg \, s^{-1} \, cm^{-2} \, \AA^{-1}}}

\newcommand{\teff}{T_{\rm eff}}
\newcommand{\met}{{\rm [M/H]}}
\newcommand{\logg}{{\rm log}(g)}

\newcommand{\rdwide}{WIDE-EGS-2974}

\newcommand{\rdcapersegs}{CAPERS-EGS-212932}
\newcommand{\rduncover}{UNCOVER-A2744-20698}

\newcommand{\rdqso}{A2744-QSO1}

\newcommand{\sfit}{{\texttt{unite}}}

\theoremstyle{thmstyleone}%
\theoremstyle{thmstyletwo}%

\theoremstyle{thmstylethree}%

\begin{document}

\title[Water in LRDs]{Water absorption confirms cool atmospheres in two little red dots}

\author*[1,2]{\fnm{Bingjie} \sur{Wang}}\email{bjwang@princeton.edu}
\author[3]{\fnm{Joel} \sur{Leja}}
\author[4]{\fnm{Ivo} \sur{Labb{\'e}}}
\author[1]{\fnm{Jenny~E.} \sur{Greene}}
\author[1]{\fnm{Hanpu} \sur{Liu}}

\author[5,6,7]{\fnm{Anna} \spfx{de} \sur{Graaff}}
\author[7]{\fnm{Raphael~E.} \sur{Hviding}}
\author[8]{\fnm{Jorryt} \sur{Matthee}}
\author[1]{\fnm{Eliot} \sur{Quataert}}

\author[9]{\fnm{Rachel} \sur{Bezanson}}
\author[10]{\fnm{Leindert A.} \sur{Boogaard}}
\author[11]{\fnm{Gabriel} \sur{Brammer}}
\author[12]{\fnm{Adam~J.} \sur{Burgasser}}
\author[1]{\fnm{Yi-Xian} \sur{Chen}}
\author[3]{\fnm{Nikko~J.} \sur{Cleri}}
\author[13]{\fnm{Sam~E.} \sur{Cutler}}
\author[14]{\fnm{Pratika} \sur{Dayal}}
\author[15]{\fnm{Lukas~J.} \sur{Furtak}}
\author[16,17]{\fnm{Seiji} \sur{Fujimoto}}
\author[4]{\fnm{Karl} \sur{Glazebrook}}
\author[1]{\fnm{Andy~D.} \sur{Goulding}}
\author[3]{\fnm{Jakob~M.} \sur{Helton}}
\author[18]{\fnm{Michaela} \sur{Hirschmann}}
\author[19]{\fnm{Yan-Fei} \sur{Jiang}}
\author[15]{\fnm{Vasily} \sur{Kokorev}}
\author[1]{\fnm{Yilun} \sur{Ma}}
\author[20]{\fnm{Tim~B.} \sur{Miller}}
\author[21,22]{\fnm{Rohan~P.} \sur{Naidu}}
\author[23]{\fnm{Pascal} \sur{Oesch}}
\author[13]{\fnm{Richard} \sur{Pan}}
\author[24]{\fnm{Casey} \sur{Papovich}}
\author[25]{\fnm{Sedona~H.} \sur{Price}}
\author[7]{\fnm{Hans-Walter} \sur{Rix}}
\author[1,26]{\fnm{David~J.} \sur{Setton}}
\author[21]{\fnm{Wendy~Q.} \sur{Sun}}
\author[21,26]{\fnm{John~R.} \sur{Weaver}}
\author[11,27]{\fnm{Katherine E.} \sur{Whitaker}}
\author[28]{\fnm{Adi} \sur{Zitrin}}

\affil*[1]{\orgdiv{Department of Astrophysical Sciences}, \orgname{Princeton University}, \city{Princeton}, \postcode{08544}, \state{NJ}, \country{USA}}
\affil*[2]{\orgdiv{NHFP Hubble Fellow}}

\affil[3]{\orgdiv{Department of Astronomy \& Astrophysics}, \orgname{The Pennsylvania State University}, \city{University Park}, \postcode{16802}, \state{PA}, \country{USA}}

\affil[4]{\orgdiv{Centre for Astrophysics and Supercomputing}, \orgname{Swinburne University of Technology}, \city{Melbourne}, \postcode{3122}, \state{Victoria}, \country{Australia}}

\affil[5]{\orgdiv{Center for Astrophysics $|$ Harvard \& Smithsonian}, \city{Cambridge}, \postcode{02138}, \state{MA}, \country{USA}}
\affil[6]{\orgdiv{Clay Fellow}}

\affil[7]{\orgdiv{Max-Planck-Institut f{\"u}r Astronomie}, \city{Heidelberg}, \postcode{69117}, \country{Germany}}

\affil[8]{\orgdiv{Institute of Science and Technology Austria (ISTA)}, \city{Klosterneuburg}, \postcode{3400}, \country{Austria}}

\affil[9]{\orgdiv{Department of Physics and Astronomy and PITT PACC}, \orgname{University of Pittsburgh}, \city{Pittsburgh}, \postcode{15260}, \state{PA}, \country{USA}}

\affil[10]{\orgdiv{Leiden Observatory}, \orgname{Leiden University}, \city{Leiden}, \postcode{NL-2300 RA}, \country{The Netherlands}}

\affil[11]{\orgname{Cosmic Dawn Center (DAWN), Niels Bohr Institute, University of Copenhagen}, \city{København N}, \postcode{DK-2200}, \country{Denmark}}

\affil[12]{\orgdiv{Department of Astronomy \& Astrophysics}, \orgname{University of California San Diego}, \city{La Jolla}, \postcode{92093}, \state{CA}, \country{USA}}

\affil[13]{\orgdiv{Department of Physics \& Astronomy}, \orgname{Tufts University}, \city{Medford}, \postcode{02155}, \state{MA}, \country{USA}}

\affil[14]{\orgdiv{Canadian Institute for Theoretical Astrophysics (CITA)}, \orgname{University of Toronto}, \city{Toronto}, \state{ON}, \country{Canada}}

\affil[15]{\orgname{University of Texas at Austin}, \city{Austin}, \state{TX}, \country{USA}}

\affil[16]{\orgdiv{David A. Dunlap Department of Astronomy and Astrophysics}, \orgname{University of Toronto}, \city{Toronto}, \state{ON}, \postcode{M5S 3H4}, \country{Canada}}

\affil[17]{\orgname{Dunlap Institute for Astronomy and Astrophysics}, \city{Toronto}, \state{ON}, \postcode{M5S 3H4}, \country{Canada}}

\affil[18]{\orgdiv{Institute of Physics, Lab for Galaxy Evolution and Spectral Modelling, EPFL, Observatory of Sauverny}, \city{Versoix}, \postcode{1290}, \country{Switzerland}}

\affil[19]{\orgdiv{Center for Computational Astrophysics}, \orgname{Flatiron Institute}, \city{New York}, \postcode{10010}, \state{NY}, \country{USA}}

\affil[20]{\orgdiv{Center for Interdisciplinary Exploration and Research in Astrophysics (CIERA)}, \orgname{Northwestern University}, \city{Evanston}, \postcode{60201}, \state{IL}, \country{USA}}

\affil[21]{\orgdiv{Kavli Institute for Astrophysics and Space Research}, \orgname{Massachusetts Institute of Technology}, \city{Cambridge}, \postcode{02139}, \state{MA}, \country{USA}}

\affil[22]{\orgdiv{Pappalardo Fellow}}

\affil[23]{\orgdiv{Department of Astronomy}, \orgname{University of Geneva}, \city{Versoix}, \postcode{1290}, \country{Switzerland}}

\affil[24]{\orgname{Texas A\&M University}, \city{College Station}, \state{TX}, \country{USA}}

\affil[25]{\orgdiv{Space Telescope Science Institute}, \city{Baltimore}, \postcode{21218}, \state{MD}, \country{USA}}

\affil[26]{Brinson Prize Fellow}

\affil[27]{\orgdiv{Department of Astronomy}, \orgname{University of Massachusetts}, \city{Amherst}, \state{MA}, \postcode{01003}, \country{USA}}

\affil[28]{\orgdiv{Department of Physics}, \orgname{Ben-Gurion University of the Negev}, \city{Be'er Sheva}, \postcode{84105}, \country{Israel}}

\abstract{
Little red dots (LRDs) are an abundant population of compact high-redshift sources with red rest-frame optical continua, discovered by the James Webb Space Telescope (JWST) \cite{Kocevski2023,Matthee2024,Labbe2025}.
Their red colors and power sources have been attributed either to dust reddening of standard hot accretion disks \cite{Furtak2024, Greene2024, Wang2024:ub} or to intrinsically cool thermal emission from dense hydrogen envelopes \cite{deGraaff2025:cliff, Naidu2025}, in both cases surrounding accreting supermassive black holes (SMBHs). These scenarios predict order-of-magnitude differences in emission temperature but have lacked decisive temperature diagnostics. Here we report a prominent absorption feature at rest-frame $\sim 1.4 \, \mu\mathrm{m}$ in two out of four LRDs at $z \sim 2$ with high signal-to-noise JWST spectra, among the coolest from a large LRD sample \cite{deGraaff:sample}. The feature matches the shape and wavelength of the water absorption band seen in cool stars. Atmosphere models require $T \lesssim 3000\, \mathrm{K}$ to reproduce it, confirming unambiguously the presence of a cool, dense gas component contributing $20-30\%$ to the emergent continuum. A composite model reproduces both the absorption and the rest-frame optical-to-infrared continuum shape and suggests a temperature range ($\sim2000\, \mathrm{K} - 4000 \, \mathrm{K}$) rather than a single blackbody predicted by some gas envelope models \cite{Begelman2025, Kido2025, Liu2025}.
Molecular absorption demonstrates that the red continua of some LRDs are intrinsic rather than dust-reddened, implying order-of-magnitude lower bolometric luminosities and black-hole masses, and providing a new diagnostic of the emitting gas.

}


\maketitle


A major puzzle in the interpretation of little red dot (LRD) spectra is the origin of their red rest-frame optical emission. If it arises from a standard hot, blue accretion disk, the red color must be produced by substantial dust reddening, implying large bolometric corrections, high bolometric luminosities, and massive black holes \cite{Labbe2024,Greene2025}. If instead the emission is intrinsically red, it may reflect thermal radiation from an optically thick, dense gas photosphere in a quasi-spherical accretion flow onto the power source \cite{Begelman2025, Kido2025, Liu2025, Nandal2025}, as proposed to explain both the blackbody-like optical continua \cite{deGraaff:sample, Umeda2025} and the unusually strong, non-stellar Balmer-limit breaks observed in some LRDs \cite{deGraaff2025:cliff, Naidu2025}. 

\vspace{0.2cm}
The problem has been that both scenarios predict similar observed continuum shapes but involve fundamentally different temperatures and inferred intrinsic luminosities for the emitting surface.
To distinguish between these scenarios we investigate rest-frame near-infrared spectroscopy with JWST of LRDs to look for temperature-sensitive molecular absorption signatures.

\vspace{0.2cm}
We adopt the compilation of LRDs \cite{deGraaff:sample} with JWST/NIRSpec \prism\ spectroscopy and robustly determined redshifts, drawn from the DAWN JWST Archive (DJA; \cite{dja}). We select LRDs based on their ``v-shape'' blue rest-UV and red rest-frame optical spectral energy distributions (SEDs) along with a compactness criterion \cite{Labbe2024,Hviding2025}. We impose a redshift $2<z<3$ and signal-to-noise $(\rm{S/N})>5$ per pixel cut to ensure sufficient rest-frame optical-to-infrared wavelength coverage and signal-to-noise to both constrain SED shape and test for absorption. More details on the sample and selection criteria are presented in the Methods section. 

\begin{figure}[t]
\centering
\includegraphics[width=\textwidth]{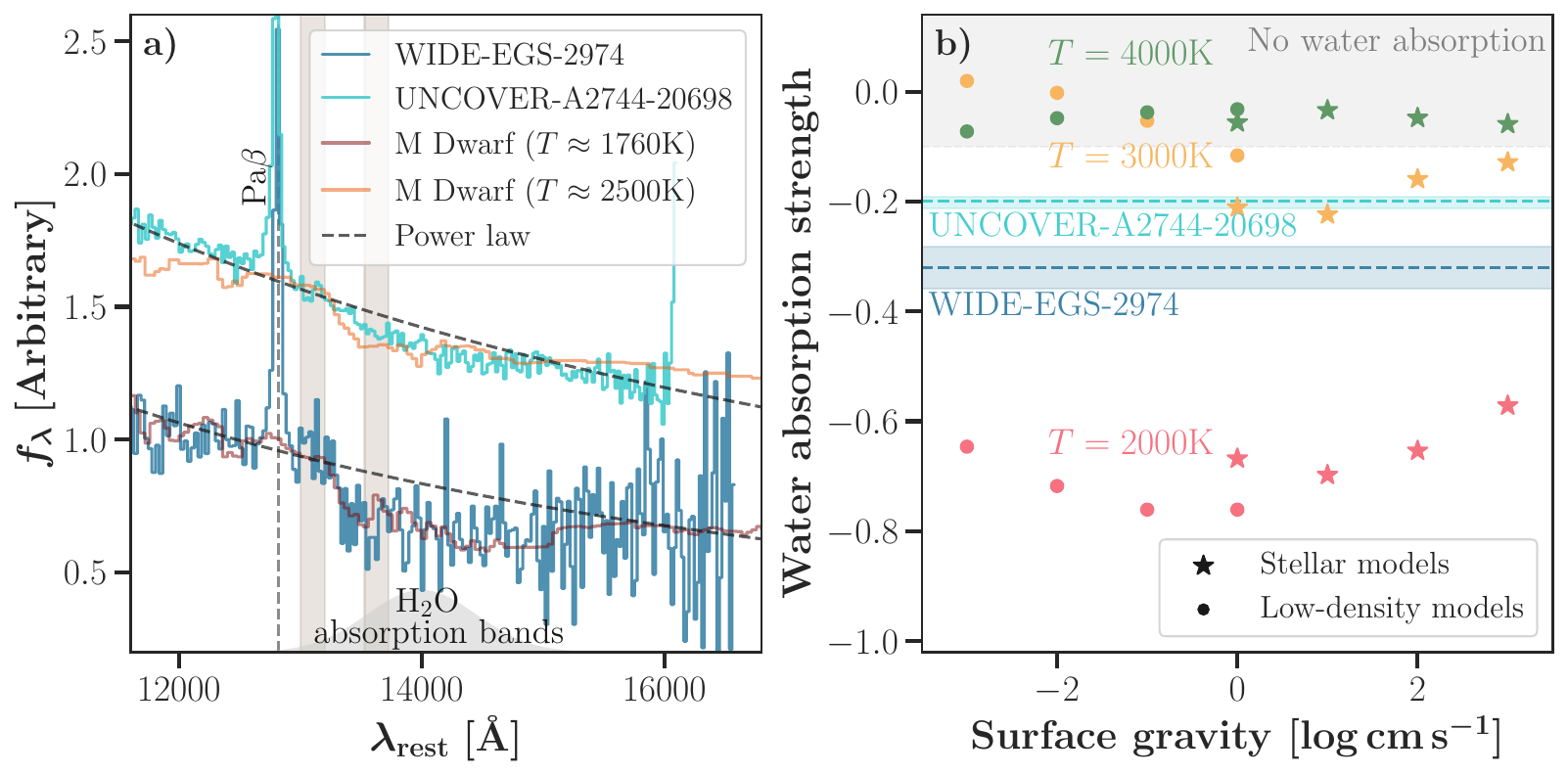}
\caption{\textbf{Detection of water absorption in two LRD spectra at $2<z<3$.}
{\textbf{a)}} The two LRD JWST/\prism\ spectra with detected water absorption are plotted in blue and cyan, respectively. Black dashed lines show power-law fits to the continuum (i.e., no water absorption) to guide the eyes. Observed M dwarf spectra with varying water absorption strength are shown in yellow and brown for comparison \cite{Filippazzo2015, Pineda2021}. We emphasize that these M dwarf spectra are not fits to the LRD data, but are included solely for illustrative purposes. The absorption features in LRDs are consistent with the characteristic water bands seen in these stellar spectra.
To quantify the absorption in a model independent way, we define a simple continuum-normalized water absorption index based on the ratio of flux densities in two narrow rest-frame windows bracketing the break at $1.310\,\mu\mathrm{m}$ and $1.363\,\mu\mathrm{m}$ (gray vertical bands): $r_{\mathrm{H_2O}} = f\lambda,1.310/f\lambda,1.363 - 1$ (i.e., more negative values indicating stronger absorption).
{\textbf{b)}} Measured water absorption strength of the LRDs (blue and cyan, shaded area indicate $1\sigma$ uncertainties) compared to $0.1\,\zsol$ \phoenix\ stellar models and low-density non-stellar models, across 7 orders of magnitude in densities spanning a range of temperatures. Models above $\sim 3{,}000\,\mathrm{K}$ fail to produce the observed water absorption, establishing firm evidence for the existence of low temperature, $\sim 2{,}000\,\mathrm{K}$, gas in LRDs.}\label{fig:water}
\end{figure}

\vspace{0.2cm}
Among the resulting sample of four, we find two sources, \rdwide\ and \rduncover, that show a broad depression at rest-frame $\lambda\approx1.4\,\mu\mathrm{m}$ (Fig.~\ref{fig:water}). The feature is present in the individual spectra at the same rest-frame location and can not be ascribed to systematics, reduction artifacts, or emission lines. Its wavelength and shape match the water absorption seen in atmospheres of cool stars, where H$_2$O produces a characteristic trough spanning $\sim1.33$ to $1.50\,\mu\mathrm{m}$. The remaining two spectra, with comparable overall continuum shape and quality, show no evidence for similar absorption features. 

\vspace{0.2cm}
We investigate stellar models spanning a wide range of gas densities, including a recent extension to low densities appropriate here (see Methods). This confirms that the strength of the water absorption depends weakly on density and moderately on metallicity, but is a strong diagnostic of low temperatures. Molecules are only expected to remain stable and imprint characteristic absorption features if the emitting gas cools below $\lesssim 3{,}000\,\mathrm{K}$ (e.g., \cite{Tsuji1964, Tsuji1973}). Models with $T\gtrsim3{,}000\,\mathrm{K}$ fail to reproduce the significant absorption, establishing a firm upper limit on the temperature of the cold gas component. The mere detection of water molecules thus provides strong evidence for a substantial gas component with thermal emission from a cold surface.

\begin{figure}[t!]
    \centering
    \includegraphics[width=\textwidth]{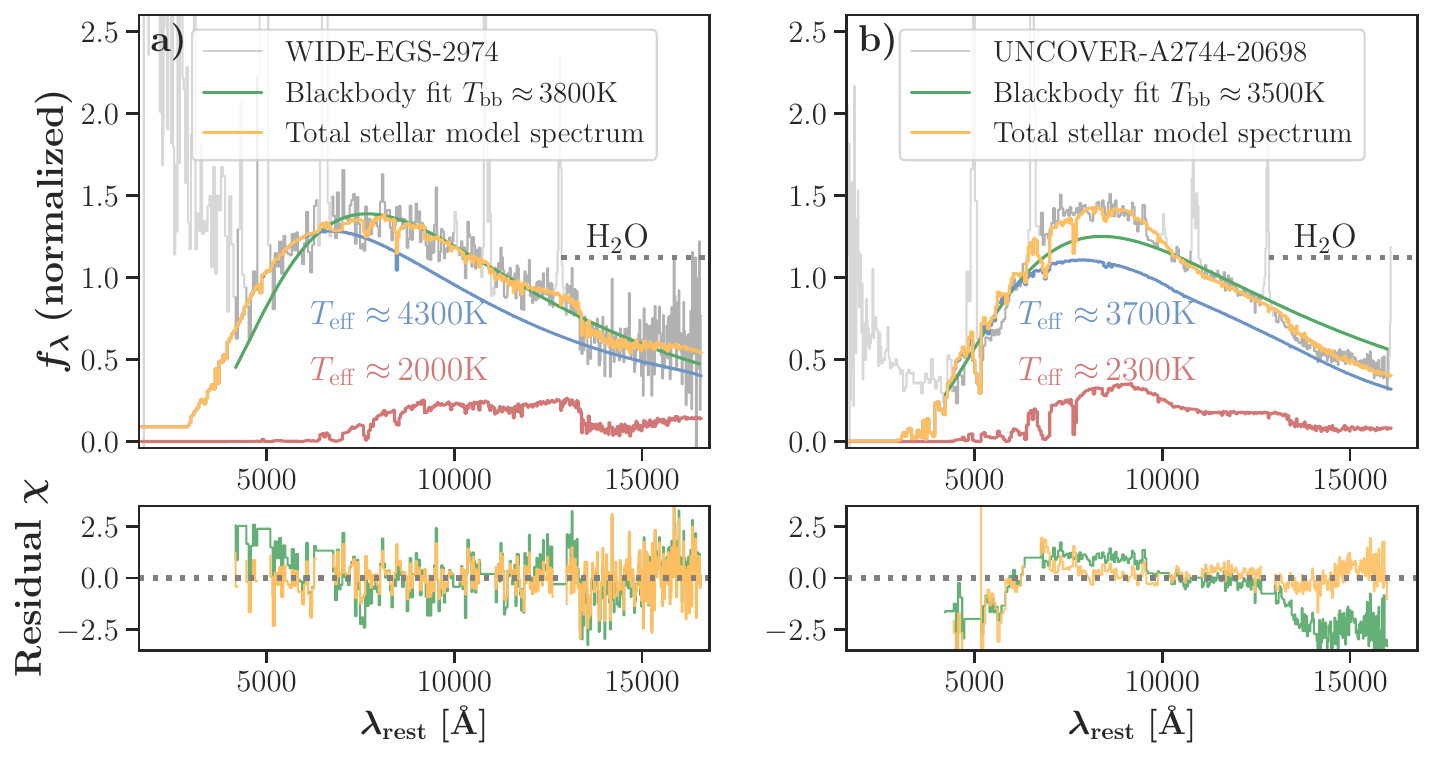} 
    \caption{{\textbf{Models fits to the optical to near-infrared continua of two LRDs with detected water absorption requires cold ($T<3{,}000\,\mathrm{K}$) gas.}}
    The observed JWST/NIRSpec \prism\ spectra are shown in dark gray, with wavelength regions excluded from the fit indicated in light gray.  Single–temperature blackbody models are shown in black.
    Best-fit single-temperature blackbody models (green) broadly reproduce the continuum shape but fail to account for the depth of the water absorption feature and yield temperatures too high for the survival of water vapor. 
    The best-fit two-component photospheric models are shown in orange, with the warm and cool components overplotted in blue and red, respectively. Fit residuals are shown below each panel. This minimal model reproduces both the overall continuum shape and the observed water absorption band. The inferred cool component has an effective temperature of $T\sim2{,}000\,\mathrm{K}$ and contributes approximately $20\%$ and $30\%$ of the total flux at rest-frame $1.4\,\mu\mathrm{m}$ in the two objects, respectively, while a very low metallicity ($Z<10^{-3}\,Z_\odot$) warmer component with $T\sim4{,}000$~K is required to match the optical continuum. 
    }
    \label{fig:model}
\end{figure}

\vspace{0.2cm}
To quantify the contribution of the cold gas component we attempt to fit the full rest-frame optical to near-infrared continuum, masking any emission lines. We do not include the rest-frame UV, as its origin is unclear (possibly reflecting young stellar populations, e.g., \cite{Baggen2025, Golubchik2025}) and the contribution of the UV to the bolometric luminosity subdominant \cite{Greene2025}. Single-temperature blackbody models broadly reproduce the overall shape (see Fig.~\ref{fig:model}), as is the case for a larger LRD population \cite{deGraaff:sample}, however the fits leave significant residuals and produce temperatures $\sim3{,}700$~K$-4{,}300\,\mathrm{K}$ that are incompatible with the presence of water vapor \cite{Tsuji1964,Tsuji1973}. 

\vspace{0.2cm}
Motivated by the close resemblance between the LRD spectra and those of cool stellar atmospheres (see Fig.~\ref{fig:water}), we construct a minimal two-temperature stellar model to mimic a broader distribution of temperatures (Fig.~\ref{fig:model}). This simple model matches both the broad continuum shape redward of the Balmer limit and the depth of the water absorption feature (see details in Methods). The best fit implies that cool gas with $T\sim2{,}000$~K contributes substantially to the emergent near-infrared emission of \rdwide\ and \rduncover, accounting for approximately $20\%$ and $30\%$ of the total flux at rest-frame $1.4\,\mu\mathrm{m}$, respectively, alongside warmer thermal emission at $T\sim4{,}000\,\mathrm{K}$ required to reproduce the optical continuum. 
Recent radiative transfer calculations suggest that an atmosphere density $2-4$ orders of magnitude below the stellar regime, $\log(g/\mathrm{cm,s^{-2}})<0$, is expected for LRDs \cite{Liu2025}, 
but we still expect water to survive at $T<3{,}000\,\mathrm{K}$ at such low density (Fig.~\ref{fig:water}). Overall, the results imply that thermal emission from cool-to-warm, low-metallicity gas reproduces the observed continuum shape and water absorption feature well.

\begin{figure}[ht]
    \centering
    \includegraphics[width=\textwidth]{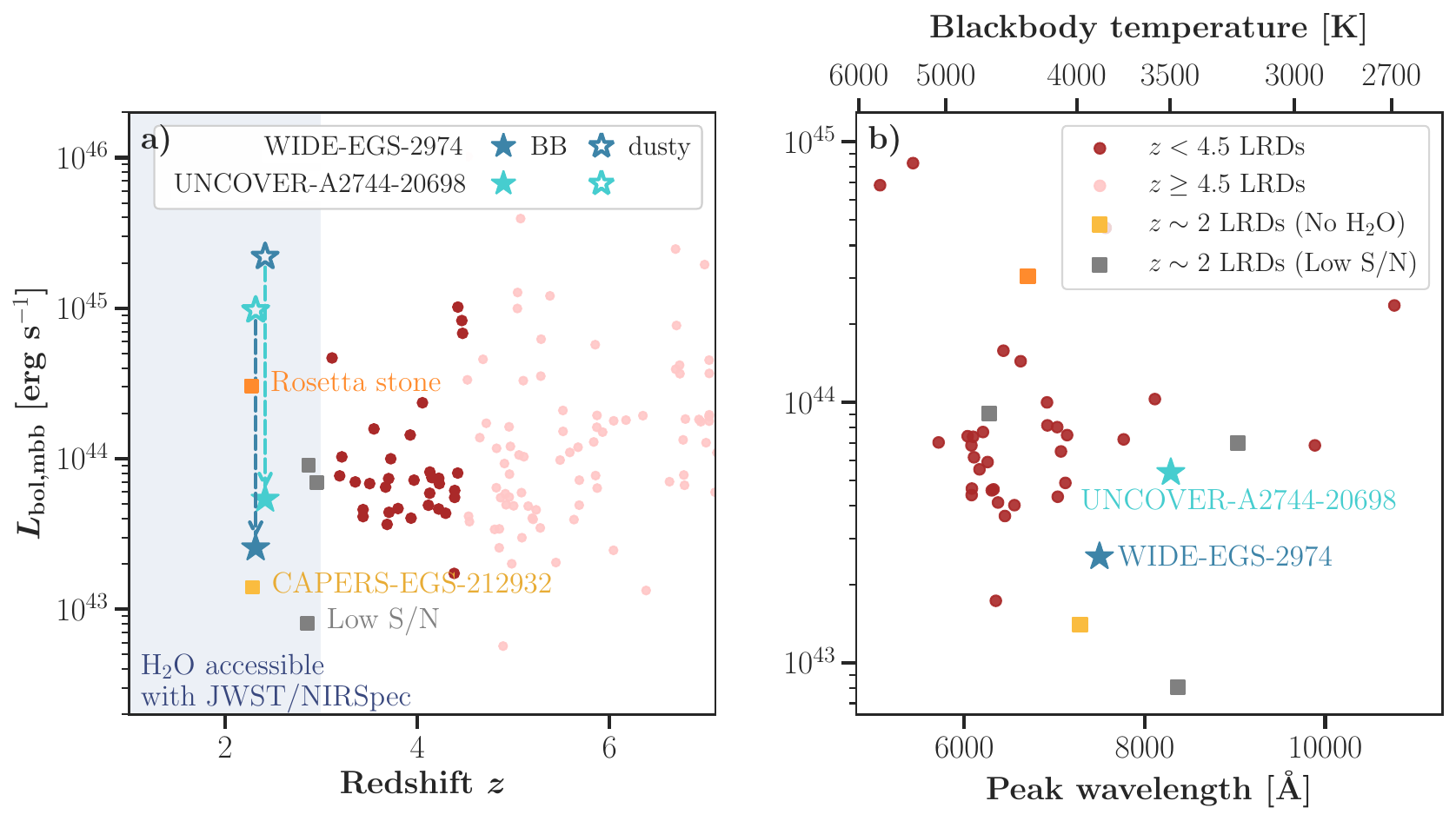}
     \caption{
     {\textbf{Bolometric luminosities and temperatures of the two cold LRDs in context of the LRD population.}}
     \textbf{a)} The bolometric luminosities of the sources studied here, inferred using the two-temperature stellar atmosphere model, are shown as blue and cyan stars. For context, we include the bolometric luminosities of a larger sample of LRDs derived from single temperature modified black body model fits \cite{deGraaff:sample}. For our sample blackbody fits produce practically identical bolometric luminosities (within 5-10\%).
     For comparison, the bolometric luminosities assuming a dust-reddened AGN interpretation are shown as unfilled stars of the same colors and would be at least an order magnitude higher.
     \textbf{b)} The bolometric luminosities versus wavelength at which the rest optical continuum peak for LRDs, showing only sources at $z<4.5$ where the blackbody model fits are well constrained. Longer peak wavelengths correspond to lower effective temperatures, and an approximate effective temperature conversion is shown on top. The LRDs of this paper, marked as stars, have bolometric luminosities within the range of the larger LRD population but occupy the colder side of the temperature distribution.
}
     \label{fig:lbol}
 \end{figure}


\vspace{0.2cm}
We now consider the results in the context of the broader sample. Within the high signal-to-noise sample at $2<z<3$, two out of four objects show clear water absorption, while the others have similar overall SEDs but appear modestly bluer (see Methods). Compared to the broader LRD population spanning $z\sim2-7$ (Fig.~\ref{fig:lbol}), the two objects studied here have consistent bolometric luminosities, but lie toward the redder end of the color distribution and colder end of the inferred temperature distribution. If the thermal interpretation of the rest-frame optical spectra of LRDs is correct, the reddest LRDs should also be the coolest \cite{Barro2025, deGraaff:sample, Umeda2025}, in tentative agreement with our findings.

\vspace{0.2cm}

The detection of water absorption provides direct evidence that at least some of the reddest LRDs are powered by intrinsically red thermal emission rather than dust-reddened hot continua.
The picture of LRDs as dusty active galactic nuclei is disfavored as redder color merely corresponds to increasing dust attenuation of an intrinsic hot surface ($>10^5\,\mathrm{K}$) which precludes the existence of molecules. The spectral properties are therefore more consistent with an optically thick envelope of gas that radiates thermally over a limited range of temperatures, surrounding a central power source (presumed to be a black hole). This aligns with an emerging set of models for LRDs, including quasi-stars \cite{Begelman2008, Begelman2025}, black hole stars \cite{deGraaff2025:cliff, Naidu2025}, or super-Eddington accretion \cite{Kido2025, Liu2025}.

\vspace{0.2cm}
The most important implication is for the inferred bolometric luminosities and SMBH masses. Thermal models predict bolometric luminosities that are at least an order of magnitude lower compared to dust reddened scenarios (Fig.~\ref{fig:lbol}; also \cite{Greene2025, Ronayne2025}).  This alleviates the problem that LRDs overproduce the black hole mass density at high redshift, and is more consistent with the clustering measurements \cite{Matthee2025, Lin2025:cluster} and their observed high number densities \cite{Pizzati2025}.  

\vspace{0.2cm}
It is unclear whether water absorption is a general feature of LRDs. The majority of known LRDs lie at $z>3$, where JWST/NIRSpec cannot access the relevant wavelength range. Among LRDs at $z<4.5$, for which blackbody temperatures can be reliably determined from NIRSpec/\prism\ spectra, inferred temperatures typically span $\sim4{,}000$–$6{,}000\,\mathrm{K}$. Existing photospheric models based on quasi-spherical inflow onto a black hole invoke Hayashi-like arguments to predict a single characteristic temperature of $\sim5{,}000\,\mathrm{K}$ \cite{Begelman2025, Kido2025}, broadly consistent with these observations but far too high for water vapor to survive. Some models can reach lower effective temperatures of $\sim2{,}000$–$3{,}000\,\mathrm{K}$ in the case of very high accretion rates \cite{Liu2025}.

\vspace{0.2cm}
It will therefore be important to determine whether hotter LRDs also harbor a low-temperature tail below $\sim3{,}000\,\mathrm{K}$, as inferred for the two objects in this paper. In the context of photospheric models, it is difficult to reconcile a gaseous, quasi-static, thermally emitting sphere in hydrostatic equilibrium with a wide range of temperatures. A single star is well described by one effective temperature. Even in red supergiants and asymptotic giant branch stars, where strong convection induces surface temperature fluctuations, the variations are typically limited to a few hundred Kelvin \cite{Young2000}. The $\gtrsim1{,}000\,\mathrm{K}$ temperature range inferred for the sources studied here may therefore indicate more complex temperature and density structures in the thermally emitting source, motivating extensions beyond simple single-temperature photospheric models.

\vspace{0.2cm}
Finally, the detection of water absorption establishes molecular spectroscopy as a new probe of LRDs, accessible with JWST/NIRSpec at $z<3$ and potentially extending to higher redshift with deep JWST/MIRI observations. Systematic measurements of molecular features across the LRD population will test whether cool thermal emission is a generic property and determine how much temperature, density, metallicity, and geometry vary among these sources, constraining the basic physical conditions of rapid black hole growth in the early Universe.

\vspace{0.2cm}

\clearpage

\counterwithin{figure}{section}
\counterwithin{table}{section}

\section*{Methods}
\label{sec:method}

\setcounter{section}{0}

\section{Sample Selection}
\label{sec:data}

\subsection{LRDs at Cosmic Noon}
\label{sec:method:z2lrds}
The parent LRD sample used in this paper is the sample compiled in \cite{deGraaff:sample}, which draws from the full DJA dataset of $\sim$17,000 low-resolution (R$\sim$100) NIRSpec/\prism\ spectra with robustly determined redshifts (grade $=$ 3). While we refer readers to \cite{deGraaff:sample} for a detailed description of the sample selection, we briefly summarize the key criteria here.

First, the spectrum must exhibit an inflection (v-shape) near the Balmer limit, quantified via broken-power-law fits. Second, the sources must be compact, either satisfying the aperture flux ratio requirement in NIRCam/F444W imaging $f_{\rm F444W} (0.2'') / f_{\rm F444W} (0.1'') < 1.7 $ \cite{Labbe2025}, or showing that the point-source component contributes $>50$\% of the F444W flux in $>95$\% of the posterior samples from S{\'e}rsic-profile fitting with \texttt{pysersic} \cite{Pasha2023}.

We additionally impose a redshift cut of $2 \leq z \leq 3$. The lower limit ensures that the Balmer limit and part of the rest-frame UV are covered, while the upper limit ensures that the water feature falls within the observed wavelength range of \prism.
Within the DJA LRD sample, a total of eight sources satisfy all the criteria listed here.

\subsection{Searching for Water Absorption}

From the sample of eight sources, we search for signatures of water absorption, particularly in the prominent band near rest-frame 1.4~$\mu$m. Although water also produces absorption features around 1.2~$\mu$m and 1.9~$\mu$m, the 1.2~$\mu$m band is easily contaminated by the nearby \pag\ emission line, while the 1.9~$\mu$m feature is shifted beyond the observed spectral range for LRDs, requiring $z < 2$ where there are no LRDs confirmed spectroscopically by JWST.
We impose the following selection criteria for water absorption:
\begin{enumerate}
  \item Water absorption, as quantified by $\rwater$. We define $\rwater$ as the difference between the median flux densities, in $f_\lambda$, between two rest-frame wavelength windows of [13000, 13200] and [13530, 13730]~\AA, normalized by the median flux density in the short wavelength window. $\rwater \leq -0.2$ is required for a significant water absorption detection. 
  \item Median S/N per pixel $>5$ in both wavelength windows, ensuring reliable detection of the water absorption feature.
\end{enumerate}

The exact threshold value of $-0.2$ for $\rwater$ is chosen to minimize contamination from sources whose apparent signal arises from a sloping continuum rather than from a genuine absorption feature. The precise numerical value is not critical for the conclusions of this paper, especially given the limited sample size.
Four out of the eight LRDs satisfy the S/N requirement.
The generally low $\mathrm{S/N}$ across the broader sample arises primarily from increased noise near the detector edge of NIRSpec.
Applying the $\rwater$ selection yields two sources, namely \rdwide\ and \rduncover.
We visually inspect both spectra to confirm the presence of water absorption. The $\rwater$ values for \rdwide\ and \rduncover\ are -0.32 and -0.20, respectively.
In summary, water absorption is detected in two out of the four LRDs for which the spectral coverage and S/N allow such a feature to be measured.

\begin{figure}[ht]
    \centering
    \begin{subfigure}[b]{\textwidth}
        \centering
        \includegraphics[width=\textwidth]{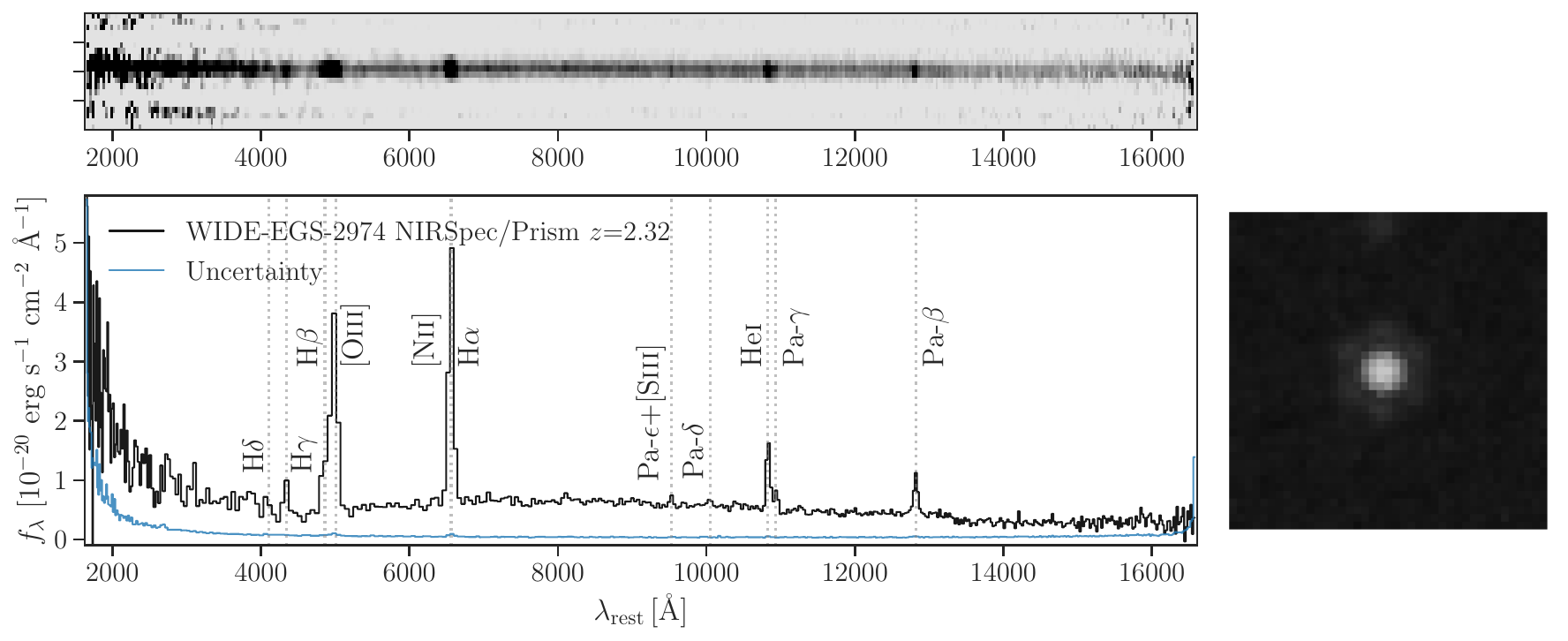}
    \end{subfigure}
    \hfill
    \begin{subfigure}[b]{\textwidth}
        \centering
        \includegraphics[width=\textwidth]{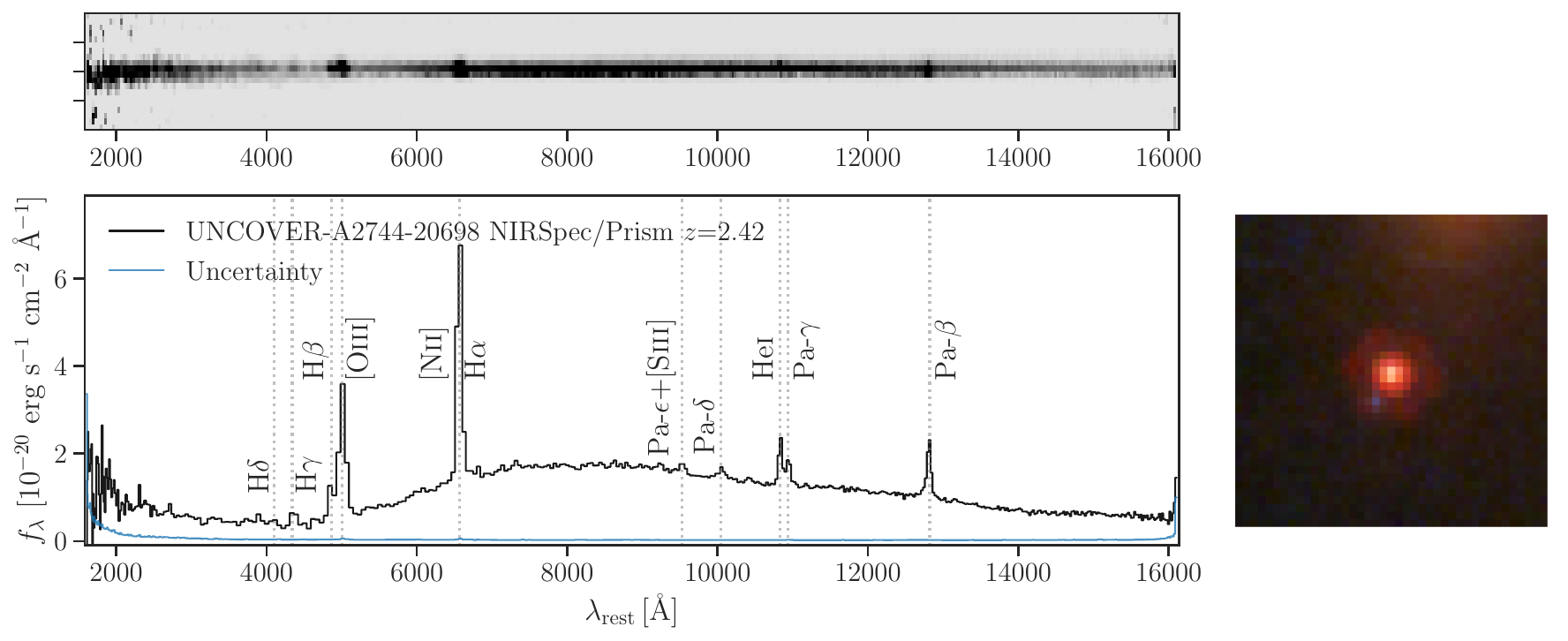}
    \end{subfigure}
    \caption{Sample of this paper. The observed JWST/\prism\ spectra are plotted in black, whereas the uncertainties are shown in blue. 2D spectra are shown on the top. Composite images on 1$''$ $\times$ 1$''$ cutouts are included to the right.
    }
    \label{fig:sample}
\end{figure}

\begin{figure}[ht]
    \centering
    \begin{subfigure}[b]{\textwidth}
        \centering
        \includegraphics[width=0.6\textwidth]{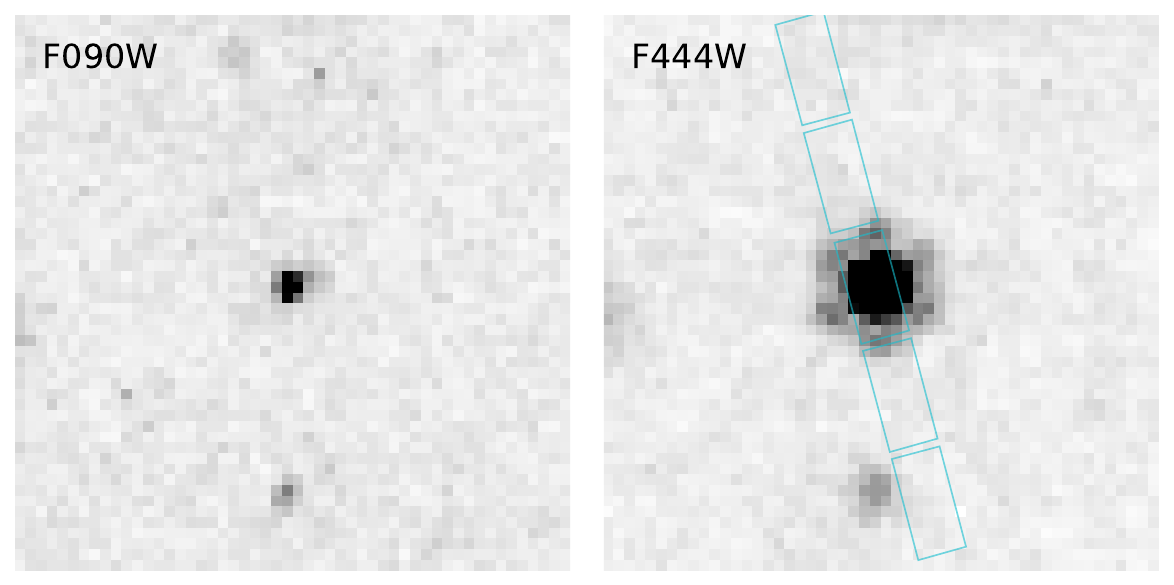}
        \caption{}
    \end{subfigure}
    \hfill
    \begin{subfigure}[b]{\textwidth}
        \centering
        \includegraphics[width=\textwidth]{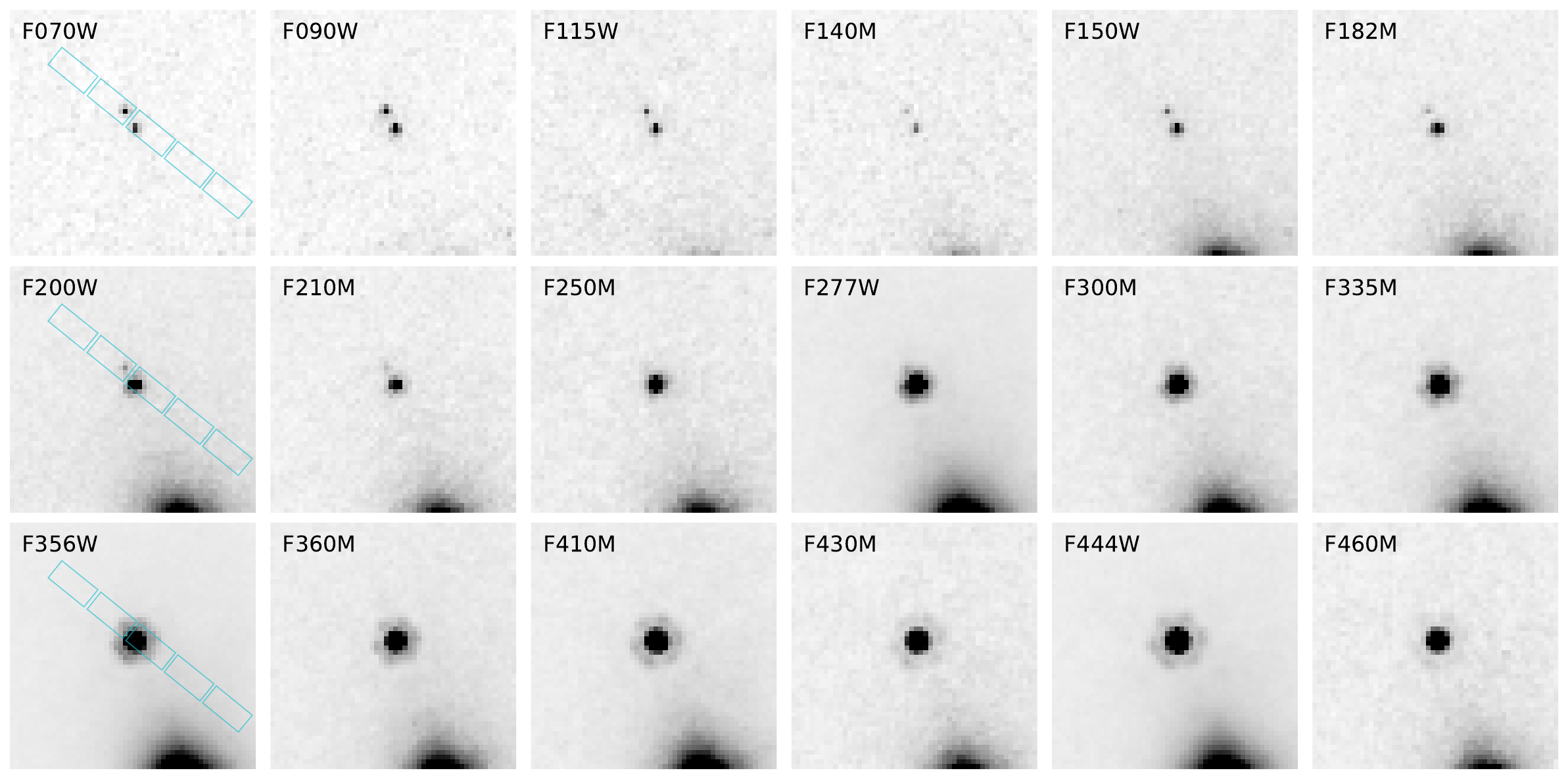}
        \caption{}
    \end{subfigure}
    \caption{Cutouts from the available JWST/NIRCam imaging for (a) \rdwide\ and (b) \rduncover. Each panel is 1.3$''$ $\times$ 1.3$''$. The NIRSpec slitlets are overlaid in cyan.}
    \label{fig:sample_stamp}
\end{figure}

The basic observational characteristics of our sample are listed in Table~\ref{tab:sample}. The cutouts across all filters and \prism\ spectra are shown in Fig.~\ref{fig:sample_stamp} and \ref{fig:sample}, respectively. Specifically, \rdwide\ was observed for 40 mins in the \prism\/CLEAR mode, 27 mins in the G235H/F170LP mode, and 29 mins in the G395H/F290LP mode, as part of the NIRSpec Wide GTO Survey (JWST-GTO-1213; PI N. Luetzgendorf; \cite{Maseda2024}).
\rduncover\ was observed for 263 mins in the \prism\/CLEAR mode, as part of the UNCOVER Treasury survey (JWST-GO-2561; PIs I. Labb\'{e} \& R. Bezanson; \cite{Bezanson2024, Price2025}).

Intriguingly, in the short-wavelength imaging, taken as part of the UNCOVER/MegaScience survey \cite{Bezanson2024, Suess2024}, a nearby companion to \rduncover\ is visible (Fig.~\ref{fig:sample_stamp}). He\,\textsc{i} emission from this source is detected in the ALT data, indicating that it lies at the same redshift as \rduncover. However, morphological analysis classifies it as a tentative companion candidate \cite{Baggen2026}.

We note that a diverse set of selection criteria for LRDs exists in the literature, and the DJA sample prioritizes purity over completeness. One LRD at $z = 2.01$ has been reported (the ``Saguaro''; \cite{Rinaldi2025}), but because it is embedded within a spiral galaxy, it fails the compactness cut and is therefore excluded from our parent sample.

\section{Measurements}

\subsection{Bolometric luminosities}
\label{sec:method:lbol}

Two definitions of bolometric luminosity are contrasted in this work: one assuming that LRDs are dusty AGNs, and one adopting the photospheric interpretation supported by the presence of water absorption. Here we describe how each estimate is obtained.

For the dusty-AGN scenario, we adopt the joint galaxy+AGN model of \cite{Wang2025:brd} and fit the spectra using the Bayesian inference code \prospector\ \cite{Johnson2021}. For both sources, this framework favors an AGN with substantial dust reddening, $\Av > 2.5$. From these fits we obtain the dust-corrected continuum luminosity at rest-frame 5100~\AA, a standard monochromatic proxy for AGN power across redshift. We then apply a canonical bolometric correction of $\sim 10$ (e.g., \cite{Richards2006, Shen2020}) to the dereddened $L_{5100}$ to estimate the bolometric luminosity, denoted $L_{\rm bol,5100}$.

However, as demonstrated in this paper and in \cite{Greene2025}, such monochromatic luminosities are likely poor tracers of the true bolometric output of LRDs. When the red colors of LRDs are intrinsic rather than due to dust extinction, applying standard AGN bolometric corrections systematically overestimates $L_{\rm bol}$. Moreover, if LRDs possess intrinsically blackbody-like continua, their SEDs peak at wavelengths set by their effective temperatures. For the coldest objects, this peak lies far redward of 5100~\AA, making $L_{5100}$ an especially biased and incomplete proxy for the total luminosity.

For the two sources analyzed in this paper, we report bolometric luminosities derived from the joint stellar-atmosphere fits, $L_{\rm bol,*}$. These values are more than an order of magnitude lower than the luminosities inferred under the dusty-AGN assumption.

For the DJA LRD sample, we adopt bolometric luminosities derived from modified-blackbody fits, $L_{\rm bol, mbb}$ \cite{deGraaff:sample}, where the rest optical continua of DJA LRDs are fit with modified blackbodies, and bolometric luminosities consistent with the revised corrections proposed in \cite{Greene2025} are obtained.
The spectrum of one of the eight LRDs selected in Methods~\S~\ref{sec:method:z2lrds} has insufficient continuum S/N for a robust blackbody fit, and is thus excluded in Fig.~\ref{fig:lbol}.
The modified-blackbody fits are well constrained only for sources at $z<4.5$, given the wavelength coverage of the NIRSpec/\prism\. Accordingly, we restrict the comparison to the $z<4.5$ DJA sample.

We compare $L_{\rm bol,mbb}$ with the bolometric luminosity derived from the joint stellar-atmosphere fits, $L_{\rm bol,*}$. The two estimates agree to within 3\% and 10\% for \rdwide\ and \rduncover, respectively. In both cases $L_{\rm bol,mbb}$ is slightly lower because it does not capture the additional flux contributed by the cool component peaking in the rest-frame infrared. The larger discrepancy for \rduncover\ is expected, as it is the colder of the two sources. Importantly, these differences are small enough that they do not impact any of our conclusions.

\subsection{Emission Lines}
\label{sec:method:lines}

All emission lines are modeled using the \sfit\ package, following a setup similar to that described in \cite{Hviding2025}. We briefly summarize the modeling components here.

The NIRSpec line-spread function (LSF) curves are empirically estimated, assuming an idealized point source \cite{deGraaff2024:msafit}. To account for the LSF under-sampling, \sfit\ integrates the model flux within each pixel, with a nuisance parameter that allows for extra broadening to capture systematic uncertainties.

Given the low spectral resolution of the NIRSpec/\prism\ data, it is challenging to robustly assess the presence of a broad emission component. For \rdwide, high-resolution grating spectra are available, which are fit jointly with the \prism\ data; however, the relatively low S/N makes the detection of broad components uncertain.
We therefore start with a simple model in which each emission line is described by a single Gaussian profile with a prior on the line width of FWHM~$\in$~[0, 5000]~$\kms$. All emission lines are tied to a common redshift and velocity dispersion. The \nii\ and \oiii\ doublet flux ratios are fixed to the theoretical values of 1:2.95, 1:2.98, respectively \cite{Galavis1997}.

Second, we perform fits that include an extra broad Gaussian component to represent possible broad Balmer and helium emission. Emission lines are grouped into narrow permitted, narrow forbidden, and broad permitted components; within each group, all lines share a common redshift and velocity dispersion. Model comparison is performed using the Widely Applicable Information Criterion (WAIC; \cite{Watanabe2010}), which balances predictive accuracy against model complexity.

For \rdwide, a broad component with ${\rm FWHM} = 1420_{-169}^{+187}~\kms$ is found, and favored at $>3\sigma$ significance.
For \rduncover, a broad component with ${\rm FWHM} = 1707_{-634}^{+985}~\kms$ is found, but the fit including only narrow components and that allowing an additional broad component yield statistically equivalent solutions. NIRCam grism spectroscopy from All the Little Things (JWST-GO-3516; PIs Matthee \& Naidu; \cite{Naidu2024}) is likewise inconclusive regarding the presence of broad emission lines in this source.
However, when performing separate fits to the \pab\ and \heib\ complex only, a broad component is favored at $>3\sigma$ significance for both sources.
As these lines are among those with the highest effective spectral resolution in the \prism\ spectra, this suggests that a broad component is likely to be present in other permitted lines.

\begin{figure}[ht]
    \centering
    \begin{subfigure}[b]{\textwidth}
        \centering
        \includegraphics[width=\textwidth]{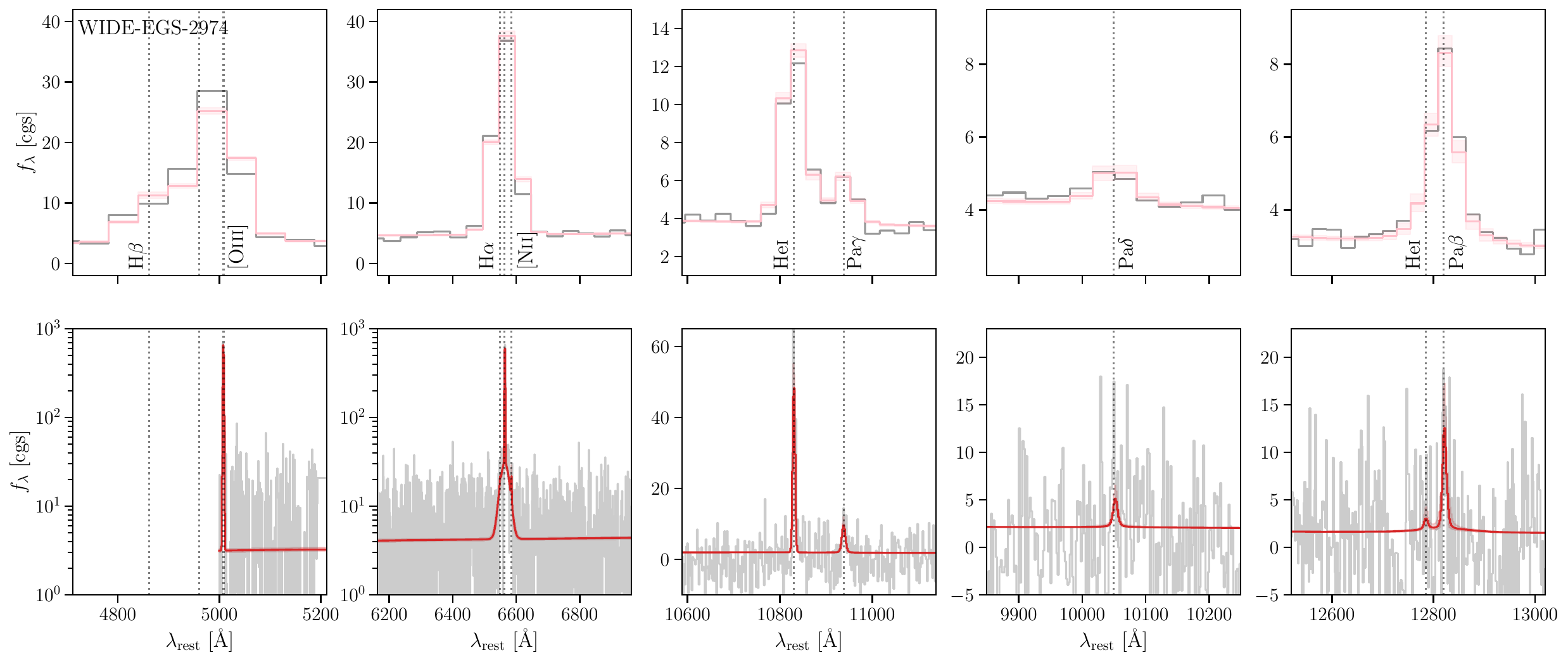}
        \caption{}
    \end{subfigure}
    \begin{subfigure}[b]{\textwidth}
        \centering
        \includegraphics[width=\textwidth]{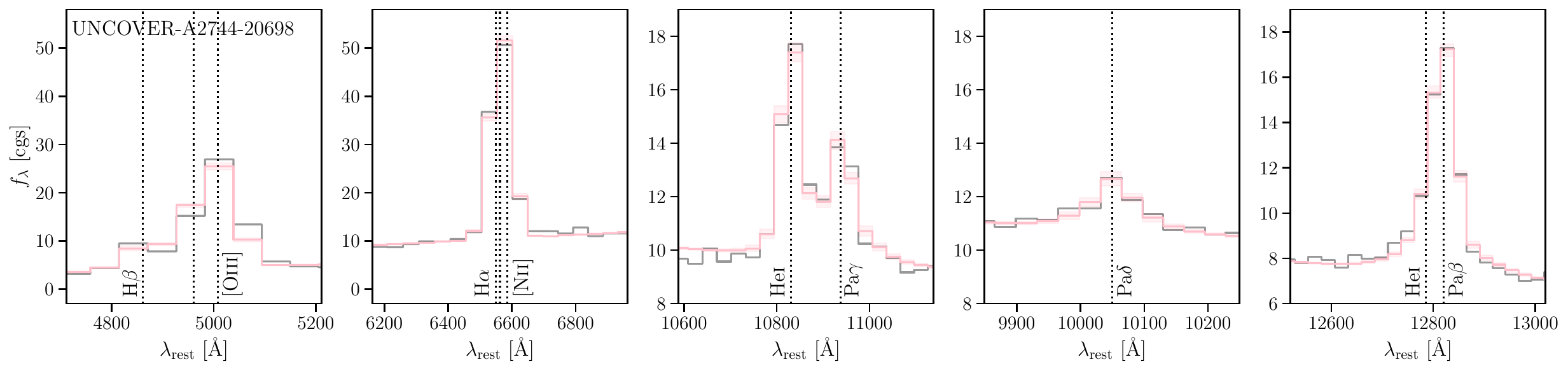}
        \caption{}
    \end{subfigure}
    \caption{Emission-line modeling. (a) The observed \prism\ and G395H spectra of \rdwide\ are shown in gray, and are fit simultaneously. Best-fit emission line models are plotted in red, with shading indicates the $1\sigma$ uncertainty. (b) The spectrum of \rduncover\ is shown in the same format.}
    \label{fig:app:lines}
\end{figure}

We show the narrow+broad fits in Fig.~\ref{fig:app:lines}, and report the FWHMs, total EWs, and total line fluxes in Table~\ref{tab:lines}.
We again caution that follow-up medium-resolution spectroscopy is necessary to confirm the presence or lack of broad-line components. 
It is worth emphasizing that a narrow-line LRD would be a remarkable discovery of its own: the largest spectroscopic study of LRDs to date finds that all point sources with v-shaped continua have broad lines \cite{Hviding2025}.

\subsection{Sample Comparison}
\label{sec:method:z2}

\begin{figure}[t!]
    \centering
    \begin{subfigure}[b]{0.45\textwidth}
        \centering
        \includegraphics[width=\textwidth]{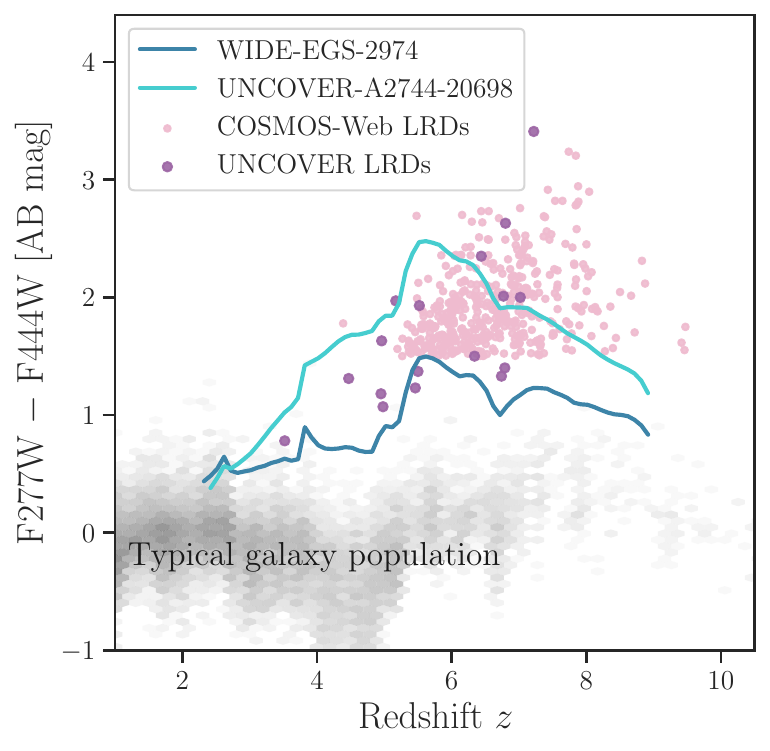}
        \caption{}
    \end{subfigure}
    \hfill
    \begin{subfigure}[b]{0.8\textwidth}
        \centering
        \includegraphics[width=\textwidth]{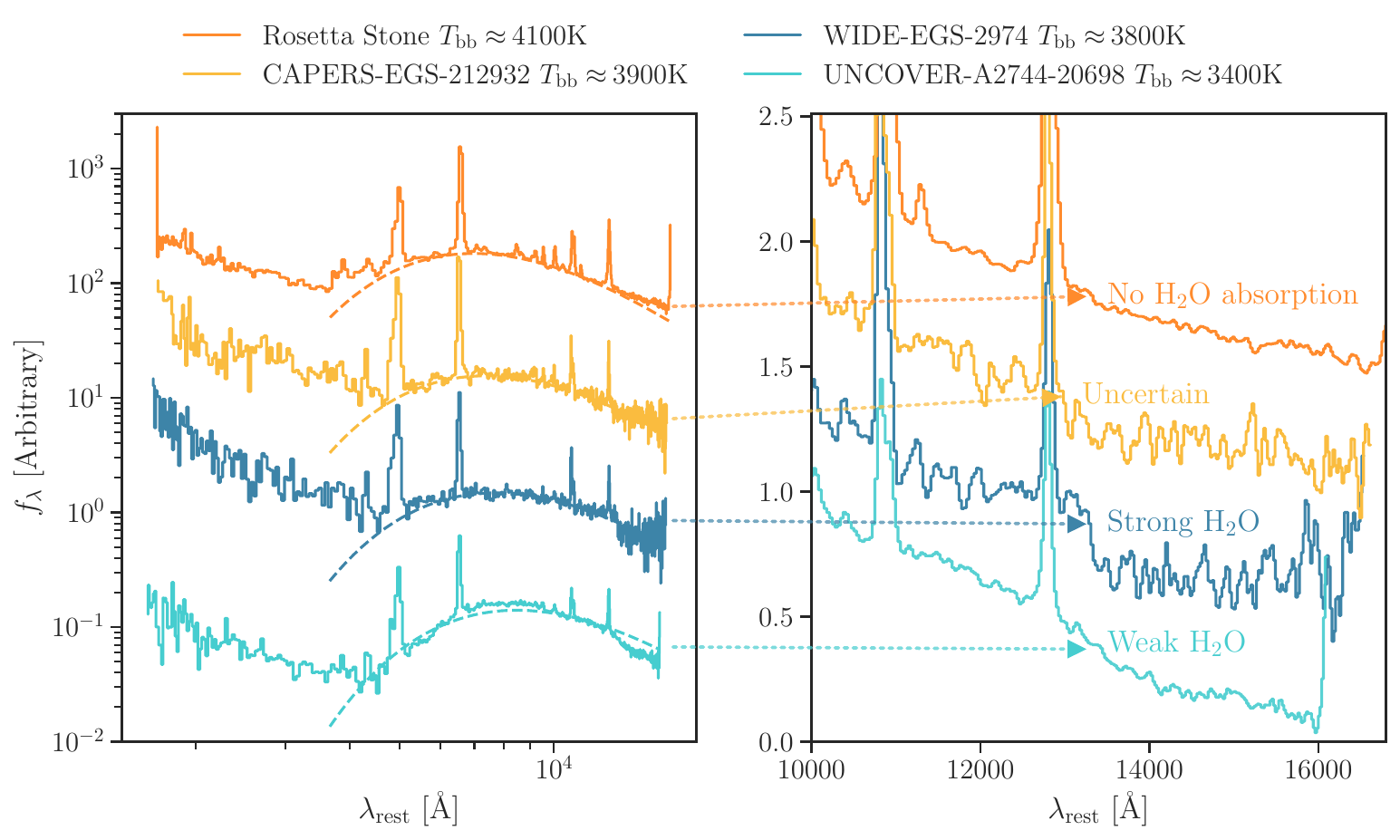}
        \caption{}
    \end{subfigure}
    \caption{The sources studied here belongs to the population of LRDs, but may represent the coldest temperature tail.
(a) The two LRDs in this paper (in blue and green) would satisfy the color-selection criteria used in the literature if observed at higher redshift. Typical galaxy populations are shown in gray for reference \cite{Weaver2024, Wang2024:sps}.
(b) Left: all the $z \sim 2$ LRDs that have sufficiently S/N spectra to determine the presence of significant water absorption are shown here. Best-fit blackbody curves are over-plotted in dashed lines. Our sample is at the colder end relative to the other $z \sim 2$ LRDs.
Right: zoom-in on the water absorption region. The bottom three spectra are smoothed with a Gaussian filter for visualization purposes.
\label{fig:color}
}
\end{figure}

High-redshift LRDs are typically selected by their red color and point-source-like morphology in long-wavelength NIRCam imaging (e.g., \cite{Kocevski2023, Kokorev2024, Matthee2024}).
Our $z \sim 2$ sources satisfy common compactness criteria \cite{Hviding2025, Labbe2025} according to \cite{deGraaff:sample}, and would meet standard LRD color selection criteria if observed at higher redshift (Fig.~\ref{fig:color}).

As for luminosities, distant LRDs generally appear more luminous (Fig.~\ref{fig:lbol}), likely reflecting selection effects, while the scarcity of luminous LRDs at lower redshift may arise from the smaller surveyed volume. Nevertheless, the bolometric luminosities of the sources in this paper are consistent with those of the broader LRD population. A systematic search for low-redshift LRDs and analogs \cite{Ma2025, Lin2025} would help map the full luminosity distribution and test any temperature--luminosity correlation at the low-temperature end.

Particularly noteworthy is \rduncover, which has a bolometric luminosity comparable to the broader LRD population but exhibits the coldest rest optical continuum temperature at $z \sim 2$. \rdwide\ is less luminous, though still consistent with the population. Even when compared to the DJA LRD sample \cite{deGraaff:sample}, the LRDs of this paper are colder than most LRDs. We illustrate this in Fig.~\ref{fig:color} also by comparing our objects with the ``Rosetta Stone'' and \rdcapersegs. To provide a more intuitive sense of temperature, we show simple blackbody fits rather than modified blackbodies. All sources except \rduncover\ are well described by a perfect blackbody; interestingly, it is also the coolest source, suggesting that deviations from a standard blackbody may encode additional temperature information, a question that future detailed modeling may resolve.
This point is illustrated in Fig.~\ref{fig:lbol}, where we show the wavelength at which the rest optical continua peak as a function of $L_{\rm bol,mbb}$. The peak wavelength acts as a simple way for quantifying the effective temperatures of LRDs. Our sample tends to peak at longer wavelengths, corresponding to lower temperatures. 

It is evident that our sample occupies the colder end of the LRD demographic space that is currently surveyed. No water absorption is detected in the hotter sources. This supports the interpretation that detectable water absorption may be a signature unique to the coldest LRDs. However, we caution against drawing firm conclusions at this stage, given the relatively limited dynamic range in temperatures probed so far.

The emission-line measurements may provide additional support for the above scenario. The widths of the hydrogen recombination lines in our sample, although uncertain without deep higher-resolution spectroscopy, appear to lie at the narrower end. In LRD photospheric models, where these line profiles are shaped by radiative transfer through a dense gaseous envelope, this trend may be a manifestation of a physical reason: when the temperature is low, the number of free electrons decreases, reducing the efficiency of electron scattering. As a result, emission-line photons are less likely to be scattered into the wings of the line profile, producing narrower emission lines.

Finally, we examine the Balmer decrements, defined as the flux ratio \ha/\hb, which is often used as a tracer of gas density in LRD studies by comparing the observed ratio to the Case B value of 2.8. 
Ratios significantly above Case B indicate collisional excitation and possibly resonant \hb\ scattering.
The Balmer decrements for \rdwide\ and \rduncover\ are $3.4 \pm 0.4$ and $7.3 \pm 1.1$, respectively, both lower than those typically reported in the literature (e.g., $10.4 \pm 0.3$ \cite{Torralba2025}, and a sample average and median of 8.7 \cite{deGraaff:sample}).
While both collisional and scattering processes are indicative of very high gas densities, the lower than average Balmer decrements of our sample may suggest that we are probing the gas conditions at the low density end of the LRD population. Intriguingly, no correlation between temperature and Balmer decrement is observed in the DJA LRD sample \cite{deGraaff:sample}. More detailed analyses of Balmer and Paschen line ratios and kinematics will be required to disentangle the dominant physical processes operating within the gas envelopes of LRDs.

\section{Molecular Absorption}
\label{sec:method:mole}

\begin{figure}[ht]
 \centering
 \includegraphics[width=0.5\textwidth]{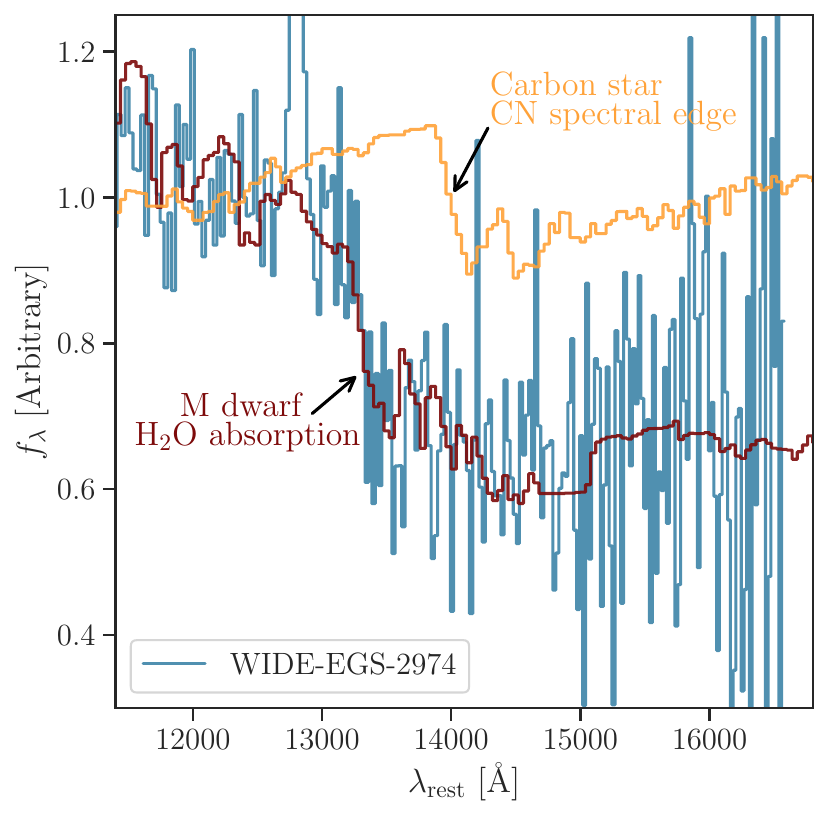}
    \caption{Identification of the molecular absorption feature. The water absorption feature, as illustrated by an M dwarf spectrum shown in brown, provides an excellent match to that observed in the LRD spectrum shown in blue.
    }
    \label{fig:mole}
\end{figure}

As an additional check on the nature of the absorption feature observed in the LRD spectra, we compare it with stellar spectra known to exhibit molecular absorption. Besides water, a potential contributor near this wavelength is the cyano radical (CN) spectral edge at $1.38~\mu$m, commonly associated with TP-AGB stars. 
Fig.~\ref{fig:mole} shows the spectrum of a carbon star \cite{Lancon2002}. While some blending with CN is possible, CN alone does not reproduce the observed feature in the LRD spectra. 
We therefore conclude that water is the most likely source of the absorption. We emphasize, however, that the primary conclusions of this work do not rely on the absorption being entirely due to water: any molecular absorption demonstrates the presence of cold, dense gas.

\section{Spectral Modeling}
\label{sec:method:model}

In this section, we present our spectral modeling for interpreting the newly detected water absorption in LRDs. We begin by placing the observed absorption strength in the context of standard stellar population synthesis (SPS) models with enhanced contributions from giants. 
The need for a significantly boosted giant component, together with the blackbody-like SEDs of LRDs, motivates us to fit the spectra with single-star atmospheric grids, which offers a quantitative handle on the fraction of the light coming from a low temperature component.
We then construct custom models at low gas densities to test whether the inferred low temperatures remain robust under non-stellar configurations. We also discuss the limitations inherent to current modeling approaches. Finally, we assess the potential for metallicity inference, pending the full construction of models with appropriate emission-line physics and populating the parameter space relevant for LRDs.

\subsection{Stellar Population Fit with Enhanced Giant Star Contributions}
\label{sec:method:prosp}

To place the strength of the observed water absorption in the context of stellar population synthesis, we generate single stellar population (SSP) models with varying contributions from giant stars using \texttt{FSPS} \cite{Conroy2010}. We then compare these models to the observed NIRSpec/\prism\ spectra of the three LRDs of this paper over the rest-frame wavelength range of 8000-15000~\AA\ to estimate the giant contribution that matches the data.

Additionally, to test whether the observed water features are consistent with those expected from giant stars, we perform stellar population modeling with enhanced giant contributions using the \prospector\ inference framework \cite{Johnson2021}, again restricting the fits to the rest-frame 8000–15000~\AA\ region.

From the SSPs, we find that the giant contribution needs to be enhanced by a factor of 10 in order to reproduce the water feature in \rdwide, the target with the strongest observed absorption in our sample.
With this significant enhancement, the composite stellar populations (CSPs) from the \prospector\ fits are indeed able to match the water feature in the LRDs.

We stress that this exercise is meant only to provide context for the strength of the water absorption in our LRDs. While a sum of stellar spectra may still provide a reasonable description of LRD continua (see discussion in the next section), the explicit up-weighting of giant stars deviates significantly from any standard initial mass function (IMF) such as \cite{Kroupa2001, Chabrier2003}, effectively adding a delta function to the IMF at a specific stellar mass.
Moreover, the time-evolution component of CSPs is not physically meaningful in this context, as parts of the LRD atmosphere are certainly not evolving into supernovae.
It is also noteworthy that, although an enhanced-giant model can reproduce the observed water feature, it simultaneously over-predicts the CaT absorption strength.

\subsection{Stellar Atmosphere Model}
\label{sec:method:stellar}

The clear detection of water absorption motivates a modeling approach that is akin to stellar atmosphere modeling; i.e., radiative transfer through a dense, warm layer of gas in hydrostatic equilibrium, specified by effective temperature, metallicity and surface gravity. 
Such an approach is, however, non-trivial for LRDs owing to e.g., uncertainties in geometry and density, as well as the complexity of molecules. Decades of development underpin stellar atmosphere modeling (see e.g., \cite{Marley2015} for a recent review), we thus aim to gain physical insights from existing stellar atmosphere models.

We use \phoenix\ stellar atmosphere models \cite{Allard2012}.
The observed spectrum is modeled as the sum of two components, which share the same priors spanning the full model parameter space:
effective temperature $\teff \in [2000, 5500]$~K, metallicity $\met \in [-4, 0.5]$, and surface gravity $\logg~/(\cms) \in [0, 5.5]$.
$\alpha$-abundance is scaled to solar.
At each likelihood evaluation, we linearly interpolate within the model grid. 
Each spectrum also has its normalization varying as a free parameter.

Uniform priors are used on all free parameters. We do not extrapolate outside of the \phoenix\ model grid, and so the boundaries are all set by the model grid.
The model spectra are smoothed with the LSF of the \prism\.
All emission lines in the observed spectra are masked and only the spectral region redward of the inflection point near the Balmer limit is included in the fit.
Finally, the best-fit is determined using the dynamic nested sampler \dynesty\ \cite{Speagle2020, Koposov2024}.

We emphasize again that this modeling is exploratory in nature. Our goal here is to use the available tools to gain insights into which physical parameters can be constrained and which spectral features are most sensitive to them, and thereby paving the way for comprehensive works in the future.

 \subsection{Potential Host Galaxy Contribution}
 \label{sec:method:host}

 A strong correlation between \oiii\ EW and Balmer-break strength is reported \cite{deGraaff:sample}, and interpreted as evidence that \oiii\ EW traces the host-galaxy contribution to the total observed light. Here we test whether taking the host contribution into account would remove the need for a multi-temperature interpretation. We do so by fitting single-temperature \phoenix\ models to spectra after applying a simple empirical host subtraction, i.e., the `black-hole star'$+$host scenario as referred to in \cite{Sun2026}.
 We assume that the \oiii\ emission arises entirely from the host galaxy, and identify candidate hosts by selecting an ensemble of real galaxies at similar redshift with comparable \oiii\ luminosities to each LRD. These spectra are then used to empirically model and subtract the host contribution.

\begin{figure}[ht]
 \centering
 \includegraphics[width=\textwidth]{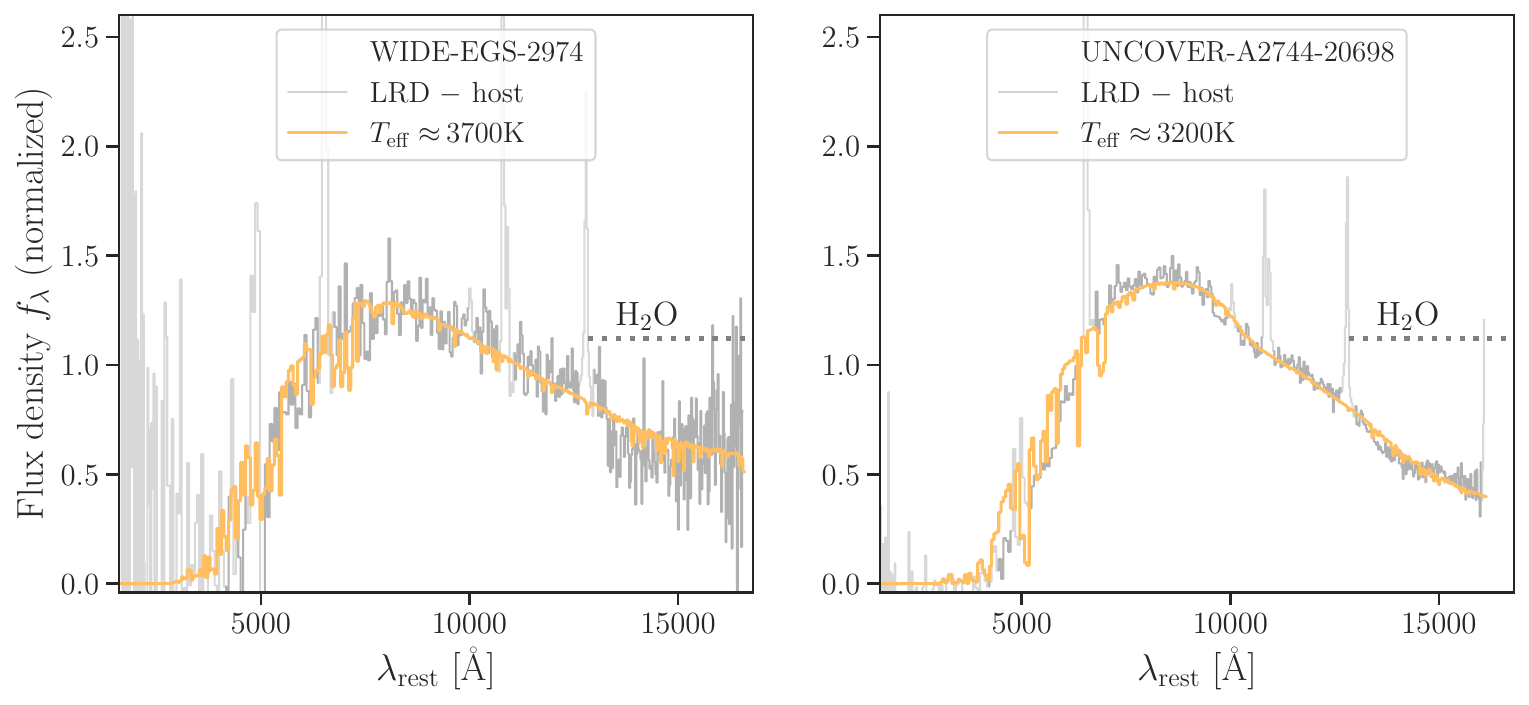}
    \caption{Single-temperature stellar atmosphere model fits to the host-subtracted LRD spectra. Notably the inferred temperatures are still too high for water to survive.
    }
    \label{fig:phoenix_host}
\end{figure}

Fig.~\ref{fig:phoenix_host} shows the resulting fits. Compared to single-temperature fits performed on the original LRD spectra, the inferred temperatures change by only $\sim$100–300~K and remain $>3000$~K at $\gg 3\sigma$ confidence---still too high for water to survive. In other words, these single-temperature models suffer from the same problem: they cannot simultaneously reproduce both the overall SED shape and the water absorption feature.
Moreover, the fits to the host-subtracted spectra are systematically worse in terms of $\chi^2$, likely reflecting the need for a multi-temperature structure and/or a more complex, non-standard host component.

Therefore, while we cannot yet rule out the presence of some exotic host contribution, adopting the reasonable assumption that \oiii\ traces the host and performing the subtraction as in \cite{Sun2026} does not alter the main conclusion: a single-temperature photosphere remains inadequate, and the data still require at least two temperature components.

\subsection{Low-density Atmosphere Model}
\label{sec:method:tlusty}

Recent radiative transfer calculations suggest that an atmosphere density far below the stellar regime is required to explain LRDs with strong Balmer breaks \cite{Liu2025}. As an order-of-magnitude estimate, we can infer the photosphere radius of the LRDs in this work from the Stefan-Boltzmann law, $R=\sqrt{L/4\pi\sigma T_{\rm eff}^4}=1100{\rm~AU}$, for $L\sim5\times10^{43}{\rm~erg~s^{-1}}$ and $T_{\rm eff}\sim4000{\rm~K}$.
If such an atmosphere is in hydrostatic equilibrium around a central black hole with mass below $2\times10^{6},M_\odot$, the implied surface gravity is very low, $\log(g/\mathrm{cm,s^{-2}})<0$.
Therefore, LRD models need to include the possibility of atmosphere densities (gravities) falling below the regime covered by current stellar atmosphere grids.

Thus, we use \tlusty\ (Version 208, \cite{Hubeny1988, Hubeny2021}) to construct atmosphere models at low $\logg$ and therefore low gas density. The open-source code has been widely used for both stellar (e.g., \cite{Lanz2007, Osorio2020}) and AGN accretion disk (e.g., \cite{Hubeny1998, Hubeny2001, Hui2005}) spectral modeling. It evaluates one-dimensional hydrostatic, radiative, and statistical equilibrium equations and outputs an atmosphere model as a function of effective temperature, surface gravity, and metallicity.
The code assumes a plane-parallel geometry, whose applicability is justified by the large LRD radius of $\sim10^3$~AU as expected from the luminosity and effective temperature.
In comparison, in all cases explored in this work, the atmosphere thickness at an optical depth of unity is limited to $\leq30$~AU. In addition, we assume local thermal equilibrium and adopt a metallicity of $Z=0.1~Z_\odot$, using the solar metallicity reported in \cite{Asplund2009}. The water line list is taken from the ExoMol database \cite{Polyansky2018} (with a rejection parameter of $-9$ as described in Appendix G in \cite{Hubeny2021}). The resulting model atmosphere is then passed to \texttt{synspec}, the companion code to \tlusty\ (Version 54, \cite{Hubeny1988, Hubeny2021}) for detailed spectral synthesis.

In this work, we explore a range of effective temperatures $T_{\rm eff}=2000, 3000, 4000$~K and surface gravities $\log(g/{\rm cm~s^{-2}})=-3, -2, -1, 0$. The temperature range is motivated by the SED shape of the LRDs and covers the destruction temperature of the water molecules. The surface gravity range extends beyond the lower limit of the \phoenix\ grid. A detailed parameter fitting to the observed spectra requires a full-scale parameter scan and will be addressed in future work \cite{Liu:inprep}. The purpose of the calculation here is to show that the water absorption feature does not change significantly with gravity over three orders of magnitude, which demonstrates that our conclusion of the low temperature inferred from the stellar atmosphere grid is robust against uncertainties in the atmosphere density of LRDs.

\subsection{Limitations}

As shown previously, a single stellar spectrum cannot reproduce the observed LRD spectra \cite{Ji2025:lord, deGraaff2025:cliff}.
However, analogous to a multi-phase ISM, if we allow a combination of stellar atmosphere models with different effective temperatures and potentially varying metallicities and surface gravities, it is possible to fit the observed LRD spectra, as we demonstrate here.
However, we note that formally (based on AIC/BIC) neither single temperature or multi-temperature fit is significantly preferred over the other; this means from the shape of the SED alone there is no need to have multiple temperature components. The key evidence, and the novel part of this paper, comes from the water absorption; water molecules can only form at $T<3000$~K, while in both single-temperature and multi-temperature fits, there exists a dominant blackbody component with $T >3000$~K. The discrepancy between the blackbody temperature and the water formation temperature comprises the core evidence requiring multi-temperature gas in these systems.

That said, we have not yet demonstrated that the entire SED of LRDs---including both the Balmer break, the rest-UV continuum, and emission lines---can be fully explained by such an approach.
The higher-temperature components inferred from our two-temperature photosphere model do not produce a break near the Balmer limit. A-type stars, initially proposed to explain the strong Balmer breaks seen in LRDs \cite{Labbe2023, Wang2024:ub}, exhibit surface temperatures of $\sim 7500-10{,}000$~K, too high for water molecules to survive. Even F-type stars, where weaker Balmer breaks can be seen, have temperatures of $\sim 6000-7500$~K, also too high.
Reproducing both the Balmer break and the rest optical-infrared continuum with stellar models alone would thus require a third, hotter component.
It remains possible, though unclear in current theoretical models, that a more appropriate non-stellar atmosphere could reproduce the full SED with a single characteristic temperature.
A thorough exploration of alternatives is beyond the scope of this work. Future progress will require constructing denser model grids extending to low densities and re-examining the underlying physics, as key assumptions of classical stellar atmospheres (e.g., hydrostatic equilibrium) may not hold in LRDs. Such model development will yield quantitative predictions testable with deep, higher-resolution spectroscopy.

\subsection{Assessment for Low Metallicity}
\label{sec:method:lowz}

A low metallicity of $< 10^{-2}~\zsol$ is reported for \rdqso, a triply imaged LRD at $z=7.04$ \cite{Furtak2024}, based on the weakness of the \oiii\ emission line relative to the narrow \hb\ emission \cite{Maiolino2025:metpoor}.
However, \oiii\ could be weak from collisional de-excitation in high gas densities. Adding to the puzzle, strong CaT absorption has been reported in a local LRD \cite{Lin2025}, pointing to metal-enriched gas envelopes. Fe\,\textsc{ii} emission appears to be ubiquitous in deep LRD spectra \cite{Labbe2024, Lambrides2025, Torralba2025}; while less useful as a metallicity tracer, it indicates the presence of iron in the atmosphere.

In our analysis, all fits favor extremely low metallicities, $< 10^{-2}~\zsol$.
To assess the cause for such preference, we perform the same fits as described in Methods \S~\ref{sec:method:stellar}, but with $\met \in [-1, 0.5]$.
Adopting this higher metallicity prior ($> 10^{-1}~\zsol$) produces systematically poorer fits across the sample, as quantified by likelihoods. We explore the driver of this preference below.

First, we look for other signs of low metallicity. CaT is a well-established metallicity tracer in stellar populations (e.g., \cite{Armandroff1988, Carrera2013}). However, the CaT region can be significantly affected by emission-line infilling, which would weaken the observed absorption and masquerade as low metallicity.
An empirical, though not definitive, way to assess the extent of line infilling is to search for \oi\ emission.
\oi\ and CaT in emission are expected to come from similar gas conditions in the broad line region \cite{Inayoshi2022}.
We find no evidence of \oi\ emission in any of the three LRDs analyzed here, in contrast to detections reported in e.g., \cite{Labbe2024, Tripodi2025, Wang2025:brd}, though similar to the lack of detection in \cite{deGraaff2025:cliff}.
Therefore, the weak CaT absorption observed in all three spectra hints at extremely low metallicity, though without dedicated models and calibrations, conclusions cannot yet be drawn.

On the modeling side, with the temperature constrained by the water absorption features, the suite of metal and molecular absorption lines \cite{Kirkpatrick1991} drives the low metallicity solutions. Models with high metallicity predict excessively strong continuum absorption at rest wavelengths $\lesssim 10{,}000$~\AA, inconsistent with observations.
In current models, strong absorption is almost exclusively associated with $Z > 10^{-1}~\zsol$, whereas weak absorption demands extremely metal-poor conditions. However, surface gravity introduces degeneracies that complicate precise metallicity determination.

While we do not think that the available data and models are sufficient for a determination of metallicity, current stellar atmosphere models suggest low metallicity, with little density dependence. However, this must be corroborated by later models. If the low metallicity of LRDs at $z \sim 2$ is confirmed, it would be surprising, given that by cosmic noon the gas reservoirs of typical star-forming galaxies are expected to have undergone substantial enrichment from multiple generations of stars. In this scenario, these LRDs would need to reside in extremely metal-poor gas clouds that have either remained largely pristine---enriched only by Population III and/or Population II stars---or have been replenished by unenriched material from the intergalactic medium.

\section{Possible Geometrical Configurations}
\label{sec:method:geo}

The detection of water absorption provides new insights into the geometries of LRDs by demanding a cold gas component.
In addition, given a reasonable expectation that a rotationally supported configuration rotates at velocities near the circular velocity at the photospheric radius (or comparable to the widths of the emission lines) of a few-thousand $\kms$, such rotation at would imprint a kinematic signature on the absorption profile. Thus, deeper, higher-resolution spectroscopy of the water feature could in principle distinguish between the fundamental modes supporting the LRD photosphere: hydrostatic pressure versus rotation.
Here, we simple conclude the discussion by postulating possible configurations based mainly on the rest optical spectra.

\begin{figure}[ht!]
 \centering
 \includegraphics[width=\textwidth]{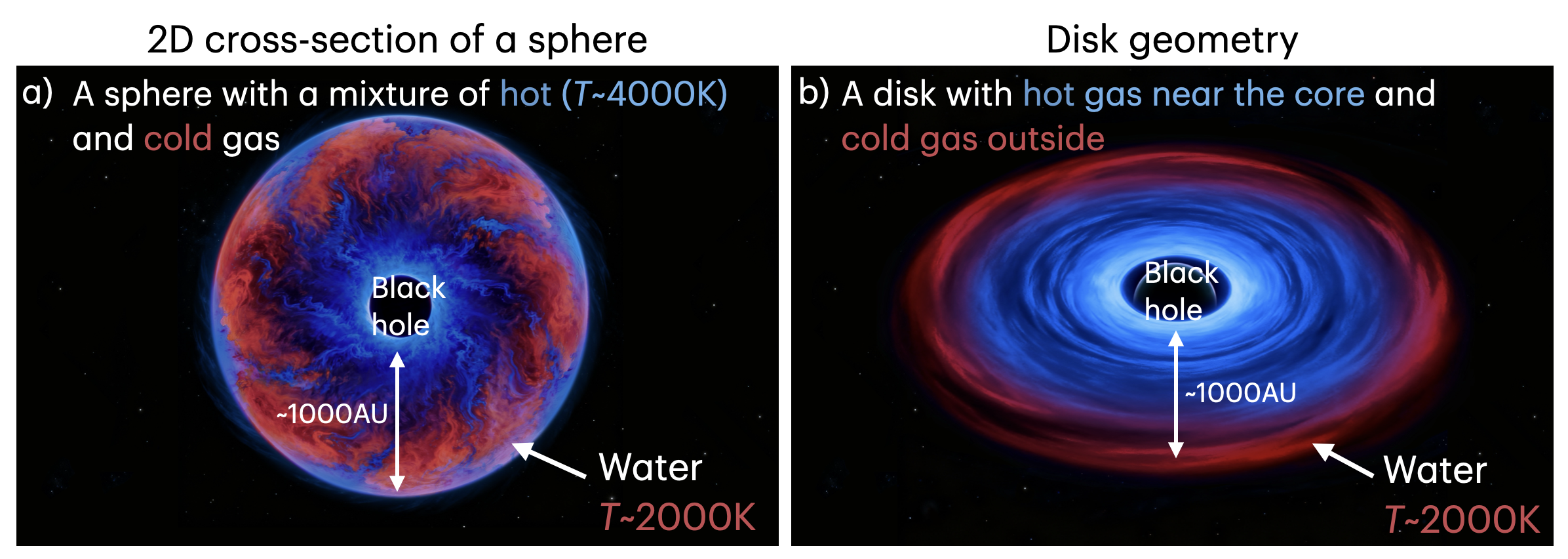}
    \caption{Schematic illustration of two of the possible geometrical configurations for LRDs. In both panels, $T \sim 2000$~K cold components are indicated by red colors, whereas $T \sim 4000$~K hot components are indicated by blue colors. 
    (a) A spherical configuration, where gas of different temperatures mixes, analogous to convection cells.
    (b) A disk geometry, with temperature decreasing from the hot inner regions to the cooler outer zones. In this scenario, the observed range of temperatures among LRDs could arise from variations in viewing angles along the line of sight. 
    }
    \label{fig:geo}
\end{figure}

Fig.~\ref{fig:geo} illustrates two of the possible geometries.
The first is a spherical structure in which hotter, $T \gtrsim 4000$~K gas resides near the center and is surrounded by cooler $T \lesssim 3000$~K outer layers. For a spherical geometry, the requirement of a hotter model component to fit the data implies that some hot gas must reach the outer regions.
Such structure could arise from processes analogous to convection, or from the highly turbulent environments inferred in previous studies, the latter being necessary to produce a smooth Balmer break with a non-stellar origin \cite{deGraaff2025:cliff, Ji2025:qso, Naidu2025}.

Stars provide an instructive analogy for the former case, and is perhaps of particular relevance in the context of quasi-star–like hydrostatic models for LRDs. In red supergiants and AGB stars, substantial temperature fluctuations at the level of several hundred kelvin are observed and are generally attributed to strong surface convection \cite{Freytag2002, Chiavassa2010}. 
The large convective cells associated with low \logg\ means that there is less geometric averaging over many cells as happens in higher \logg\ objects.
Interferometric observations have resolved surface inhomogeneities in such stars, for instance, Betelgeuse, which experienced significant dimming recently, suggesting a contrasted and rapidly changing photosphere \cite{Young2000, Montarges2021}. 
Sunspots provide another interesting case: strong magnetic fields inhibit convection, preventing heat from reaching the solar surface. As a result, sunspots exhibit temperatures of only $\sim$3000–4000 K---far cooler than the surrounding photosphere---and are known to have water absorption in their spectra \cite{Hall1970, Polyansky1997}.

The right panel of Fig.~\ref{fig:geo} shows LRDs as having a disk-like geometry, as proposed in e.g., \cite{Inayoshi2025:binary, Zhang2025, Zwick2025}. 
The presence of water absorption features potentially points to a strong radial temperature gradient, wherein the temperature drops sharply from the hot interior to cooler outer zones. In this scenario, the observed range of temperatures among LRDs could result from variations in viewing angle along different lines of sight.
However, it is unclear how disk models can be tuned to produce a continuum that peaks at $\sim 4000-5000$~K while simultaneously hosting a subdominant component at $T \sim 2000$~K. For gravitationally unstable disks, the emergent emission is expected to be dominated by the coldest annulus; naively, the presence of a $\sim 2000$~K component would therefore imply a continuum brighter than that from a $\sim 4000$~K region, contrary to what is observed. One may instead imagine a disk coexisting with an envelope, analogous to Be stars \cite{Hillenbrand1992, Rivinius2013}.

In addition, we note that water absorption can arise in models with extended atmospheres. 
For red giant and supergiant stars, several studies have argued for an empirical concept called MOLsphere---warm, semidetached, molecular layers surrounding the photosphere \cite{Matsuura1999, Tsuji2000}. This model predicts water in emission at $\lambda_{\rm rest}> 6 ~\mu$m, in addition to the absorption bands in the near-infrared as observed in the two LRDs of this work.
Observational evidence for such structures has been reported in stars, although results are mixed \cite{Ryde2006}.
In this framework, or more broadly, in models with smooth temperature gradients, one might expect titanium oxide (TiO) absorption bands to appear prior to detectable water absorption, since TiO can survive at higher temperatures. Yet, we find no evidence for TiO features in the current LRD spectra.
This absence is likely not problematic for the configurations discussed above, as the optical regime is dominated by emission at effective temperatures exceeding those at which molecules can survive. It is worth emphasizing that the lack of prominent molecular absorption in the rest optical appears to suggest a steep temperature gradient: the $\sim$2000 K component contributes substantially to the emergent flux in order for water absorption to be detectable, while any intermediate-temperature ($\sim$3000~K) region likely occupies a much smaller emitting area compared to the $\sim 2000$~K and $\sim 4000$~K regions. Such a configuration is physically possible, but it is currently not straightforward to realize from a theoretical perspective.

Finally, we leave open the possibility of a single-temperature LRD scenario, in which the apparent multi-temperature SED arises from additional components rather than from a multi-temperature photosphere. We consider two broad classes of configurations as follows.

A significant contribution from stellar populations may be present. The UV emission can originate from hot ($>10^4$~K), young stars and extends into the rest optical, while a emission from a $T~\sim2000$~K LRD photosphere dominates $\gtrsim 1~\mu$m so that water absorption is detectable. Depending on the density, such a configuration can point to a host galaxy, or be reminiscent of a nuclear star cluster scenario (e.g., \cite{Kritos2025, Inayoshi2025:stars, Nandal2025}). Dense and massive stellar systems are commonly observed at the centers of galaxies (see, e.g., \cite{Neumayer2020} for a review). A notable example is the nuclear star cluster surrounding the central black hole Sgr A* (e.g., \cite{Bartko2010, Genzel2010}). Theoretical studies further suggest that the outer regions of a standard steady-state accretion disk around a massive black hole can become gravitationally unstable and fragment into stars, forming massive stellar populations \cite{Goodman2003}. This naturally produces a softer ionizing spectrum, in line with recent study of the ionization properties of LRDs \cite{Wang2025:qion}.
However, while this scenario can plausibly account for the UV emission, it remains unclear how it would also generate the additional component peaking at $T \sim 4000$~K required by the observed rest optical continuum of LRDs.

Alternatively, the observed SED could in principle arise from two LRDs, one characterized by $T \sim 4000$~K and the other by $T \sim 2000$~K. This could correspond to either a binary system or two LRDs being embedded within the same host galaxy. Dual LRD candidates are reported in \cite{Tanaka2024}, and a pair of LRDs separated by $\sim 70$~pc is found to reside in a star-forming galaxy \cite{Yanagisawa2026}. A close binary, however, would likely be accompanied by additional, potentially exotic signatures, such as variability or interaction effects (e.g., Roche-lobe overflow).

All these geometries lead to different physical interpretations of LRDs including host galaxy luminosity and structure of the central engine and the gas envelope. We can potentially test these scenarios by searching for correlations between inferred temperatures and viewing angle indicators such as line kinematics and polarization, by examining the velocity structure and spatial extent of emission and absorption lines with higher-resolution spectroscopy, and by probing for variability that may distinguish compact accretion-driven configurations from more extended stellar-like ones.

\clearpage

\counterwithin{figure}{section}
\counterwithin{table}{section}
\renewcommand{\thesection}{\Alph{section}}

\section*{Extended Data}

\setcounter{section}{5} 

\clearpage

\begin{table*}[t]
\caption{Key Parameters}\label{tab:sample_mea}
\centering
\begin{tabular}{@{}llll@{}}
\toprule
 & \rdwide & \rduncover \\
\midrule
Measurements\\
\midrule
RA (deg)  & 215.126 & 3.557 \\
Dec (deg) & 53.048  & $-30.408$ \\
$\zspec$  & 2.32    & 2.42 \\
$\rwater$ & $-0.32 \pm 0.04$ & $-0.20 \pm 0.01$\\
\midrule

Model inferred parameters \\
\midrule
$T_{\rm eff, cold}$ (K) 
& $2036.82^{+60.91}_{-22.75}$ 
& $2299.89^{+3.02}_{-3.65}$ \\

$[\mathrm{M/H}]_{\rm cold}$ ($\mathrm{log}\,\zsol$) 
& $-2.51^{+0.14}_{-0.20}$ 
& $-3.30^{+0.01}_{-0.01}$ \\

$\log g_{\rm cold}$ ($\mathrm{log}\, \,\mathrm{cm}\,\mathrm{s}^{-2}$) 
& $1.37^{+2.17}_{-0.23}$ 
& $0.50^{+0.01}_{-0.01}$ \\

$T_{\rm eff, hot}$ (K) 
& $4297.54^{+65.10}_{-56.23}$ 
& $3676.76^{+8.49}_{-8.91}$ \\

$[\mathrm{M/H}]_{\rm hot}$ ($\mathrm{log}\,\zsol$) 
& $-3.70^{+0.32}_{-0.26}$ 
& $-4.00^{+0.01}_{-0.00}$ \\

$\log g_{\rm hot}$ ($\mathrm{log}\, \,\mathrm{cm}\,\mathrm{s}^{-2}$) 
& $3.50^{+1.61}_{-1.64}$ 
& $3.06^{+0.02}_{-0.03}$ \\
\botrule
\end{tabular}
\end{table*}

\begin{table}[h]
\caption{Emission Lines.}\label{tab:lines}%
\begin{tabular}{@{}llllll@{}}
\toprule
 & \rdwide & \rduncover \\
\midrule
EW [\AA] (total) \\
\ha\ &     $609_{-25}^{+24}$ &  $348_{-15}^{+11}$ &  \\
\hb\ &     $219_{-21}^{+24}$ &  $123_{-14}^{+16}$ &  \\
\niia\ &   $9_{-2}^{+2}$ &      $2_{-2}^{+4}$ &      \\
\niib\ &   $26_{-6}^{+6}$ &     $7_{-5}^{+11}$ &     \\
\oiiia\ &  $198_{-10}^{+9}$ &   $136_{-7}^{+8}$ &    \\
\oiiib\ &  $582_{-28}^{+28}$ &  $392_{-20}^{+22}$ &  \\
\midrule
Flux [$10^{-20} \, \cgs$] (total) \\
\ha\ &     $2551_{-68}^{+60}$ &  $3627_{-161}^{+98}$ &  \\
\hb\ &     $747_{-61}^{+63}$ &   $492_{-47}^{+49}$ &    \\
\niia\ &   $44_{-9}^{+10}$ &     $23_{-18}^{+39}$ &     \\
\niib\ &   $129_{-27}^{+29}$ &   $68_{-54}^{+116}$ &    \\
\oiiia\ &  $685_{-16}^{+15}$ &   $583_{-18}^{+19}$ &    \\
\oiiib\ &  $2042_{-48}^{+44}$ &  $1738_{-54}^{+55}$ &   \\
\midrule
FWHM [$\kms$]\\
Broad lines &  $1416_{-164}^{+164}$ &  $1664_{-624}^{+772}$ &  \\
\botrule
\end{tabular}
\end{table}

\clearpage

\bmhead{Data availability} The JWST/NIRSpec data are available in the DAWN JWST Archive \cite{dja}.
The sample of this paper was observed as part of the following programs: JWST-GTO-1213 (PI N. Luetzgendorf; \cite{Maseda2024}), and JWST-GO-2561 (PIs I. Labb\'{e} \& R. Bezanson; \cite{Bezanson2024, Price2025}).

\bmhead{Code availability} Publicly available codes and standard data reduction tools in the Python environments were used: msaexp \cite{Brammer2022}, prospector \cite{Johnson2021}, unite \cite{Hviding2025}.

\bmhead{Acknowledgements}
B.W. thanks Adam Burrows and Christian Jespersen for helpful discussions on molecular absorption.
B.W. also thanks Yiwei Sun for help with the schematic diagrams.
H.L. thanks Ivan Hubeny for providing the water line list for the low-density atmosphere models.
B.W. acknowledges support provided by NASA through Hubble Fellowship grant HST-HF2-51592.001 awarded by the Space Telescope Science Institute, which is operated by the Association of Universities for Research in Astronomy, In., for NASA, under the contract NAS 5-26555. IL acknowledges support by the Australian Research Council through Future Fellowship FT220100798.
S.F. acknowledges support from the Dunlap Institute, funded through an endowment established by the David Dunlap family and the University of Toronto.
J.M.H. acknowledges support from the JWST Near Infrared Camera (NIRCam) Science Team Lead, NAS5-02105, from NASA Goddard Space Flight Center to the University of Arizona; J.M.H. also acknowledges support from JWST Programs \#3215 and \#8544.
D.S. and J.R.W. acknowledges that support for this work was provided by The Brinson Foundation through a Brinson Prize Fellowship grant.
A.Z. acknowledges support by the Israel Science Foundation Grant No. 864/23.
Computations for this research were performed on the Pennsylvania State University's Institute for Computational and Data Sciences' Roar supercomputer;
and on computational resources managed and supported by Princeton Research Computing, a consortium of groups including the Princeton Institute for Computational Science and Engineering (PICSciE) and Research Computing at Princeton University.
This publication made use of the NASA Astrophysical Data System for bibliographic information.

\bmhead{Author contributions}
B.W. led the analysis, made the figures, and wrote the manuscript. J.L., I.L., and J.E.G. offered high-level advice throughout the work. I.L. first identified the UNCOVER source, revised and improved the manuscript. 
H.L. developed the low-density atmosphere models. A.d.G. and R.E.H. compiled the parent sample. All authors provided feedback and helped shape the analysis and manuscript.

\bmhead{Competing interests}
The authors declare that they have no competing financial interests.

\bmhead{Materials \& correspondence}
Correspondence and requests for materials should be addressed to B.W. (email: bjwang@princeton.edu).

\clearpage

\bibliography{lrd_h2o_wang.bib}


\begin{thebibliography}{104}
\ifx \bisbn   \undefined \def \bisbn  #1{ISBN #1}\fi
\ifx \binits  \undefined \def \binits#1{#1}\fi
\ifx \bauthor  \undefined \def \bauthor#1{#1}\fi
\ifx \batitle  \undefined \def \batitle#1{#1}\fi
\ifx \bjtitle  \undefined \def \bjtitle#1{#1}\fi
\ifx \bvolume  \undefined \def \bvolume#1{\textbf{#1}}\fi
\ifx \byear  \undefined \def \byear#1{#1}\fi
\ifx \bissue  \undefined \def \bissue#1{#1}\fi
\ifx \bfpage  \undefined \def \bfpage#1{#1}\fi
\ifx \blpage  \undefined \def \blpage #1{#1}\fi
\ifx \burl  \undefined \def \burl#1{\textsf{#1}}\fi
\ifx \doiurl  \undefined \def \doiurl#1{\url{https://doi.org/#1}}\fi
\ifx \betal  \undefined \def \betal{\textit{et al.}}\fi
\ifx \binstitute  \undefined \def \binstitute#1{#1}\fi
\ifx \binstitutionaled  \undefined \def \binstitutionaled#1{#1}\fi
\ifx \bctitle  \undefined \def \bctitle#1{#1}\fi
\ifx \beditor  \undefined \def \beditor#1{#1}\fi
\ifx \bpublisher  \undefined \def \bpublisher#1{#1}\fi
\ifx \bbtitle  \undefined \def \bbtitle#1{#1}\fi
\ifx \bedition  \undefined \def \bedition#1{#1}\fi
\ifx \bseriesno  \undefined \def \bseriesno#1{#1}\fi
\ifx \blocation  \undefined \def \blocation#1{#1}\fi
\ifx \bsertitle  \undefined \def \bsertitle#1{#1}\fi
\ifx \bsnm \undefined \def \bsnm#1{#1}\fi
\ifx \bsuffix \undefined \def \bsuffix#1{#1}\fi
\ifx \bparticle \undefined \def \bparticle#1{#1}\fi
\ifx \barticle \undefined \def \barticle#1{#1}\fi
\bibcommenthead
\ifx \bconfdate \undefined \def \bconfdate #1{#1}\fi
\ifx \botherref \undefined \def \botherref #1{#1}\fi
\ifx \url \undefined \def \url#1{\textsf{#1}}\fi
\ifx \bchapter \undefined \def \bchapter#1{#1}\fi
\ifx \bbook \undefined \def \bbook#1{#1}\fi
\ifx \bcomment \undefined \def \bcomment#1{#1}\fi
\ifx \oauthor \undefined \def \oauthor#1{#1}\fi
\ifx \citeauthoryear \undefined \def \citeauthoryear#1{#1}\fi
\ifx \endbibitem  \undefined \def \endbibitem {}\fi
\ifx \bconflocation  \undefined \def \bconflocation#1{#1}\fi
\ifx \arxivurl  \undefined \def \arxivurl#1{\textsf{#1}}\fi
\csname PreBibitemsHook\endcsname

\bibitem[\protect\citeauthoryear{{Kocevski} et~al.}{2023}]{Kocevski2023}
\begin{barticle}
\bauthor{\bsnm{{Kocevski}}, \binits{D.D.}},
\bauthor{\bsnm{{Onoue}}, \binits{M.}},
\bauthor{\bsnm{{Inayoshi}}, \binits{K.}},
\bauthor{\bsnm{{Trump}}, \binits{J.R.}},
\bauthor{\bsnm{{Arrabal Haro}}, \binits{P.}},
\bauthor{\bsnm{{Grazian}}, \binits{A.}},
\bauthor{\bsnm{{Dickinson}}, \binits{M.}},
\bauthor{\bsnm{{Finkelstein}}, \binits{S.L.}},
\bauthor{\bsnm{{Kartaltepe}}, \binits{J.S.}},
\bauthor{\bsnm{{Hirschmann}}, \binits{M.}},
\bauthor{\bsnm{{Aird}}, \binits{J.}},
\bauthor{\bsnm{{Holwerda}}, \binits{B.W.}},
\bauthor{\bsnm{{Fujimoto}}, \binits{S.}},
\bauthor{\bsnm{{Juneau}}, \binits{S.}},
\bauthor{\bsnm{{Amor{\'\i}n}}, \binits{R.O.}},
\bauthor{\bsnm{{Backhaus}}, \binits{B.E.}},
\bauthor{\bsnm{{Bagley}}, \binits{M.B.}},
\bauthor{\bsnm{{Barro}}, \binits{G.}},
\bauthor{\bsnm{{Bell}}, \binits{E.F.}},
\bauthor{\bsnm{{Bisigello}}, \binits{L.}},
\bauthor{\bsnm{{Calabr{\`o}}}, \binits{A.}},
\bauthor{\bsnm{{Cleri}}, \binits{N.J.}},
\bauthor{\bsnm{{Cooper}}, \binits{M.C.}},
\bauthor{\bsnm{{Ding}}, \binits{X.}},
\bauthor{\bsnm{{Grogin}}, \binits{N.A.}},
\bauthor{\bsnm{{Ho}}, \binits{L.C.}},
\bauthor{\bsnm{{Hutchison}}, \binits{T.A.}},
\bauthor{\bsnm{{Inoue}}, \binits{A.K.}},
\bauthor{\bsnm{{Jiang}}, \binits{L.}},
\bauthor{\bsnm{{Jones}}, \binits{B.}},
\bauthor{\bsnm{{Koekemoer}}, \binits{A.M.}},
\bauthor{\bsnm{{Li}}, \binits{W.}},
\bauthor{\bsnm{{Li}}, \binits{Z.}},
\bauthor{\bsnm{{McGrath}}, \binits{E.J.}},
\bauthor{\bsnm{{Molina}}, \binits{J.}},
\bauthor{\bsnm{{Papovich}}, \binits{C.}},
\bauthor{\bsnm{{P{\'e}rez-Gonz{\'a}lez}}, \binits{P.G.}},
\bauthor{\bsnm{{Pirzkal}}, \binits{N.}},
\bauthor{\bsnm{{Wilkins}}, \binits{S.M.}},
\bauthor{\bsnm{{Yang}}, \binits{G.}},
\bauthor{\bsnm{{Yung}}, \binits{L.Y.A.}}:
\batitle{{Hidden Little Monsters: Spectroscopic Identification of Low-mass,
  Broad-line AGNs at z > 5 with CEERS}}.
\bjtitle{\apjl}
\bvolume{954}(\bissue{1}),
\bfpage{4}
(\byear{2023})
\doiurl{10.3847/2041-8213/ace5a0}
{\href{https://arxiv.org/abs/2302.00012}{{arXiv:2302.00012}}}
{[astro-ph.GA]}
\end{barticle}
\endbibitem

\bibitem[\protect\citeauthoryear{{Matthee} et~al.}{2024}]{Matthee2024}
\begin{barticle}
\bauthor{\bsnm{{Matthee}}, \binits{J.}},
\bauthor{\bsnm{{Naidu}}, \binits{R.P.}},
\bauthor{\bsnm{{Brammer}}, \binits{G.}},
\bauthor{\bsnm{{Chisholm}}, \binits{J.}},
\bauthor{\bsnm{{Eilers}}, \binits{A.-C.}},
\bauthor{\bsnm{{Goulding}}, \binits{A.}},
\bauthor{\bsnm{{Greene}}, \binits{J.}},
\bauthor{\bsnm{{Kashino}}, \binits{D.}},
\bauthor{\bsnm{{Labbe}}, \binits{I.}},
\bauthor{\bsnm{{Lilly}}, \binits{S.J.}},
\bauthor{\bsnm{{Mackenzie}}, \binits{R.}},
\bauthor{\bsnm{{Oesch}}, \binits{P.A.}},
\bauthor{\bsnm{{Weibel}}, \binits{A.}},
\bauthor{\bsnm{{Wuyts}}, \binits{S.}},
\bauthor{\bsnm{{Xiao}}, \binits{M.}},
\bauthor{\bsnm{{Bordoloi}}, \binits{R.}},
\bauthor{\bsnm{{Bouwens}}, \binits{R.}},
\bauthor{\bsnm{{van Dokkum}}, \binits{P.}},
\bauthor{\bsnm{{Illingworth}}, \binits{G.}},
\bauthor{\bsnm{{Kramarenko}}, \binits{I.}},
\bauthor{\bsnm{{Maseda}}, \binits{M.V.}},
\bauthor{\bsnm{{Mason}}, \binits{C.}},
\bauthor{\bsnm{{Meyer}}, \binits{R.A.}},
\bauthor{\bsnm{{Nelson}}, \binits{E.J.}},
\bauthor{\bsnm{{Reddy}}, \binits{N.A.}},
\bauthor{\bsnm{{Shivaei}}, \binits{I.}},
\bauthor{\bsnm{{Simcoe}}, \binits{R.A.}},
\bauthor{\bsnm{{Yue}}, \binits{M.}}:
\batitle{{Little Red Dots: An Abundant Population of Faint Active Galactic
  Nuclei at z {\ensuremath{\sim}} 5 Revealed by the EIGER and FRESCO JWST
  Surveys}}.
\bjtitle{\apj}
\bvolume{963}(\bissue{2}),
\bfpage{129}
(\byear{2024})
\doiurl{10.3847/1538-4357/ad2345}
{\href{https://arxiv.org/abs/2306.05448}{{arXiv:2306.05448}}}
{[astro-ph.GA]}
\end{barticle}
\endbibitem

\bibitem[\protect\citeauthoryear{{Labbe} et~al.}{2025}]{Labbe2025}
\begin{barticle}
\bauthor{\bsnm{{Labbe}}, \binits{I.}},
\bauthor{\bsnm{{Greene}}, \binits{J.E.}},
\bauthor{\bsnm{{Bezanson}}, \binits{R.}},
\bauthor{\bsnm{{Fujimoto}}, \binits{S.}},
\bauthor{\bsnm{{Furtak}}, \binits{L.J.}},
\bauthor{\bsnm{{Goulding}}, \binits{A.D.}},
\bauthor{\bsnm{{Matthee}}, \binits{J.}},
\bauthor{\bsnm{{Naidu}}, \binits{R.P.}},
\bauthor{\bsnm{{Oesch}}, \binits{P.A.}},
\bauthor{\bsnm{{Atek}}, \binits{H.}},
\bauthor{\bsnm{{Brammer}}, \binits{G.}},
\bauthor{\bsnm{{Chemerynska}}, \binits{I.}},
\bauthor{\bsnm{{Coe}}, \binits{D.}},
\bauthor{\bsnm{{Cutler}}, \binits{S.E.}},
\bauthor{\bsnm{{Dayal}}, \binits{P.}},
\bauthor{\bsnm{{Feldmann}}, \binits{R.}},
\bauthor{\bsnm{{Franx}}, \binits{M.}},
\bauthor{\bsnm{{Glazebrook}}, \binits{K.}},
\bauthor{\bsnm{{Leja}}, \binits{J.}},
\bauthor{\bsnm{{Maseda}}, \binits{M.}},
\bauthor{\bsnm{{Marchesini}}, \binits{D.}},
\bauthor{\bsnm{{Nanayakkara}}, \binits{T.}},
\bauthor{\bsnm{{Nelson}}, \binits{E.J.}},
\bauthor{\bsnm{{Pan}}, \binits{R.}},
\bauthor{\bsnm{{Papovich}}, \binits{C.}},
\bauthor{\bsnm{{Price}}, \binits{S.H.}},
\bauthor{\bsnm{{Suess}}, \binits{K.A.}},
\bauthor{\bsnm{{Wang}}, \binits{B.}},
\bauthor{\bsnm{{Weaver}}, \binits{J.R.}},
\bauthor{\bsnm{{Whitaker}}, \binits{K.E.}},
\bauthor{\bsnm{{Williams}}, \binits{C.C.}},
\bauthor{\bsnm{{Zitrin}}, \binits{A.}}:
\batitle{{UNCOVER: Candidate Red Active Galactic Nuclei at 3 < z < 7 with JWST
  and ALMA}}.
\bjtitle{\apj}
\bvolume{978}(\bissue{1}),
\bfpage{92}
(\byear{2025})
\doiurl{10.3847/1538-4357/ad3551}
{\href{https://arxiv.org/abs/2306.07320}{{arXiv:2306.07320}}}
{[astro-ph.GA]}
\end{barticle}
\endbibitem

\bibitem[\protect\citeauthoryear{{Furtak} et~al.}{2024}]{Furtak2024}
\begin{barticle}
\bauthor{\bsnm{{Furtak}}, \binits{L.J.}},
\bauthor{\bsnm{{Labb{\'e}}}, \binits{I.}},
\bauthor{\bsnm{{Zitrin}}, \binits{A.}},
\bauthor{\bsnm{{Greene}}, \binits{J.E.}},
\bauthor{\bsnm{{Dayal}}, \binits{P.}},
\bauthor{\bsnm{{Chemerynska}}, \binits{I.}},
\bauthor{\bsnm{{Kokorev}}, \binits{V.}},
\bauthor{\bsnm{{Miller}}, \binits{T.B.}},
\bauthor{\bsnm{{Goulding}}, \binits{A.D.}},
\bauthor{\bsnm{{de Graaff}}, \binits{A.}},
\bauthor{\bsnm{{Bezanson}}, \binits{R.}},
\bauthor{\bsnm{{Brammer}}, \binits{G.B.}},
\bauthor{\bsnm{{Cutler}}, \binits{S.E.}},
\bauthor{\bsnm{{Leja}}, \binits{J.}},
\bauthor{\bsnm{{Pan}}, \binits{R.}},
\bauthor{\bsnm{{Price}}, \binits{S.H.}},
\bauthor{\bsnm{{Wang}}, \binits{B.}},
\bauthor{\bsnm{{Weaver}}, \binits{J.R.}},
\bauthor{\bsnm{{Whitaker}}, \binits{K.E.}},
\bauthor{\bsnm{{Atek}}, \binits{H.}},
\bauthor{\bsnm{{Bogd{\'a}n}}, \binits{{\'A}.}},
\bauthor{\bsnm{{Charlot}}, \binits{S.}},
\bauthor{\bsnm{{Curtis-Lake}}, \binits{E.}},
\bauthor{\bsnm{{van Dokkum}}, \binits{P.}},
\bauthor{\bsnm{{Endsley}}, \binits{R.}},
\bauthor{\bsnm{{Feldmann}}, \binits{R.}},
\bauthor{\bsnm{{Fudamoto}}, \binits{Y.}},
\bauthor{\bsnm{{Fujimoto}}, \binits{S.}},
\bauthor{\bsnm{{Glazebrook}}, \binits{K.}},
\bauthor{\bsnm{{Juneau}}, \binits{S.}},
\bauthor{\bsnm{{Marchesini}}, \binits{D.}},
\bauthor{\bsnm{{Maseda}}, \binits{M.V.}},
\bauthor{\bsnm{{Nelson}}, \binits{E.}},
\bauthor{\bsnm{{Oesch}}, \binits{P.A.}},
\bauthor{\bsnm{{Plat}}, \binits{A.}},
\bauthor{\bsnm{{Setton}}, \binits{D.J.}},
\bauthor{\bsnm{{Stark}}, \binits{D.P.}},
\bauthor{\bsnm{{Williams}}, \binits{C.C.}}:
\batitle{{A high black-hole-to-host mass ratio in a lensed AGN in the early
  Universe}}.
\bjtitle{\nat}
\bvolume{628}(\bissue{8006}),
\bfpage{57}--\blpage{61}
(\byear{2024})
\doiurl{10.1038/s41586-024-07184-8}
{\href{https://arxiv.org/abs/2308.05735}{{arXiv:2308.05735}}}
{[astro-ph.GA]}
\end{barticle}
\endbibitem

\bibitem[\protect\citeauthoryear{{Greene} et~al.}{2024}]{Greene2024}
\begin{barticle}
\bauthor{\bsnm{{Greene}}, \binits{J.E.}},
\bauthor{\bsnm{{Labbe}}, \binits{I.}},
\bauthor{\bsnm{{Goulding}}, \binits{A.D.}},
\bauthor{\bsnm{{Furtak}}, \binits{L.J.}},
\bauthor{\bsnm{{Chemerynska}}, \binits{I.}},
\bauthor{\bsnm{{Kokorev}}, \binits{V.}},
\bauthor{\bsnm{{Dayal}}, \binits{P.}},
\bauthor{\bsnm{{Volonteri}}, \binits{M.}},
\bauthor{\bsnm{{Williams}}, \binits{C.C.}},
\bauthor{\bsnm{{Wang}}, \binits{B.}},
\bauthor{\bsnm{{Setton}}, \binits{D.J.}},
\bauthor{\bsnm{{Burgasser}}, \binits{A.J.}},
\bauthor{\bsnm{{Bezanson}}, \binits{R.}},
\bauthor{\bsnm{{Atek}}, \binits{H.}},
\bauthor{\bsnm{{Brammer}}, \binits{G.}},
\bauthor{\bsnm{{Cutler}}, \binits{S.E.}},
\bauthor{\bsnm{{Feldmann}}, \binits{R.}},
\bauthor{\bsnm{{Fujimoto}}, \binits{S.}},
\bauthor{\bsnm{{Glazebrook}}, \binits{K.}},
\bauthor{\bsnm{{de Graaff}}, \binits{A.}},
\bauthor{\bsnm{{Khullar}}, \binits{G.}},
\bauthor{\bsnm{{Leja}}, \binits{J.}},
\bauthor{\bsnm{{Marchesini}}, \binits{D.}},
\bauthor{\bsnm{{Maseda}}, \binits{M.V.}},
\bauthor{\bsnm{{Matthee}}, \binits{J.}},
\bauthor{\bsnm{{Miller}}, \binits{T.B.}},
\bauthor{\bsnm{{Naidu}}, \binits{R.P.}},
\bauthor{\bsnm{{Nanayakkara}}, \binits{T.}},
\bauthor{\bsnm{{Oesch}}, \binits{P.A.}},
\bauthor{\bsnm{{Pan}}, \binits{R.}},
\bauthor{\bsnm{{Papovich}}, \binits{C.}},
\bauthor{\bsnm{{Price}}, \binits{S.H.}},
\bauthor{\bsnm{{van Dokkum}}, \binits{P.}},
\bauthor{\bsnm{{Weaver}}, \binits{J.R.}},
\bauthor{\bsnm{{Whitaker}}, \binits{K.E.}},
\bauthor{\bsnm{{Zitrin}}, \binits{A.}}:
\batitle{{UNCOVER Spectroscopy Confirms the Surprising Ubiquity of Active
  Galactic Nuclei in Red Sources at z > 5}}.
\bjtitle{\apj}
\bvolume{964}(\bissue{1}),
\bfpage{39}
(\byear{2024})
\doiurl{10.3847/1538-4357/ad1e5f}
{\href{https://arxiv.org/abs/2309.05714}{{arXiv:2309.05714}}}
{[astro-ph.GA]}
\end{barticle}
\endbibitem

\bibitem[\protect\citeauthoryear{{Wang} et~al.}{2024}]{Wang2024:ub}
\begin{barticle}
\bauthor{\bsnm{{Wang}}, \binits{B.}},
\bauthor{\bsnm{{Leja}}, \binits{J.}},
\bauthor{\bsnm{{de Graaff}}, \binits{A.}},
\bauthor{\bsnm{{Brammer}}, \binits{G.B.}},
\bauthor{\bsnm{{Weibel}}, \binits{A.}},
\bauthor{\bsnm{{van Dokkum}}, \binits{P.}},
\bauthor{\bsnm{{Baggen}}, \binits{J.F.W.}},
\bauthor{\bsnm{{Suess}}, \binits{K.A.}},
\bauthor{\bsnm{{Greene}}, \binits{J.E.}},
\bauthor{\bsnm{{Bezanson}}, \binits{R.}},
\bauthor{\bsnm{{Cleri}}, \binits{N.J.}},
\bauthor{\bsnm{{Hirschmann}}, \binits{M.}},
\bauthor{\bsnm{{Labb{\'e}}}, \binits{I.}},
\bauthor{\bsnm{{Matthee}}, \binits{J.}},
\bauthor{\bsnm{{McConachie}}, \binits{I.}},
\bauthor{\bsnm{{Naidu}}, \binits{R.P.}},
\bauthor{\bsnm{{Nelson}}, \binits{E.}},
\bauthor{\bsnm{{Oesch}}, \binits{P.A.}},
\bauthor{\bsnm{{Setton}}, \binits{D.J.}},
\bauthor{\bsnm{{Williams}}, \binits{C.C.}}:
\batitle{{RUBIES: Evolved Stellar Populations with Extended Formation Histories
  at z {\ensuremath{\sim}} 7{\textendash}8 in Candidate Massive Galaxies
  Identified with JWST/NIRSpec}}.
\bjtitle{\apjl}
\bvolume{969}(\bissue{1}),
\bfpage{13}
(\byear{2024})
\doiurl{10.3847/2041-8213/ad55f7}
{\href{https://arxiv.org/abs/2405.01473}{{arXiv:2405.01473}}}
{[astro-ph.GA]}
\end{barticle}
\endbibitem

\bibitem[\protect\citeauthoryear{{de Graaff} et~al.}{2025}]{deGraaff2025:cliff}
\begin{barticle}
\bauthor{\bsnm{{de Graaff}}, \binits{A.}},
\bauthor{\bsnm{{Rix}}, \binits{H.-W.}},
\bauthor{\bsnm{{Naidu}}, \binits{R.P.}},
\bauthor{\bsnm{{Labb{\'e}}}, \binits{I.}},
\bauthor{\bsnm{{Wang}}, \binits{B.}},
\bauthor{\bsnm{{Leja}}, \binits{J.}},
\bauthor{\bsnm{{Matthee}}, \binits{J.}},
\bauthor{\bsnm{{Katz}}, \binits{H.}},
\bauthor{\bsnm{{Greene}}, \binits{J.E.}},
\bauthor{\bsnm{{Hviding}}, \binits{R.E.}},
\bauthor{\bsnm{{Baggen}}, \binits{J.}},
\bauthor{\bsnm{{Bezanson}}, \binits{R.}},
\bauthor{\bsnm{{Boogaard}}, \binits{L.A.}},
\bauthor{\bsnm{{Brammer}}, \binits{G.}},
\bauthor{\bsnm{{Dayal}}, \binits{P.}},
\bauthor{\bsnm{{van Dokkum}}, \binits{P.}},
\bauthor{\bsnm{{Goulding}}, \binits{A.D.}},
\bauthor{\bsnm{{Hirschmann}}, \binits{M.}},
\bauthor{\bsnm{{Maseda}}, \binits{M.V.}},
\bauthor{\bsnm{{McConachie}}, \binits{I.}},
\bauthor{\bsnm{{Miller}}, \binits{T.B.}},
\bauthor{\bsnm{{Nelson}}, \binits{E.}},
\bauthor{\bsnm{{Oesch}}, \binits{P.A.}},
\bauthor{\bsnm{{Setton}}, \binits{D.J.}},
\bauthor{\bsnm{{Shivaei}}, \binits{I.}},
\bauthor{\bsnm{{Weibel}}, \binits{A.}},
\bauthor{\bsnm{{Whitaker}}, \binits{K.E.}},
\bauthor{\bsnm{{Williams}}, \binits{C.C.}}:
\batitle{{A remarkable ruby: Absorption in dense gas, rather than evolved
  stars, drives the extreme Balmer break of a little red dot at z = 3.5}}.
\bjtitle{\aap}
\bvolume{701},
\bfpage{168}
(\byear{2025})
\doiurl{10.1051/0004-6361/202554681}
{\href{https://arxiv.org/abs/2503.16600}{{arXiv:2503.16600}}}
{[astro-ph.GA]}
\end{barticle}
\endbibitem

\bibitem[\protect\citeauthoryear{{Naidu} et~al.}{2025}]{Naidu2025}
\begin{botherref}
\oauthor{\bsnm{{Naidu}}, \binits{R.P.}},
\oauthor{\bsnm{{Matthee}}, \binits{J.}},
\oauthor{\bsnm{{Katz}}, \binits{H.}},
\oauthor{\bsnm{{de Graaff}}, \binits{A.}},
\oauthor{\bsnm{{Oesch}}, \binits{P.}},
\oauthor{\bsnm{{Smith}}, \binits{A.}},
\oauthor{\bsnm{{Greene}}, \binits{J.E.}},
\oauthor{\bsnm{{Brammer}}, \binits{G.}},
\oauthor{\bsnm{{Weibel}}, \binits{A.}},
\oauthor{\bsnm{{Hviding}}, \binits{R.}},
\oauthor{\bsnm{{Chisholm}}, \binits{J.}},
\oauthor{\bsnm{{Labb\textbackslash'e}}, \binits{I.}},
\oauthor{\bsnm{{Simcoe}}, \binits{R.A.}},
\oauthor{\bsnm{{Witten}}, \binits{C.}},
\oauthor{\bsnm{{Atek}}, \binits{H.}},
\oauthor{\bsnm{{Baggen}}, \binits{J.F.W.}},
\oauthor{\bsnm{{Belli}}, \binits{S.}},
\oauthor{\bsnm{{Bezanson}}, \binits{R.}},
\oauthor{\bsnm{{Boogaard}}, \binits{L.A.}},
\oauthor{\bsnm{{Bose}}, \binits{S.}},
\oauthor{\bsnm{{Covelo-Paz}}, \binits{A.}},
\oauthor{\bsnm{{Dayal}}, \binits{P.}},
\oauthor{\bsnm{{Fudamoto}}, \binits{Y.}},
\oauthor{\bsnm{{Furtak}}, \binits{L.J.}},
\oauthor{\bsnm{{Giovinazzo}}, \binits{E.}},
\oauthor{\bsnm{{Goulding}}, \binits{A.}},
\oauthor{\bsnm{{Gronke}}, \binits{M.}},
\oauthor{\bsnm{{Heintz}}, \binits{K.E.}},
\oauthor{\bsnm{{Hirschmann}}, \binits{M.}},
\oauthor{\bsnm{{Illingworth}}, \binits{G.}},
\oauthor{\bsnm{{Inoue}}, \binits{A.K.}},
\oauthor{\bsnm{{Johnson}}, \binits{B.D.}},
\oauthor{\bsnm{{Leja}}, \binits{J.}},
\oauthor{\bsnm{{Leonova}}, \binits{E.}},
\oauthor{\bsnm{{McConachie}}, \binits{I.}},
\oauthor{\bsnm{{Maseda}}, \binits{M.V.}},
\oauthor{\bsnm{{Natarajan}}, \binits{P.}},
\oauthor{\bsnm{{Nelson}}, \binits{E.}},
\oauthor{\bsnm{{Setton}}, \binits{D.J.}},
\oauthor{\bsnm{{Shivaei}}, \binits{I.}},
\oauthor{\bsnm{{Sobral}}, \binits{D.}},
\oauthor{\bsnm{{Stefanon}}, \binits{M.}},
\oauthor{\bsnm{{Tacchella}}, \binits{S.}},
\oauthor{\bsnm{{Toft}}, \binits{S.}},
\oauthor{\bsnm{{Torralba}}, \binits{A.}},
\oauthor{\bsnm{{van Dokkum}}, \binits{P.}},
\oauthor{\bsnm{{van der Wel}}, \binits{A.}},
\oauthor{\bsnm{{Volonteri}}, \binits{M.}},
\oauthor{\bsnm{{Walter}}, \binits{F.}},
\oauthor{\bsnm{{Wang}}, \binits{B.}},
\oauthor{\bsnm{{Watson}}, \binits{D.}}:
{A ``Black Hole Star'' Reveals the Remarkable Gas-Enshrouded Hearts of the
  Little Red Dots}.
arXiv e-prints,
2503--16596
(2025)
\doiurl{10.48550/arXiv.2503.16596}
{\href{https://arxiv.org/abs/2503.16596}{{arXiv:2503.16596}}}
{[astro-ph.GA]}
\end{botherref}
\endbibitem

\bibitem[\protect\citeauthoryear{{de Graaff} et~al.}{2025}]{deGraaff:sample}
\begin{botherref}
\oauthor{\bsnm{{de Graaff}}, \binits{A.}},
\oauthor{\bsnm{{Hviding}}, \binits{R.E.}},
\oauthor{\bsnm{{Naidu}}, \binits{R.P.}},
\oauthor{\bsnm{{Greene}}, \binits{J.E.}},
\oauthor{\bsnm{{Miller}}, \binits{T.B.}},
\oauthor{\bsnm{{Leja}}, \binits{J.}},
\oauthor{\bsnm{{Matthee}}, \binits{J.}},
\oauthor{\bsnm{{Brammer}}, \binits{G.}},
\oauthor{\bsnm{{Katz}}, \binits{H.}},
\oauthor{\bsnm{{Bezanson}}, \binits{R.}},
\oauthor{\bsnm{{Boogaard}}, \binits{L.A.}},
\oauthor{\bsnm{{Bose}}, \binits{S.}},
\oauthor{\bsnm{{Chisholm}}, \binits{J.}},
\oauthor{\bsnm{{Cleri}}, \binits{N.J.}},
\oauthor{\bsnm{{Dayal}}, \binits{P.}},
\oauthor{\bsnm{{Feldmann}}, \binits{R.}},
\oauthor{\bsnm{{Fudamoto}}, \binits{Y.}},
\oauthor{\bsnm{{Fujimoto}}, \binits{S.}},
\oauthor{\bsnm{{Furtak}}, \binits{L.J.}},
\oauthor{\bsnm{{Glazebrook}}, \binits{K.}},
\oauthor{\bsnm{{Gottumukkala}}, \binits{R.}},
\oauthor{\bsnm{{Heintz}}, \binits{K.E.}},
\oauthor{\bsnm{{Kokorev}}, \binits{V.}},
\oauthor{\bsnm{{Labbe}}, \binits{I.}},
\oauthor{\bsnm{{Maseda}}, \binits{M.V.}},
\oauthor{\bsnm{{McConachie}}, \binits{I.}},
\oauthor{\bsnm{{Nanayakkara}}, \binits{T.}},
\oauthor{\bsnm{{Nelson}}, \binits{E.}},
\oauthor{\bsnm{{Nowaczyk}}, \binits{P.}},
\oauthor{\bsnm{{Oesch}}, \binits{P.A.}},
\oauthor{\bsnm{{Rix}}, \binits{H.-W.}},
\oauthor{\bsnm{{Setton}}, \binits{D.J.}},
\oauthor{\bsnm{{Torralba}}, \binits{A.}},
\oauthor{\bsnm{{Walter}}, \binits{F.}},
\oauthor{\bsnm{{Wang}}, \binits{B.}},
\oauthor{\bsnm{{Weibel}}, \binits{A.}},
\oauthor{\bsnm{{van der Wel}}, \binits{A.}}:
{Little Red Dots host Black Hole Stars: A unified family of gas-reddened AGN
  revealed by JWST/NIRSpec spectroscopy}.
arXiv e-prints,
2511--21820
(2025)
\doiurl{10.48550/arXiv.2511.21820}
{\href{https://arxiv.org/abs/2511.21820}{{arXiv:2511.21820}}}
{[astro-ph.GA]}
\end{botherref}
\endbibitem

\bibitem[\protect\citeauthoryear{{Begelman} and {Dexter}}{2026}]{Begelman2025}
\begin{barticle}
\bauthor{\bsnm{{Begelman}}, \binits{M.C.}},
\bauthor{\bsnm{{Dexter}}, \binits{J.}}:
\batitle{{Little Red Dots as Late-stage Quasi-stars}}.
\bjtitle{\apj}
\bvolume{996}(\bissue{1}),
\bfpage{48}
(\byear{2026})
\doiurl{10.3847/1538-4357/ae274a}
{\href{https://arxiv.org/abs/2507.09085}{{arXiv:2507.09085}}}
{[astro-ph.GA]}
\end{barticle}
\endbibitem

\bibitem[\protect\citeauthoryear{{Kido} et~al.}{2025}]{Kido2025}
\begin{barticle}
\bauthor{\bsnm{{Kido}}, \binits{D.}},
\bauthor{\bsnm{{Ioka}}, \binits{K.}},
\bauthor{\bsnm{{Hotokezaka}}, \binits{K.}},
\bauthor{\bsnm{{Inayoshi}}, \binits{K.}},
\bauthor{\bsnm{{Irwin}}, \binits{C.M.}}:
\batitle{{Black hole envelopes in Little Red Dots}}.
\bjtitle{\mnras}
\bvolume{544}(\bissue{4}),
\bfpage{3407}--\blpage{3416}
(\byear{2025})
\doiurl{10.1093/mnras/staf1898}
{\href{https://arxiv.org/abs/2505.06965}{{arXiv:2505.06965}}}
{[astro-ph.HE]}
\end{barticle}
\endbibitem

\bibitem[\protect\citeauthoryear{{Liu} et~al.}{2025}]{Liu2025}
\begin{barticle}
\bauthor{\bsnm{{Liu}}, \binits{H.}},
\bauthor{\bsnm{{Jiang}}, \binits{Y.-F.}},
\bauthor{\bsnm{{Quataert}}, \binits{E.}},
\bauthor{\bsnm{{Greene}}, \binits{J.E.}},
\bauthor{\bsnm{{Ma}}, \binits{Y.}}:
\batitle{{The Balmer Break and Optical Continuum of Little Red Dots from
  Super-Eddington Accretion}}.
\bjtitle{\apj}
\bvolume{994}(\bissue{1}),
\bfpage{113}
(\byear{2025})
\doiurl{10.3847/1538-4357/ae0c19}
{\href{https://arxiv.org/abs/2507.07190}{{arXiv:2507.07190}}}
{[astro-ph.GA]}
\end{barticle}
\endbibitem

\bibitem[\protect\citeauthoryear{{Labbe} et~al.}{2024}]{Labbe2024}
\begin{botherref}
\oauthor{\bsnm{{Labbe}}, \binits{I.}},
\oauthor{\bsnm{{Greene}}, \binits{J.E.}},
\oauthor{\bsnm{{Matthee}}, \binits{J.}},
\oauthor{\bsnm{{Treiber}}, \binits{H.}},
\oauthor{\bsnm{{Kokorev}}, \binits{V.}},
\oauthor{\bsnm{{Miller}}, \binits{T.B.}},
\oauthor{\bsnm{{Kramarenko}}, \binits{I.}},
\oauthor{\bsnm{{Setton}}, \binits{D.J.}},
\oauthor{\bsnm{{Ma}}, \binits{Y.}},
\oauthor{\bsnm{{Goulding}}, \binits{A.D.}},
\oauthor{\bsnm{{Bezanson}}, \binits{R.}},
\oauthor{\bsnm{{Naidu}}, \binits{R.P.}},
\oauthor{\bsnm{{Williams}}, \binits{C.C.}},
\oauthor{\bsnm{{Atek}}, \binits{H.}},
\oauthor{\bsnm{{Brammer}}, \binits{G.}},
\oauthor{\bsnm{{Cutler}}, \binits{S.E.}},
\oauthor{\bsnm{{Chemerynska}}, \binits{I.}},
\oauthor{\bsnm{{Cloonan}}, \binits{A.P.}},
\oauthor{\bsnm{{Dayal}}, \binits{P.}},
\oauthor{\bsnm{{de Graaff}}, \binits{A.}},
\oauthor{\bsnm{{Fudamoto}}, \binits{Y.}},
\oauthor{\bsnm{{Fujimoto}}, \binits{S.}},
\oauthor{\bsnm{{Furtak}}, \binits{L.J.}},
\oauthor{\bsnm{{Glazebrook}}, \binits{K.}},
\oauthor{\bsnm{{Heintz}}, \binits{K.E.}},
\oauthor{\bsnm{{Leja}}, \binits{J.}},
\oauthor{\bsnm{{Marchesini}}, \binits{D.}},
\oauthor{\bsnm{{Nanayakkara}}, \binits{T.}},
\oauthor{\bsnm{{Nelson}}, \binits{E.J.}},
\oauthor{\bsnm{{Oesch}}, \binits{P.A.}},
\oauthor{\bsnm{{Pan}}, \binits{R.}},
\oauthor{\bsnm{{Price}}, \binits{S.H.}},
\oauthor{\bsnm{{Shivaei}}, \binits{I.}},
\oauthor{\bsnm{{Sobral}}, \binits{D.}},
\oauthor{\bsnm{{Suess}}, \binits{K.A.}},
\oauthor{\bsnm{{van Dokkum}}, \binits{P.}},
\oauthor{\bsnm{{Wang}}, \binits{B.}},
\oauthor{\bsnm{{Weaver}}, \binits{J.R.}},
\oauthor{\bsnm{{Whitaker}}, \binits{K.E.}},
\oauthor{\bsnm{{Zitrin}}, \binits{A.}}:
{An unambiguous AGN and a Balmer break in an Ultraluminous Little Red Dot at
  z=4.47 from Ultradeep UNCOVER and All the Little Things Spectroscopy}.
arXiv e-prints,
2412--04557
(2024)
\doiurl{10.48550/arXiv.2412.04557}
{\href{https://arxiv.org/abs/2412.04557}{{arXiv:2412.04557}}}
{[astro-ph.GA]}
\end{botherref}
\endbibitem

\bibitem[\protect\citeauthoryear{{Greene} et~al.}{2026}]{Greene2025}
\begin{barticle}
\bauthor{\bsnm{{Greene}}, \binits{J.E.}},
\bauthor{\bsnm{{Setton}}, \binits{D.J.}},
\bauthor{\bsnm{{Furtak}}, \binits{L.J.}},
\bauthor{\bsnm{{Naidu}}, \binits{R.P.}},
\bauthor{\bsnm{{Volonteri}}, \binits{M.}},
\bauthor{\bsnm{{Dayal}}, \binits{P.}},
\bauthor{\bsnm{{Labbe}}, \binits{I.}},
\bauthor{\bsnm{{van Dokkum}}, \binits{P.}},
\bauthor{\bsnm{{Bezanson}}, \binits{R.}},
\bauthor{\bsnm{{Brammer}}, \binits{G.}},
\bauthor{\bsnm{{Cutler}}, \binits{S.E.}},
\bauthor{\bsnm{{Glazebrook}}, \binits{K.}},
\bauthor{\bsnm{{de Graaff}}, \binits{A.}},
\bauthor{\bsnm{{Hirschmann}}, \binits{M.}},
\bauthor{\bsnm{{Hviding}}, \binits{R.E.}},
\bauthor{\bsnm{{Kokorev}}, \binits{V.}},
\bauthor{\bsnm{{Leja}}, \binits{J.}},
\bauthor{\bsnm{{Liu}}, \binits{H.}},
\bauthor{\bsnm{{Ma}}, \binits{Y.}},
\bauthor{\bsnm{{Matthee}}, \binits{J.}},
\bauthor{\bsnm{{Nanayakkara}}, \binits{T.}},
\bauthor{\bsnm{{Oesch}}, \binits{P.A.}},
\bauthor{\bsnm{{Pan}}, \binits{R.}},
\bauthor{\bsnm{{Price}}, \binits{S.H.}},
\bauthor{\bsnm{{Spilker}}, \binits{J.S.}},
\bauthor{\bsnm{{Wang}}, \binits{B.}},
\bauthor{\bsnm{{Weaver}}, \binits{J.R.}},
\bauthor{\bsnm{{Whitaker}}, \binits{K.E.}},
\bauthor{\bsnm{{Williams}}, \binits{C.C.}},
\bauthor{\bsnm{{Zitrin}}, \binits{A.}}:
\batitle{{What You See Is What You Get: Empirically Measured Bolometric
  Luminosities of Little Red Dots}}.
\bjtitle{\apj}
\bvolume{996}(\bissue{2}),
\bfpage{129}
(\byear{2026})
\doiurl{10.3847/1538-4357/ae1836}
{\href{https://arxiv.org/abs/2509.05434}{{arXiv:2509.05434}}}
{[astro-ph.GA]}
\end{barticle}
\endbibitem

\bibitem[\protect\citeauthoryear{{Nandal} and {Loeb}}{2025}]{Nandal2025}
\begin{botherref}
\oauthor{\bsnm{{Nandal}}, \binits{D.}},
\oauthor{\bsnm{{Loeb}}, \binits{A.}}:
{Supermassive Stars Match the Spectral Signatures of JWST's Little Red Dots}.
arXiv e-prints,
2507--12618
(2025)
\doiurl{10.48550/arXiv.2507.12618}
{\href{https://arxiv.org/abs/2507.12618}{{arXiv:2507.12618}}}
{[astro-ph.GA]}
\end{botherref}
\endbibitem

\bibitem[\protect\citeauthoryear{{Umeda} et~al.}{2025}]{Umeda2025}
\begin{botherref}
\oauthor{\bsnm{{Umeda}}, \binits{H.}},
\oauthor{\bsnm{{Inayoshi}}, \binits{K.}},
\oauthor{\bsnm{{Harikane}}, \binits{Y.}},
\oauthor{\bsnm{{Murase}}, \binits{K.}}:
{A Black-Hole Envelope Interpretation for Cosmological Demographics of Little
  Red Dots}.
arXiv e-prints,
2512--04208
(2025)
\doiurl{10.48550/arXiv.2512.04208}
{\href{https://arxiv.org/abs/2512.04208}{{arXiv:2512.04208}}}
{[astro-ph.GA]}
\end{botherref}
\endbibitem

\bibitem[\protect\citeauthoryear{{Brammer} and {Valentino}}{2025}]{dja}
\begin{botherref}
\oauthor{\bsnm{{Brammer}}, \binits{G.}},
\oauthor{\bsnm{{Valentino}}, \binits{F.}}:
{The DAWN JWST Archive: Compilation of Public NIRSpec Spectra}.
Zenodo
(2025).
\doiurl{10.5281/zenodo.15472354}
\end{botherref}
\endbibitem

\bibitem[\protect\citeauthoryear{{Hviding} et~al.}{2025}]{Hviding2025}
\begin{barticle}
\bauthor{\bsnm{{Hviding}}, \binits{R.E.}},
\bauthor{\bsnm{{de Graaff}}, \binits{A.}},
\bauthor{\bsnm{{Miller}}, \binits{T.B.}},
\bauthor{\bsnm{{Setton}}, \binits{D.J.}},
\bauthor{\bsnm{{Greene}}, \binits{J.E.}},
\bauthor{\bsnm{{Labb{\'e}}}, \binits{I.}},
\bauthor{\bsnm{{Brammer}}, \binits{G.}},
\bauthor{\bsnm{{Bezanson}}, \binits{R.}},
\bauthor{\bsnm{{Boogaard}}, \binits{L.A.}},
\bauthor{\bsnm{{Cleri}}, \binits{N.J.}},
\bauthor{\bsnm{{Leja}}, \binits{J.}},
\bauthor{\bsnm{{Maseda}}, \binits{M.V.}},
\bauthor{\bsnm{{McConachie}}, \binits{I.}},
\bauthor{\bsnm{{Matthee}}, \binits{J.}},
\bauthor{\bsnm{{Naidu}}, \binits{R.P.}},
\bauthor{\bsnm{{Oesch}}, \binits{P.A.}},
\bauthor{\bsnm{{Wang}}, \binits{B.}},
\bauthor{\bsnm{{Whitaker}}, \binits{K.E.}},
\bauthor{\bsnm{{Williams}}, \binits{C.C.}}:
\batitle{{RUBIES: A spectroscopic census of little red dots: All point sources
  with v-shaped continua have broad lines}}.
\bjtitle{\aap}
\bvolume{702},
\bfpage{57}
(\byear{2025})
\doiurl{10.1051/0004-6361/202555816}
{\href{https://arxiv.org/abs/2506.05459}{{arXiv:2506.05459}}}
{[astro-ph.GA]}
\end{barticle}
\endbibitem

\bibitem[\protect\citeauthoryear{{Tsuji}}{1964}]{Tsuji1964}
\begin{barticle}
\bauthor{\bsnm{{Tsuji}}, \binits{T.}}:
\batitle{{Molecular abundance in stellar atmospheres}}.
\bjtitle{Annals of the Tokyo Astronomical Observatory}
\bvolume{9}(\bissue{1}),
\bfpage{1}--\blpage{110}
(\byear{1964})
\end{barticle}
\endbibitem

\bibitem[\protect\citeauthoryear{{Tsuji}}{1973}]{Tsuji1973}
\begin{barticle}
\bauthor{\bsnm{{Tsuji}}, \binits{T.}}:
\batitle{{Molecular abundances in stellar atmospheres. II.}}
\bjtitle{\aap}
\bvolume{23},
\bfpage{411}
(\byear{1973})
\end{barticle}
\endbibitem

\bibitem[\protect\citeauthoryear{{Filippazzo} et~al.}{2015}]{Filippazzo2015}
\begin{barticle}
\bauthor{\bsnm{{Filippazzo}}, \binits{J.C.}},
\bauthor{\bsnm{{Rice}}, \binits{E.L.}},
\bauthor{\bsnm{{Faherty}}, \binits{J.}},
\bauthor{\bsnm{{Cruz}}, \binits{K.L.}},
\bauthor{\bsnm{{Van Gordon}}, \binits{M.M.}},
\bauthor{\bsnm{{Looper}}, \binits{D.L.}}:
\batitle{{Fundamental Parameters and Spectral Energy Distributions of Young and
  Field Age Objects with Masses Spanning the Stellar to Planetary Regime}}.
\bjtitle{\apj}
\bvolume{810}(\bissue{2}),
\bfpage{158}
(\byear{2015})
\doiurl{10.1088/0004-637X/810/2/158}
{\href{https://arxiv.org/abs/1508.01767}{{arXiv:1508.01767}}}
{[astro-ph.SR]}
\end{barticle}
\endbibitem

\bibitem[\protect\citeauthoryear{{Pineda} et~al.}{2021}]{Pineda2021}
\begin{barticle}
\bauthor{\bsnm{{Pineda}}, \binits{J.S.}},
\bauthor{\bsnm{{Youngblood}}, \binits{A.}},
\bauthor{\bsnm{{France}}, \binits{K.}}:
\batitle{{The M-dwarf Ultraviolet Spectroscopic Sample. I. Determining Stellar
  Parameters for Field Stars}}.
\bjtitle{\apj}
\bvolume{918}(\bissue{1}),
\bfpage{40}
(\byear{2021})
\doiurl{10.3847/1538-4357/ac0aea}
{\href{https://arxiv.org/abs/2106.07656}{{arXiv:2106.07656}}}
{[astro-ph.SR]}
\end{barticle}
\endbibitem

\bibitem[\protect\citeauthoryear{{Baggen} et~al.}{2025}]{Baggen2025}
\begin{botherref}
\oauthor{\bsnm{{Baggen}}, \binits{J.F.W.}},
\oauthor{\bsnm{{van Dokkum}}, \binits{P.}},
\oauthor{\bsnm{{Labb{\'e}}}, \binits{I.}},
\oauthor{\bsnm{{Brammer}}, \binits{G.}}:
{(Re)solving the complex multi-scale morphology and V-shaped SED of a newly
  discovered strongly-lensed Little Red Dot in Abell 383}.
arXiv e-prints,
2512--03239
(2025)
\doiurl{10.48550/arXiv.2512.03239}
{\href{https://arxiv.org/abs/2512.03239}{{arXiv:2512.03239}}}
{[astro-ph.GA]}
\end{botherref}
\endbibitem

\bibitem[\protect\citeauthoryear{{Golubchik} et~al.}{2025}]{Golubchik2025}
\begin{botherref}
\oauthor{\bsnm{{Golubchik}}, \binits{M.}},
\oauthor{\bsnm{{Furtak}}, \binits{L.J.}},
\oauthor{\bsnm{{Allingham}}, \binits{J.F.V.}},
\oauthor{\bsnm{{Zitrin}}, \binits{A.}},
\oauthor{\bsnm{{Akins}}, \binits{H.B.}},
\oauthor{\bsnm{{Kokorev}}, \binits{V.}},
\oauthor{\bsnm{{Fujimoto}}, \binits{S.}},
\oauthor{\bsnm{{Abdurro'uf}}},
\oauthor{\bsnm{{Amor{\'\i}n}}, \binits{R.O.}},
\oauthor{\bsnm{{Bauer}}, \binits{F.E.}},
\oauthor{\bsnm{{Bezanson}}, \binits{R.}},
\oauthor{\bsnm{{Brada{\v{c}}}}, \binits{M.}},
\oauthor{\bsnm{{Bradley}}, \binits{L.D.}},
\oauthor{\bsnm{{Brammer}}, \binits{G.B.}},
\oauthor{\bsnm{{Chisholm}}, \binits{J.}},
\oauthor{\bsnm{{Coe}}, \binits{D.}},
\oauthor{\bsnm{{Conselice}}, \binits{C.J.}},
\oauthor{\bsnm{{Dayal}}, \binits{P.}},
\oauthor{\bsnm{{Dessauges-Zavadsky}}, \binits{M.}},
\oauthor{\bsnm{{Diego}}, \binits{J.M.}},
\oauthor{\bsnm{{Faisst}}, \binits{A.L.}},
\oauthor{\bsnm{{Fei}}, \binits{Q.}},
\oauthor{\bsnm{{Ferguson}}, \binits{H.C.}},
\oauthor{\bsnm{{Finkelstein}}, \binits{S.L.}},
\oauthor{\bsnm{{Frye}}, \binits{B.L.}},
\oauthor{\bsnm{{Gonz{\'a}lez-Otero}}, \binits{M.}},
\oauthor{\bsnm{{Greene}}, \binits{J.E.}},
\oauthor{\bsnm{{Harikane}}, \binits{Y.}},
\oauthor{\bsnm{{Hsiao}}, \binits{T.Y.-Y.}},
\oauthor{\bsnm{{Inayoshi}}, \binits{K.}},
\oauthor{\bsnm{{Jim{\'e}nez-Teja}}, \binits{Y.}},
\oauthor{\bsnm{{Knudsen}}, \binits{K.}},
\oauthor{\bsnm{{Koekemoer}}, \binits{A.M.}},
\oauthor{\bsnm{{Labb{\'e}}}, \binits{I.}},
\oauthor{\bsnm{{Lucas}}, \binits{R.A.}},
\oauthor{\bsnm{{Magdis}}, \binits{G.E.}},
\oauthor{\bsnm{{Matthee}}, \binits{J.}},
\oauthor{\bsnm{{Messa}}, \binits{M.}},
\oauthor{\bsnm{{Naidu}}, \binits{R.P.}},
\oauthor{\bsnm{{Nakane}}, \binits{M.}},
\oauthor{\bsnm{{Noirot}}, \binits{G.}},
\oauthor{\bsnm{{Pan}}, \binits{R.}},
\oauthor{\bsnm{{Papovich}}, \binits{C.}},
\oauthor{\bsnm{{Richard}}, \binits{J.}},
\oauthor{\bsnm{{Ricotti}}, \binits{M.}},
\oauthor{\bsnm{{Robbins}}, \binits{L.}},
\oauthor{\bsnm{{Stark}}, \binits{D.P.}},
\oauthor{\bsnm{{Sun}}, \binits{F.}},
\oauthor{\bsnm{{Treu}}, \binits{T.}},
\oauthor{\bsnm{{Tripodi}}, \binits{R.}},
\oauthor{\bsnm{{Vanzella}}, \binits{E.}},
\oauthor{\bsnm{{Willott}}, \binits{C.}},
\oauthor{\bsnm{{Windhorst}}, \binits{R.A.}}:
{VENUS: When Red meets Blue -- A multiply imaged Little Red Dot with an
  apparent blue companion behind the galaxy cluster Abell 383}.
arXiv e-prints,
2512--02117
(2025)
\doiurl{10.48550/arXiv.2512.02117}
{\href{https://arxiv.org/abs/2512.02117}{{arXiv:2512.02117}}}
{[astro-ph.GA]}
\end{botherref}
\endbibitem

\bibitem[\protect\citeauthoryear{{Barro} et~al.}{2025}]{Barro2025}
\begin{botherref}
\oauthor{\bsnm{{Barro}}, \binits{G.}},
\oauthor{\bsnm{{Perez-Gonzalez}}, \binits{P.G.}},
\oauthor{\bsnm{{Kocevski}}, \binits{D.}},
\oauthor{\bsnm{{Trump}}, \binits{J.R.}},
\oauthor{\bsnm{{Dickinson}}, \binits{M.}},
\oauthor{\bsnm{{Arrabal Haro}}, \binits{P.}},
\oauthor{\bsnm{{Brooks}}, \binits{M.}},
\oauthor{\bsnm{{Donnan}}, \binits{C.T.}},
\oauthor{\bsnm{{Dunlop}}, \binits{J.S.}},
\oauthor{\bsnm{{Finkelstein}}, \binits{S.L.}},
\oauthor{\bsnm{{Franco}}, \binits{M.}},
\oauthor{\bsnm{{Gandolfi}}, \binits{G.}},
\oauthor{\bsnm{{Giavalisco}}, \binits{M.}},
\oauthor{\bsnm{{Grogin}}, \binits{N.A.}},
\oauthor{\bsnm{{Hirschmann}}, \binits{M.}},
\oauthor{\bsnm{{Kartaltepe}}, \binits{J.S.}},
\oauthor{\bsnm{{Koekemoer}}, \binits{A.M.}},
\oauthor{\bsnm{{Larson}}, \binits{R.L.}},
\oauthor{\bsnm{{Leung}}, \binits{G.C.K.}},
\oauthor{\bsnm{{Lucas}}, \binits{R.A.}},
\oauthor{\bsnm{{McGrath}}, \binits{E.J.}},
\oauthor{\bsnm{{Papovich}}, \binits{C.}},
\oauthor{\bsnm{{Perez-Diaz}}, \binits{B.}},
\oauthor{\bsnm{{Somerville}}, \binits{R.S.}},
\oauthor{\bsnm{{Taylor}}, \binits{E.}},
\oauthor{\bsnm{{Taylor}}, \binits{A.J.}},
\oauthor{\bsnm{{Tripodi}}, \binits{R.}},
\oauthor{\bsnm{{Yung}}, \binits{L.Y.A.}},
\oauthor{\bsnm{{Wang}}, \binits{X.}}:
{From ``The Cliff'' to ``Virgil'': Mapping the Spectral Diversity of Little Red
  Dots with JWST/NIRSpec}.
arXiv e-prints,
2512--15853
(2025)
\doiurl{10.48550/arXiv.2512.15853}
{\href{https://arxiv.org/abs/2512.15853}{{arXiv:2512.15853}}}
{[astro-ph.GA]}
\end{botherref}
\endbibitem

\bibitem[\protect\citeauthoryear{{Begelman} et~al.}{2008}]{Begelman2008}
\begin{barticle}
\bauthor{\bsnm{{Begelman}}, \binits{M.C.}},
\bauthor{\bsnm{{Rossi}}, \binits{E.M.}},
\bauthor{\bsnm{{Armitage}}, \binits{P.J.}}:
\batitle{{Quasi-stars: accreting black holes inside massive envelopes}}.
\bjtitle{\mnras}
\bvolume{387}(\bissue{4}),
\bfpage{1649}--\blpage{1659}
(\byear{2008})
\doiurl{10.1111/j.1365-2966.2008.13344.x}
{\href{https://arxiv.org/abs/0711.4078}{{arXiv:0711.4078}}}
{[astro-ph]}
\end{barticle}
\endbibitem

\bibitem[\protect\citeauthoryear{{Ronayne} et~al.}{2025}]{Ronayne2025}
\begin{botherref}
\oauthor{\bsnm{{Ronayne}}, \binits{K.}},
\oauthor{\bsnm{{Papovich}}, \binits{C.}},
\oauthor{\bsnm{{Kirkpatrick}}, \binits{A.}},
\oauthor{\bsnm{{Backhaus}}, \binits{B.E.}},
\oauthor{\bsnm{{Cullen}}, \binits{F.}},
\oauthor{\bsnm{{Shen}}, \binits{L.}},
\oauthor{\bsnm{{Bagley}}, \binits{M.B.}},
\oauthor{\bsnm{{Barro}}, \binits{G.}},
\oauthor{\bsnm{{Finkelstein}}, \binits{S.L.}},
\oauthor{\bsnm{{Hamblin}}, \binits{K.}},
\oauthor{\bsnm{{Kartaltepe}}, \binits{J.S.}},
\oauthor{\bsnm{{Kocevski}}, \binits{D.D.}},
\oauthor{\bsnm{{Koekemoer}}, \binits{A.M.}},
\oauthor{\bsnm{{Lambrides}}, \binits{E.}},
\oauthor{\bsnm{{Pacucci}}, \binits{F.}},
\oauthor{\bsnm{{Yang}}, \binits{G.}}:
{MEGA: Spectrophotometric SED Fitting of Little Red Dots Detected in JWST
  MIRI}.
arXiv e-prints,
2508--20177
(2025)
\doiurl{10.48550/arXiv.2508.20177}
{\href{https://arxiv.org/abs/2508.20177}{{arXiv:2508.20177}}}
{[astro-ph.GA]}
\end{botherref}
\endbibitem

\bibitem[\protect\citeauthoryear{{Matthee} et~al.}{2025}]{Matthee2025}
\begin{barticle}
\bauthor{\bsnm{{Matthee}}, \binits{J.}},
\bauthor{\bsnm{{Naidu}}, \binits{R.P.}},
\bauthor{\bsnm{{Kotiwale}}, \binits{G.}},
\bauthor{\bsnm{{Furtak}}, \binits{L.J.}},
\bauthor{\bsnm{{Kramarenko}}, \binits{I.}},
\bauthor{\bsnm{{Mackenzie}}, \binits{R.}},
\bauthor{\bsnm{{Greene}}, \binits{J.}},
\bauthor{\bsnm{{Adamo}}, \binits{A.}},
\bauthor{\bsnm{{Bouwens}}, \binits{R.J.}},
\bauthor{\bsnm{{Di Cesare}}, \binits{C.}},
\bauthor{\bsnm{{Eilers}}, \binits{A.-C.}},
\bauthor{\bsnm{{de Graaff}}, \binits{A.}},
\bauthor{\bsnm{{Heintz}}, \binits{K.E.}},
\bauthor{\bsnm{{Kashino}}, \binits{D.}},
\bauthor{\bsnm{{Maseda}}, \binits{M.V.}},
\bauthor{\bsnm{{Tacchella}}, \binits{S.}},
\bauthor{\bsnm{{Torralba}}, \binits{A.}}:
\batitle{{Environmental Evidence for Overly Massive Black Holes in Low-mass
  Galaxies and a Black Hole{\textendash}Halo Mass Relation at z
  {\ensuremath{\sim}} 5}}.
\bjtitle{\apj}
\bvolume{988}(\bissue{2}),
\bfpage{246}
(\byear{2025})
\doiurl{10.3847/1538-4357/ade886}
{\href{https://arxiv.org/abs/2412.02846}{{arXiv:2412.02846}}}
{[astro-ph.GA]}
\end{barticle}
\endbibitem

\bibitem[\protect\citeauthoryear{{Lin} et~al.}{2026}]{Lin2025:cluster}
\begin{barticle}
\bauthor{\bsnm{{Lin}}, \binits{X.}},
\bauthor{\bsnm{{Fan}}, \binits{X.}},
\bauthor{\bsnm{{Sun}}, \binits{F.}},
\bauthor{\bsnm{{Zhang}}, \binits{J.}},
\bauthor{\bsnm{{Egami}}, \binits{E.}},
\bauthor{\bsnm{{Helton}}, \binits{J.M.}},
\bauthor{\bsnm{{Wang}}, \binits{F.}},
\bauthor{\bsnm{{Zhang}}, \binits{H.}},
\bauthor{\bsnm{{Bunker}}, \binits{A.J.}},
\bauthor{\bsnm{{Cai}}, \binits{Z.}},
\bauthor{\bsnm{{Ji}}, \binits{Z.}},
\bauthor{\bsnm{{Jin}}, \binits{X.}},
\bauthor{\bsnm{{Maiolino}}, \binits{R.}},
\bauthor{\bsnm{{Pudoka}}, \binits{M.A.}},
\bauthor{\bsnm{{Rinaldi}}, \binits{P.}},
\bauthor{\bsnm{{Robertson}}, \binits{B.}},
\bauthor{\bsnm{{Tacchella}}, \binits{S.}},
\bauthor{\bsnm{{Tee}}, \binits{W.L.}},
\bauthor{\bsnm{{Sun}}, \binits{Y.}},
\bauthor{\bsnm{{Willmer}}, \binits{C.N.A.}},
\bauthor{\bsnm{{Willott}}, \binits{C.}},
\bauthor{\bsnm{{Zhu}}, \binits{Y.}}:
\batitle{{The Large-scale Environments of Low-luminosity AGNs at 3.9 < z < 6
  and Implications for Their Host Dark Matter Halos from a Complete NIRCam
  Grism Redshift Survey}}.
\bjtitle{\apj}
\bvolume{997}(\bissue{1}),
\bfpage{61}
(\byear{2026})
\doiurl{10.3847/1538-4357/ae1eef}
{\href{https://arxiv.org/abs/2505.02896}{{arXiv:2505.02896}}}
{[astro-ph.GA]}
\end{barticle}
\endbibitem

\bibitem[\protect\citeauthoryear{{Pizzati} et~al.}{2025}]{Pizzati2025}
\begin{barticle}
\bauthor{\bsnm{{Pizzati}}, \binits{E.}},
\bauthor{\bsnm{{Hennawi}}, \binits{J.F.}},
\bauthor{\bsnm{{Schaye}}, \binits{J.}},
\bauthor{\bsnm{{Eilers}}, \binits{A.-C.}},
\bauthor{\bsnm{{Huang}}, \binits{J.}},
\bauthor{\bsnm{{Schindler}}, \binits{J.-T.}},
\bauthor{\bsnm{{Wang}}, \binits{F.}}:
\batitle{{'Little red dots' cannot reside in the same dark matter haloes as
  comparably luminous unobscured quasars}}.
\bjtitle{\mnras}
\bvolume{539}(\bissue{4}),
\bfpage{2910}--\blpage{2925}
(\byear{2025})
\doiurl{10.1093/mnras/staf660}
{\href{https://arxiv.org/abs/2409.18208}{{arXiv:2409.18208}}}
{[astro-ph.GA]}
\end{barticle}
\endbibitem

\bibitem[\protect\citeauthoryear{{Young} et~al.}{2000}]{Young2000}
\begin{barticle}
\bauthor{\bsnm{{Young}}, \binits{J.S.}},
\bauthor{\bsnm{{Baldwin}}, \binits{J.E.}},
\bauthor{\bsnm{{Boysen}}, \binits{R.C.}},
\bauthor{\bsnm{{Haniff}}, \binits{C.A.}},
\bauthor{\bsnm{{Lawson}}, \binits{P.R.}},
\bauthor{\bsnm{{Mackay}}, \binits{C.D.}},
\bauthor{\bsnm{{Pearson}}, \binits{D.}},
\bauthor{\bsnm{{Rogers}}, \binits{J.}},
\bauthor{\bsnm{{St.-Jacques}}, \binits{D.}},
\bauthor{\bsnm{{Warner}}, \binits{P.J.}},
\bauthor{\bsnm{{Wilson}}, \binits{D.M.A.}},
\bauthor{\bsnm{{Wilson}}, \binits{R.W.}}:
\batitle{{New views of Betelgeuse: multi-wavelength surface imaging and
  implications for models of hotspot generation}}.
\bjtitle{\mnras}
\bvolume{315}(\bissue{3}),
\bfpage{635}--\blpage{645}
(\byear{2000})
\doiurl{10.1046/j.1365-8711.2000.03438.x}
\end{barticle}
\endbibitem

\bibitem[\protect\citeauthoryear{{Pasha} and {Miller}}{2023}]{Pasha2023}
\begin{barticle}
\bauthor{\bsnm{{Pasha}}, \binits{I.}},
\bauthor{\bsnm{{Miller}}, \binits{T.B.}}:
\batitle{{pysersic: A Python package for determining galaxy structural
  properties via Bayesian inference, accelerated with jax}}.
\bjtitle{The Journal of Open Source Software}
\bvolume{8}(\bissue{89}),
\bfpage{5703}
(\byear{2023})
\doiurl{10.21105/joss.05703}
{\href{https://arxiv.org/abs/2306.05454}{{arXiv:2306.05454}}}
{[astro-ph.GA]}
\end{barticle}
\endbibitem

\bibitem[\protect\citeauthoryear{{Maseda} et~al.}{2024}]{Maseda2024}
\begin{barticle}
\bauthor{\bsnm{{Maseda}}, \binits{M.V.}},
\bauthor{\bsnm{{de Graaff}}, \binits{A.}},
\bauthor{\bsnm{{Franx}}, \binits{M.}},
\bauthor{\bsnm{{Rix}}, \binits{H.-W.}},
\bauthor{\bsnm{{Carniani}}, \binits{S.}},
\bauthor{\bsnm{{Laseter}}, \binits{I.}},
\bauthor{\bsnm{{Dudzevi{\v{c}}i{\={u}}t{\.{e}}}}, \binits{U.}},
\bauthor{\bsnm{{Rawle}}, \binits{T.}},
\bauthor{\bsnm{{Parlanti}}, \binits{E.}},
\bauthor{\bsnm{{Arribas}}, \binits{S.}},
\bauthor{\bsnm{{Bunker}}, \binits{A.J.}},
\bauthor{\bsnm{{Cameron}}, \binits{A.J.}},
\bauthor{\bsnm{{Charlot}}, \binits{S.}},
\bauthor{\bsnm{{Curti}}, \binits{M.}},
\bauthor{\bsnm{{D'Eugenio}}, \binits{F.}},
\bauthor{\bsnm{{Jones}}, \binits{G.C.}},
\bauthor{\bsnm{{Kumari}}, \binits{N.}},
\bauthor{\bsnm{{Maiolino}}, \binits{R.}},
\bauthor{\bsnm{{{\"U}bler}}, \binits{H.}},
\bauthor{\bsnm{{Saxena}}, \binits{A.}},
\bauthor{\bsnm{{Smit}}, \binits{R.}},
\bauthor{\bsnm{{Willott}}, \binits{C.}},
\bauthor{\bsnm{{Witstok}}, \binits{J.}}:
\batitle{{The NIRSpec Wide GTO Survey}}.
\bjtitle{\aap}
\bvolume{689},
\bfpage{73}
(\byear{2024})
\doiurl{10.1051/0004-6361/202449914}
{\href{https://arxiv.org/abs/2403.05506}{{arXiv:2403.05506}}}
{[astro-ph.GA]}
\end{barticle}
\endbibitem

\bibitem[\protect\citeauthoryear{{Bezanson} et~al.}{2024}]{Bezanson2024}
\begin{barticle}
\bauthor{\bsnm{{Bezanson}}, \binits{R.}},
\bauthor{\bsnm{{Labbe}}, \binits{I.}},
\bauthor{\bsnm{{Whitaker}}, \binits{K.E.}},
\bauthor{\bsnm{{Leja}}, \binits{J.}},
\bauthor{\bsnm{{Price}}, \binits{S.H.}},
\bauthor{\bsnm{{Franx}}, \binits{M.}},
\bauthor{\bsnm{{Brammer}}, \binits{G.}},
\bauthor{\bsnm{{Marchesini}}, \binits{D.}},
\bauthor{\bsnm{{Zitrin}}, \binits{A.}},
\bauthor{\bsnm{{Wang}}, \binits{B.}},
\bauthor{\bsnm{{Weaver}}, \binits{J.R.}},
\bauthor{\bsnm{{Furtak}}, \binits{L.J.}},
\bauthor{\bsnm{{Atek}}, \binits{H.}},
\bauthor{\bsnm{{Coe}}, \binits{D.}},
\bauthor{\bsnm{{Cutler}}, \binits{S.E.}},
\bauthor{\bsnm{{Dayal}}, \binits{P.}},
\bauthor{\bsnm{{van Dokkum}}, \binits{P.}},
\bauthor{\bsnm{{Feldmann}}, \binits{R.}},
\bauthor{\bsnm{{F{\"o}rster Schreiber}}, \binits{N.M.}},
\bauthor{\bsnm{{Fujimoto}}, \binits{S.}},
\bauthor{\bsnm{{Geha}}, \binits{M.}},
\bauthor{\bsnm{{Glazebrook}}, \binits{K.}},
\bauthor{\bsnm{{de Graaff}}, \binits{A.}},
\bauthor{\bsnm{{Greene}}, \binits{J.E.}},
\bauthor{\bsnm{{Juneau}}, \binits{S.}},
\bauthor{\bsnm{{Kassin}}, \binits{S.}},
\bauthor{\bsnm{{Kriek}}, \binits{M.}},
\bauthor{\bsnm{{Khullar}}, \binits{G.}},
\bauthor{\bsnm{{Maseda}}, \binits{M.}},
\bauthor{\bsnm{{Mowla}}, \binits{L.A.}},
\bauthor{\bsnm{{Muzzin}}, \binits{A.}},
\bauthor{\bsnm{{Nanayakkara}}, \binits{T.}},
\bauthor{\bsnm{{Nelson}}, \binits{E.J.}},
\bauthor{\bsnm{{Oesch}}, \binits{P.A.}},
\bauthor{\bsnm{{Pacifici}}, \binits{C.}},
\bauthor{\bsnm{{Pan}}, \binits{R.}},
\bauthor{\bsnm{{Papovich}}, \binits{C.}},
\bauthor{\bsnm{{Setton}}, \binits{D.J.}},
\bauthor{\bsnm{{Shapley}}, \binits{A.E.}},
\bauthor{\bsnm{{Smit}}, \binits{R.}},
\bauthor{\bsnm{{Stefanon}}, \binits{M.}},
\bauthor{\bsnm{{Taylor}}, \binits{E.N.}},
\bauthor{\bsnm{{Williams}}, \binits{C.C.}}:
\batitle{{The JWST UNCOVER Treasury Survey: Ultradeep NIRSpec and NIRCam
  Observations before the Epoch of Reionization}}.
\bjtitle{\apj}
\bvolume{974}(\bissue{1}),
\bfpage{92}
(\byear{2024})
\doiurl{10.3847/1538-4357/ad66cf}
{\href{https://arxiv.org/abs/2212.04026}{{arXiv:2212.04026}}}
{[astro-ph.GA]}
\end{barticle}
\endbibitem

\bibitem[\protect\citeauthoryear{{Price} et~al.}{2025}]{Price2025}
\begin{barticle}
\bauthor{\bsnm{{Price}}, \binits{S.H.}},
\bauthor{\bsnm{{Bezanson}}, \binits{R.}},
\bauthor{\bsnm{{Labbe}}, \binits{I.}},
\bauthor{\bsnm{{Furtak}}, \binits{L.J.}},
\bauthor{\bsnm{{de Graaff}}, \binits{A.}},
\bauthor{\bsnm{{Greene}}, \binits{J.E.}},
\bauthor{\bsnm{{Kokorev}}, \binits{V.}},
\bauthor{\bsnm{{Setton}}, \binits{D.J.}},
\bauthor{\bsnm{{Suess}}, \binits{K.A.}},
\bauthor{\bsnm{{Brammer}}, \binits{G.}},
\bauthor{\bsnm{{Cutler}}, \binits{S.E.}},
\bauthor{\bsnm{{Leja}}, \binits{J.}},
\bauthor{\bsnm{{Pan}}, \binits{R.}},
\bauthor{\bsnm{{Wang}}, \binits{B.}},
\bauthor{\bsnm{{Weaver}}, \binits{J.R.}},
\bauthor{\bsnm{{Whitaker}}, \binits{K.E.}},
\bauthor{\bsnm{{Atek}}, \binits{H.}},
\bauthor{\bsnm{{Burgasser}}, \binits{A.J.}},
\bauthor{\bsnm{{Chemerynska}}, \binits{I.}},
\bauthor{\bsnm{{Dayal}}, \binits{P.}},
\bauthor{\bsnm{{Feldmann}}, \binits{R.}},
\bauthor{\bsnm{{F{\"o}rster Schreiber}}, \binits{N.M.}},
\bauthor{\bsnm{{Fudamoto}}, \binits{Y.}},
\bauthor{\bsnm{{Fujimoto}}, \binits{S.}},
\bauthor{\bsnm{{Glazebrook}}, \binits{K.}},
\bauthor{\bsnm{{Goulding}}, \binits{A.D.}},
\bauthor{\bsnm{{Khullar}}, \binits{G.}},
\bauthor{\bsnm{{Kriek}}, \binits{M.}},
\bauthor{\bsnm{{Marchesini}}, \binits{D.}},
\bauthor{\bsnm{{Maseda}}, \binits{M.V.}},
\bauthor{\bsnm{{Miller}}, \binits{T.B.}},
\bauthor{\bsnm{{Muzzin}}, \binits{A.}},
\bauthor{\bsnm{{Nanayakkara}}, \binits{T.}},
\bauthor{\bsnm{{Nelson}}, \binits{E.}},
\bauthor{\bsnm{{Oesch}}, \binits{P.A.}},
\bauthor{\bsnm{{Shipley}}, \binits{H.}},
\bauthor{\bsnm{{Smit}}, \binits{R.}},
\bauthor{\bsnm{{Taylor}}, \binits{E.N.}},
\bauthor{\bsnm{{Dokkum}}, \binits{P.v.}},
\bauthor{\bsnm{{Williams}}, \binits{C.C.}},
\bauthor{\bsnm{{Zitrin}}, \binits{A.}}:
\batitle{{The UNCOVER Survey: First Release of Ultradeep JWST/NIRSpec PRISM
  Spectra for {\ensuremath{\sim}}700 Galaxies from z {\ensuremath{\sim}}
  0.3{\textendash}13 in A2744}}.
\bjtitle{\apj}
\bvolume{982}(\bissue{1}),
\bfpage{51}
(\byear{2025})
\doiurl{10.3847/1538-4357/adaec1}
{\href{https://arxiv.org/abs/2408.03920}{{arXiv:2408.03920}}}
{[astro-ph.GA]}
\end{barticle}
\endbibitem

\bibitem[\protect\citeauthoryear{{Suess} et~al.}{2024}]{Suess2024}
\begin{barticle}
\bauthor{\bsnm{{Suess}}, \binits{K.A.}},
\bauthor{\bsnm{{Weaver}}, \binits{J.R.}},
\bauthor{\bsnm{{Price}}, \binits{S.H.}},
\bauthor{\bsnm{{Pan}}, \binits{R.}},
\bauthor{\bsnm{{Wang}}, \binits{B.}},
\bauthor{\bsnm{{Bezanson}}, \binits{R.}},
\bauthor{\bsnm{{Brammer}}, \binits{G.}},
\bauthor{\bsnm{{Cutler}}, \binits{S.E.}},
\bauthor{\bsnm{{Labb{\'e}}}, \binits{I.}},
\bauthor{\bsnm{{Leja}}, \binits{J.}},
\bauthor{\bsnm{{Williams}}, \binits{C.C.}},
\bauthor{\bsnm{{Whitaker}}, \binits{K.E.}},
\bauthor{\bsnm{{Atek}}, \binits{H.}},
\bauthor{\bsnm{{Dayal}}, \binits{P.}},
\bauthor{\bsnm{{de Graaff}}, \binits{A.}},
\bauthor{\bsnm{{Feldmann}}, \binits{R.}},
\bauthor{\bsnm{{Franx}}, \binits{M.}},
\bauthor{\bsnm{{Fudamoto}}, \binits{Y.}},
\bauthor{\bsnm{{Fujimoto}}, \binits{S.}},
\bauthor{\bsnm{{Furtak}}, \binits{L.J.}},
\bauthor{\bsnm{{Goulding}}, \binits{A.D.}},
\bauthor{\bsnm{{Greene}}, \binits{J.E.}},
\bauthor{\bsnm{{Khullar}}, \binits{G.}},
\bauthor{\bsnm{{Kokorev}}, \binits{V.}},
\bauthor{\bsnm{{Kriek}}, \binits{M.}},
\bauthor{\bsnm{{Lorenz}}, \binits{B.}},
\bauthor{\bsnm{{Marchesini}}, \binits{D.}},
\bauthor{\bsnm{{Maseda}}, \binits{M.V.}},
\bauthor{\bsnm{{Matthee}}, \binits{J.}},
\bauthor{\bsnm{{Miller}}, \binits{T.B.}},
\bauthor{\bsnm{{Mitsuhashi}}, \binits{I.}},
\bauthor{\bsnm{{Mowla}}, \binits{L.A.}},
\bauthor{\bsnm{{Muzzin}}, \binits{A.}},
\bauthor{\bsnm{{Naidu}}, \binits{R.P.}},
\bauthor{\bsnm{{Nanayakkara}}, \binits{T.}},
\bauthor{\bsnm{{Nelson}}, \binits{E.J.}},
\bauthor{\bsnm{{Oesch}}, \binits{P.A.}},
\bauthor{\bsnm{{Setton}}, \binits{D.J.}},
\bauthor{\bsnm{{Shipley}}, \binits{H.}},
\bauthor{\bsnm{{Smit}}, \binits{R.}},
\bauthor{\bsnm{{Spilker}}, \binits{J.S.}},
\bauthor{\bsnm{{van Dokkum}}, \binits{P.}},
\bauthor{\bsnm{{Zitrin}}, \binits{A.}}:
\batitle{{Medium Bands, Mega Science: A JWST/NIRCam Medium-band Imaging Survey
  of A2744}}.
\bjtitle{\apj}
\bvolume{976}(\bissue{1}),
\bfpage{101}
(\byear{2024})
\doiurl{10.3847/1538-4357/ad75fe}
{\href{https://arxiv.org/abs/2404.13132}{{arXiv:2404.13132}}}
{[astro-ph.GA]}
\end{barticle}
\endbibitem

\bibitem[\protect\citeauthoryear{{Baggen} et~al.}{2026}]{Baggen2026}
\begin{botherref}
\oauthor{\bsnm{{Baggen}}, \binits{J.F.W.}},
\oauthor{\bsnm{{Scoggins}}, \binits{M.T.}},
\oauthor{\bsnm{{van Dokkum}}, \binits{P.}},
\oauthor{\bsnm{{Haiman}}, \binits{Z.}},
\oauthor{\bsnm{{Torralba}}, \binits{A.}},
\oauthor{\bsnm{{Matthee}}, \binits{J.}}:
{Connecting the Dots: UV-Bright Companions of Little Red Dots as Lyman-Werner
  Sources Enabling Direct Collapse Black Hole Formation}.
arXiv e-prints,
2602--02702
(2026)
{\href{https://arxiv.org/abs/2602.02702}{{arXiv:2602.02702}}}
{[astro-ph.GA]}
\end{botherref}
\endbibitem

\bibitem[\protect\citeauthoryear{{Rinaldi} et~al.}{2025}]{Rinaldi2025}
\begin{botherref}
\oauthor{\bsnm{{Rinaldi}}, \binits{P.}},
\oauthor{\bsnm{{Rieke}}, \binits{G.H.}},
\oauthor{\bsnm{{Wu}}, \binits{Z.}},
\oauthor{\bsnm{{Gilbert}}, \binits{C.J.E.}},
\oauthor{\bsnm{{Pacucci}}, \binits{F.}},
\oauthor{\bsnm{{Barchiesi}}, \binits{L.}},
\oauthor{\bsnm{{Alberts}}, \binits{S.}},
\oauthor{\bsnm{{Carniani}}, \binits{S.}},
\oauthor{\bsnm{{Bunker}}, \binits{A.J.}},
\oauthor{\bsnm{{Bhatawdekar}}, \binits{R.}},
\oauthor{\bsnm{{D'Eugenio}}, \binits{F.}},
\oauthor{\bsnm{{Ji}}, \binits{Z.}},
\oauthor{\bsnm{{Johnson}}, \binits{B.D.}},
\oauthor{\bsnm{{Hainline}}, \binits{K.}},
\oauthor{\bsnm{{Kokorev}}, \binits{V.}},
\oauthor{\bsnm{{Kumari}}, \binits{N.}},
\oauthor{\bsnm{{Iani}}, \binits{E.}},
\oauthor{\bsnm{{Lyu}}, \binits{J.}},
\oauthor{\bsnm{{Maiolino}}, \binits{R.}},
\oauthor{\bsnm{{Parlanti}}, \binits{E.}},
\oauthor{\bsnm{{Robertson}}, \binits{B.E.}},
\oauthor{\bsnm{{Sun}}, \binits{Y.}},
\oauthor{\bsnm{{Vignali}}, \binits{C.}},
\oauthor{\bsnm{{Williams}}, \binits{C.C.}},
\oauthor{\bsnm{{Willmer}}, \binits{C.N.A.}},
\oauthor{\bsnm{{Zhu}}, \binits{Y.}}:
{Beyond the Dot: an LRD-like nucleus at the Heart of an IR-Bright Galaxy and
  its implications for high-redshift LRDs}.
arXiv e-prints,
2507--17738
(2025)
\doiurl{10.48550/arXiv.2507.17738}
{\href{https://arxiv.org/abs/2507.17738}{{arXiv:2507.17738}}}
{[astro-ph.GA]}
\end{botherref}
\endbibitem

\bibitem[\protect\citeauthoryear{{Wang} et~al.}{2025}]{Wang2025:brd}
\begin{barticle}
\bauthor{\bsnm{{Wang}}, \binits{B.}},
\bauthor{\bsnm{{de Graaff}}, \binits{A.}},
\bauthor{\bsnm{{Davies}}, \binits{R.L.}},
\bauthor{\bsnm{{Greene}}, \binits{J.E.}},
\bauthor{\bsnm{{Leja}}, \binits{J.}},
\bauthor{\bsnm{{Brammer}}, \binits{G.B.}},
\bauthor{\bsnm{{Goulding}}, \binits{A.D.}},
\bauthor{\bsnm{{Miller}}, \binits{T.B.}},
\bauthor{\bsnm{{Suess}}, \binits{K.A.}},
\bauthor{\bsnm{{Weibel}}, \binits{A.}},
\bauthor{\bsnm{{Williams}}, \binits{C.C.}},
\bauthor{\bsnm{{Bezanson}}, \binits{R.}},
\bauthor{\bsnm{{Boogaard}}, \binits{L.A.}},
\bauthor{\bsnm{{Cleri}}, \binits{N.J.}},
\bauthor{\bsnm{{Hirschmann}}, \binits{M.}},
\bauthor{\bsnm{{Katz}}, \binits{H.}},
\bauthor{\bsnm{{Labb{\'e}}}, \binits{I.}},
\bauthor{\bsnm{{Maseda}}, \binits{M.V.}},
\bauthor{\bsnm{{Matthee}}, \binits{J.}},
\bauthor{\bsnm{{McConachie}}, \binits{I.}},
\bauthor{\bsnm{{Naidu}}, \binits{R.P.}},
\bauthor{\bsnm{{Oesch}}, \binits{P.A.}},
\bauthor{\bsnm{{Rix}}, \binits{H.-W.}},
\bauthor{\bsnm{{Setton}}, \binits{D.J.}},
\bauthor{\bsnm{{Whitaker}}, \binits{K.E.}}:
\batitle{{RUBIES: JWST/NIRSpec Confirmation of an Infrared-luminous, Broad-line
  Little Red Dot with an Ionized Outflow}}.
\bjtitle{\apj}
\bvolume{984}(\bissue{2}),
\bfpage{121}
(\byear{2025})
\doiurl{10.3847/1538-4357/adc1ca}
{\href{https://arxiv.org/abs/2403.02304}{{arXiv:2403.02304}}}
{[astro-ph.GA]}
\end{barticle}
\endbibitem

\bibitem[\protect\citeauthoryear{{Johnson} et~al.}{2021}]{Johnson2021}
\begin{barticle}
\bauthor{\bsnm{{Johnson}}, \binits{B.D.}},
\bauthor{\bsnm{{Leja}}, \binits{J.}},
\bauthor{\bsnm{{Conroy}}, \binits{C.}},
\bauthor{\bsnm{{Speagle}}, \binits{J.S.}}:
\batitle{{Stellar Population Inference with Prospector}}.
\bjtitle{\apjs}
\bvolume{254}(\bissue{2}),
\bfpage{22}
(\byear{2021})
\doiurl{10.3847/1538-4365/abef67}
{\href{https://arxiv.org/abs/2012.01426}{{arXiv:2012.01426}}}
{[astro-ph.GA]}
\end{barticle}
\endbibitem

\bibitem[\protect\citeauthoryear{{Richards} et~al.}{2006}]{Richards2006}
\begin{barticle}
\bauthor{\bsnm{{Richards}}, \binits{G.T.}},
\bauthor{\bsnm{{Lacy}}, \binits{M.}},
\bauthor{\bsnm{{Storrie-Lombardi}}, \binits{L.J.}},
\bauthor{\bsnm{{Hall}}, \binits{P.B.}},
\bauthor{\bsnm{{Gallagher}}, \binits{S.C.}},
\bauthor{\bsnm{{Hines}}, \binits{D.C.}},
\bauthor{\bsnm{{Fan}}, \binits{X.}},
\bauthor{\bsnm{{Papovich}}, \binits{C.}},
\bauthor{\bsnm{{Vanden Berk}}, \binits{D.E.}},
\bauthor{\bsnm{{Trammell}}, \binits{G.B.}},
\bauthor{\bsnm{{Schneider}}, \binits{D.P.}},
\bauthor{\bsnm{{Vestergaard}}, \binits{M.}},
\bauthor{\bsnm{{York}}, \binits{D.G.}},
\bauthor{\bsnm{{Jester}}, \binits{S.}},
\bauthor{\bsnm{{Anderson}}, \binits{S.F.}},
\bauthor{\bsnm{{Budav{\'a}ri}}, \binits{T.}},
\bauthor{\bsnm{{Szalay}}, \binits{A.S.}}:
\batitle{{Spectral Energy Distributions and Multiwavelength Selection of Type 1
  Quasars}}.
\bjtitle{\apjs}
\bvolume{166}(\bissue{2}),
\bfpage{470}--\blpage{497}
(\byear{2006})
\doiurl{10.1086/506525}
{\href{https://arxiv.org/abs/astro-ph/0601558}{{arXiv:astro-ph/0601558}}}
{[astro-ph]}
\end{barticle}
\endbibitem

\bibitem[\protect\citeauthoryear{{Shen} et~al.}{2020}]{Shen2020}
\begin{barticle}
\bauthor{\bsnm{{Shen}}, \binits{X.}},
\bauthor{\bsnm{{Hopkins}}, \binits{P.F.}},
\bauthor{\bsnm{{Faucher-Gigu{\`e}re}}, \binits{C.-A.}},
\bauthor{\bsnm{{Alexander}}, \binits{D.M.}},
\bauthor{\bsnm{{Richards}}, \binits{G.T.}},
\bauthor{\bsnm{{Ross}}, \binits{N.P.}},
\bauthor{\bsnm{{Hickox}}, \binits{R.C.}}:
\batitle{{The bolometric quasar luminosity function at z = 0-7}}.
\bjtitle{\mnras}
\bvolume{495}(\bissue{3}),
\bfpage{3252}--\blpage{3275}
(\byear{2020})
\doiurl{10.1093/mnras/staa1381}
{\href{https://arxiv.org/abs/2001.02696}{{arXiv:2001.02696}}}
{[astro-ph.GA]}
\end{barticle}
\endbibitem

\bibitem[\protect\citeauthoryear{{de Graaff}
  et~al.}{2024}]{deGraaff2024:msafit}
\begin{barticle}
\bauthor{\bsnm{{de Graaff}}, \binits{A.}},
\bauthor{\bsnm{{Rix}}, \binits{H.-W.}},
\bauthor{\bsnm{{Carniani}}, \binits{S.}},
\bauthor{\bsnm{{Suess}}, \binits{K.A.}},
\bauthor{\bsnm{{Charlot}}, \binits{S.}},
\bauthor{\bsnm{{Curtis-Lake}}, \binits{E.}},
\bauthor{\bsnm{{Arribas}}, \binits{S.}},
\bauthor{\bsnm{{Baker}}, \binits{W.M.}},
\bauthor{\bsnm{{Boyett}}, \binits{K.}},
\bauthor{\bsnm{{Bunker}}, \binits{A.J.}},
\bauthor{\bsnm{{Cameron}}, \binits{A.J.}},
\bauthor{\bsnm{{Chevallard}}, \binits{J.}},
\bauthor{\bsnm{{Curti}}, \binits{M.}},
\bauthor{\bsnm{{Eisenstein}}, \binits{D.J.}},
\bauthor{\bsnm{{Franx}}, \binits{M.}},
\bauthor{\bsnm{{Hainline}}, \binits{K.}},
\bauthor{\bsnm{{Hausen}}, \binits{R.}},
\bauthor{\bsnm{{Ji}}, \binits{Z.}},
\bauthor{\bsnm{{Johnson}}, \binits{B.D.}},
\bauthor{\bsnm{{Jones}}, \binits{G.C.}},
\bauthor{\bsnm{{Maiolino}}, \binits{R.}},
\bauthor{\bsnm{{Maseda}}, \binits{M.V.}},
\bauthor{\bsnm{{Nelson}}, \binits{E.}},
\bauthor{\bsnm{{Parlanti}}, \binits{E.}},
\bauthor{\bsnm{{Rawle}}, \binits{T.}},
\bauthor{\bsnm{{Robertson}}, \binits{B.}},
\bauthor{\bsnm{{Tacchella}}, \binits{S.}},
\bauthor{\bsnm{{{\"U}bler}}, \binits{H.}},
\bauthor{\bsnm{{Williams}}, \binits{C.C.}},
\bauthor{\bsnm{{Willmer}}, \binits{C.N.A.}},
\bauthor{\bsnm{{Willott}}, \binits{C.}}:
\batitle{{Ionised gas kinematics and dynamical masses of z
  {\ensuremath{\gtrsim}} 6 galaxies from JADES/NIRSpec high-resolution
  spectroscopy}}.
\bjtitle{\aap}
\bvolume{684},
\bfpage{87}
(\byear{2024})
\doiurl{10.1051/0004-6361/202347755}
{\href{https://arxiv.org/abs/2308.09742}{{arXiv:2308.09742}}}
{[astro-ph.GA]}
\end{barticle}
\endbibitem

\bibitem[\protect\citeauthoryear{{Galavis} et~al.}{1997}]{Galavis1997}
\begin{barticle}
\bauthor{\bsnm{{Galavis}}, \binits{M.E.}},
\bauthor{\bsnm{{Mendoza}}, \binits{C.}},
\bauthor{\bsnm{{Zeippen}}, \binits{C.J.}}:
\batitle{{Atomic data from the IRON Project. XXII. Radiative rates for
  forbidden transitions within the ground configuration of ions in the carbon
  and oxygen isoelectronic sequences}}.
\bjtitle{\aaps}
\bvolume{123},
\bfpage{159}--\blpage{171}
(\byear{1997})
\doiurl{10.1051/aas:1997344}
\end{barticle}
\endbibitem

\bibitem[\protect\citeauthoryear{Watanabe}{2010}]{Watanabe2010}
\begin{barticle}
\bauthor{\bsnm{Watanabe}, \binits{S.}}:
\batitle{Asymptotic equivalence of bayes cross validation and widely applicable
  information criterion in singular learning theory}.
\bjtitle{Journal of Machine Learning Research}
\bvolume{11}(\bissue{116}),
\bfpage{3571}--\blpage{3594}
(\byear{2010})
\end{barticle}
\endbibitem

\bibitem[\protect\citeauthoryear{{Naidu} et~al.}{2024}]{Naidu2024}
\begin{botherref}
\oauthor{\bsnm{{Naidu}}, \binits{R.P.}},
\oauthor{\bsnm{{Matthee}}, \binits{J.}},
\oauthor{\bsnm{{Kramarenko}}, \binits{I.}},
\oauthor{\bsnm{{Weibel}}, \binits{A.}},
\oauthor{\bsnm{{Brammer}}, \binits{G.}},
\oauthor{\bsnm{{Oesch}}, \binits{P.A.}},
\oauthor{\bsnm{{Lechner}}, \binits{P.}},
\oauthor{\bsnm{{Furtak}}, \binits{L.J.}},
\oauthor{\bsnm{{Di Cesare}}, \binits{C.}},
\oauthor{\bsnm{{Torralba}}, \binits{A.}},
\oauthor{\bsnm{{Kotiwale}}, \binits{G.}},
\oauthor{\bsnm{{Bezanson}}, \binits{R.}},
\oauthor{\bsnm{{Bouwens}}, \binits{R.J.}},
\oauthor{\bsnm{{Chandra}}, \binits{V.}},
\oauthor{\bsnm{{Claeyssens}}, \binits{A.}},
\oauthor{\bsnm{{Danhaive}}, \binits{A.L.}},
\oauthor{\bsnm{{Frebel}}, \binits{A.}},
\oauthor{\bsnm{{de Graaff}}, \binits{A.}},
\oauthor{\bsnm{{Greene}}, \binits{J.E.}},
\oauthor{\bsnm{{Heintz}}, \binits{K.E.}},
\oauthor{\bsnm{{Ji}}, \binits{A.P.}},
\oauthor{\bsnm{{Kashino}}, \binits{D.}},
\oauthor{\bsnm{{Katz}}, \binits{H.}},
\oauthor{\bsnm{{Labbe}}, \binits{I.}},
\oauthor{\bsnm{{Leja}}, \binits{J.}},
\oauthor{\bsnm{{Li}}, \binits{Y.}},
\oauthor{\bsnm{{Maseda}}, \binits{M.V.}},
\oauthor{\bsnm{{Richard}}, \binits{J.}},
\oauthor{\bsnm{{Shivaei}}, \binits{I.}},
\oauthor{\bsnm{{Simcoe}}, \binits{R.A.}},
\oauthor{\bsnm{{Sobral}}, \binits{D.}},
\oauthor{\bsnm{{Suess}}, \binits{K.A.}},
\oauthor{\bsnm{{Tacchella}}, \binits{S.}},
\oauthor{\bsnm{{Williams}}, \binits{C.C.}}:
{All the Little Things in Abell 2744: $>$1000 Gravitationally Lensed Dwarf
  Galaxies at $z=0-9$ from JWST NIRCam Grism Spectroscopy}.
arXiv e-prints,
2410--01874
(2024)
\doiurl{10.48550/arXiv.2410.01874}
{\href{https://arxiv.org/abs/2410.01874}{{arXiv:2410.01874}}}
{[astro-ph.GA]}
\end{botherref}
\endbibitem

\bibitem[\protect\citeauthoryear{{Kokorev} et~al.}{2024}]{Kokorev2024}
\begin{barticle}
\bauthor{\bsnm{{Kokorev}}, \binits{V.}},
\bauthor{\bsnm{{Caputi}}, \binits{K.I.}},
\bauthor{\bsnm{{Greene}}, \binits{J.E.}},
\bauthor{\bsnm{{Dayal}}, \binits{P.}},
\bauthor{\bsnm{{Trebitsch}}, \binits{M.}},
\bauthor{\bsnm{{Cutler}}, \binits{S.E.}},
\bauthor{\bsnm{{Fujimoto}}, \binits{S.}},
\bauthor{\bsnm{{Labb{\'e}}}, \binits{I.}},
\bauthor{\bsnm{{Miller}}, \binits{T.B.}},
\bauthor{\bsnm{{Iani}}, \binits{E.}},
\bauthor{\bsnm{{Navarro-Carrera}}, \binits{R.}},
\bauthor{\bsnm{{Rinaldi}}, \binits{P.}}:
\batitle{{A Census of Photometrically Selected Little Red Dots at 4 < z < 9 in
  JWST Blank Fields}}.
\bjtitle{\apj}
\bvolume{968}(\bissue{1}),
\bfpage{38}
(\byear{2024})
\doiurl{10.3847/1538-4357/ad4265}
{\href{https://arxiv.org/abs/2401.09981}{{arXiv:2401.09981}}}
{[astro-ph.GA]}
\end{barticle}
\endbibitem

\bibitem[\protect\citeauthoryear{{Ma} et~al.}{2025}]{Ma2025}
\begin{botherref}
\oauthor{\bsnm{{Ma}}, \binits{Y.}},
\oauthor{\bsnm{{Greene}}, \binits{J.E.}},
\oauthor{\bsnm{{Setton}}, \binits{D.J.}},
\oauthor{\bsnm{{Goulding}}, \binits{A.D.}},
\oauthor{\bsnm{{Annunziatella}}, \binits{M.}},
\oauthor{\bsnm{{Fan}}, \binits{X.}},
\oauthor{\bsnm{{Kokorev}}, \binits{V.}},
\oauthor{\bsnm{{Labbe}}, \binits{I.}},
\oauthor{\bsnm{{Li}}, \binits{J.}},
\oauthor{\bsnm{{Lin}}, \binits{X.}},
\oauthor{\bsnm{{Marchesini}}, \binits{D.}},
\oauthor{\bsnm{{Matthee}}, \binits{J.}},
\oauthor{\bsnm{{Robbins}}, \binits{L.}},
\oauthor{\bsnm{{Sajina}}, \binits{A.}},
\oauthor{\bsnm{{Sawicki}}, \binits{M.}},
\oauthor{\bsnm{{Telford}}, \binits{O.G.}}:
{Counting Little Red Dots at $z<4$ with Ground-based Surveys and Spectroscopic
  Follow-up}.
arXiv e-prints,
2504--08032
(2025)
\doiurl{10.48550/arXiv.2504.08032}
{\href{https://arxiv.org/abs/2504.08032}{{arXiv:2504.08032}}}
{[astro-ph.GA]}
\end{botherref}
\endbibitem

\bibitem[\protect\citeauthoryear{{Lin} et~al.}{2025}]{Lin2025}
\begin{botherref}
\oauthor{\bsnm{{Lin}}, \binits{X.}},
\oauthor{\bsnm{{Fan}}, \binits{X.}},
\oauthor{\bsnm{{Cai}}, \binits{Z.}},
\oauthor{\bsnm{{Bian}}, \binits{F.}},
\oauthor{\bsnm{{Liu}}, \binits{H.}},
\oauthor{\bsnm{{Sun}}, \binits{F.}},
\oauthor{\bsnm{{Ma}}, \binits{Y.}},
\oauthor{\bsnm{{Greene}}, \binits{J.E.}},
\oauthor{\bsnm{{Strauss}}, \binits{M.A.}},
\oauthor{\bsnm{{Green}}, \binits{R.}},
\oauthor{\bsnm{{Lyu}}, \binits{J.}},
\oauthor{\bsnm{{Champagne}}, \binits{J.B.}},
\oauthor{\bsnm{{Goulding}}, \binits{A.D.}},
\oauthor{\bsnm{{Inayoshi}}, \binits{K.}},
\oauthor{\bsnm{{Jin}}, \binits{X.}},
\oauthor{\bsnm{{Leung}}, \binits{G.C.K.}},
\oauthor{\bsnm{{Li}}, \binits{M.}},
\oauthor{\bsnm{{Liu}}, \binits{Y.}},
\oauthor{\bsnm{{Mao}}, \binits{J.}},
\oauthor{\bsnm{{Pudoka}}, \binits{M.A.}},
\oauthor{\bsnm{{Tee}}, \binits{W.L.}},
\oauthor{\bsnm{{Wang}}, \binits{B.}},
\oauthor{\bsnm{{Wang}}, \binits{F.}},
\oauthor{\bsnm{{Wu}}, \binits{Y.}},
\oauthor{\bsnm{{Yang}}, \binits{J.}},
\oauthor{\bsnm{{Zhang}}, \binits{H.}},
\oauthor{\bsnm{{Zhu}}, \binits{Y.}}:
{The Discovery of Little Red Dots in the Local Universe: Signatures of Cool Gas
  Envelopes}.
arXiv e-prints,
2507--10659
(2025)
\doiurl{10.48550/arXiv.2507.10659}
{\href{https://arxiv.org/abs/2507.10659}{{arXiv:2507.10659}}}
{[astro-ph.GA]}
\end{botherref}
\endbibitem

\bibitem[\protect\citeauthoryear{{Torralba} et~al.}{2025}]{Torralba2025}
\begin{botherref}
\oauthor{\bsnm{{Torralba}}, \binits{A.}},
\oauthor{\bsnm{{Matthee}}, \binits{J.}},
\oauthor{\bsnm{{Pezzulli}}, \binits{G.}},
\oauthor{\bsnm{{Naidu}}, \binits{R.P.}},
\oauthor{\bsnm{{Ishikawa}}, \binits{Y.}},
\oauthor{\bsnm{{Brammer}}, \binits{G.B.}},
\oauthor{\bsnm{{Chang}}, \binits{S.-J.}},
\oauthor{\bsnm{{Chisholm}}, \binits{J.}},
\oauthor{\bsnm{{de Graaff}}, \binits{A.}},
\oauthor{\bsnm{{D'Eugenio}}, \binits{F.}},
\oauthor{\bsnm{{Di Cesare}}, \binits{C.}},
\oauthor{\bsnm{{Eilers}}, \binits{A.-C.}},
\oauthor{\bsnm{{Greene}}, \binits{J.E.}},
\oauthor{\bsnm{{Gronke}}, \binits{M.}},
\oauthor{\bsnm{{Iani}}, \binits{E.}},
\oauthor{\bsnm{{Kokorev}}, \binits{V.}},
\oauthor{\bsnm{{Kotiwale}}, \binits{G.}},
\oauthor{\bsnm{{Kramarenko}}, \binits{I.}},
\oauthor{\bsnm{{Ma}}, \binits{Y.}},
\oauthor{\bsnm{{Mascia}}, \binits{S.}},
\oauthor{\bsnm{{Navarrete}}, \binits{B.}},
\oauthor{\bsnm{{Nelson}}, \binits{E.}},
\oauthor{\bsnm{{Oesch}}, \binits{P.}},
\oauthor{\bsnm{{Simcoe}}, \binits{R.A.}},
\oauthor{\bsnm{{Wuyts}}, \binits{S.}}:
{The warm outer layer of a Little Red Dot as the source of [Fe II] and
  collisional Balmer lines with scattering wings}.
arXiv e-prints,
2510--00103
(2025)
\doiurl{10.48550/arXiv.2510.00103}
{\href{https://arxiv.org/abs/2510.00103}{{arXiv:2510.00103}}}
{[astro-ph.GA]}
\end{botherref}
\endbibitem

\bibitem[\protect\citeauthoryear{{Lan{\c{c}}on} and
  {Mouhcine}}{2002}]{Lancon2002}
\begin{barticle}
\bauthor{\bsnm{{Lan{\c{c}}on}}, \binits{A.}},
\bauthor{\bsnm{{Mouhcine}}, \binits{M.}}:
\batitle{{The modelling of intermediate-age stellar populations. II. Average
  spectra for upper AGB stars, and their use}}.
\bjtitle{\aap}
\bvolume{393},
\bfpage{167}--\blpage{181}
(\byear{2002})
\doiurl{10.1051/0004-6361:20020585}
{\href{https://arxiv.org/abs/astro-ph/0206252}{{arXiv:astro-ph/0206252}}}
{[astro-ph]}
\end{barticle}
\endbibitem

\bibitem[\protect\citeauthoryear{{Conroy} and {Gunn}}{2010}]{Conroy2010}
\begin{barticle}
\bauthor{\bsnm{{Conroy}}, \binits{C.}},
\bauthor{\bsnm{{Gunn}}, \binits{J.E.}}:
\batitle{{The Propagation of Uncertainties in Stellar Population Synthesis
  Modeling. III. Model Calibration, Comparison, and Evaluation}}.
\bjtitle{\apj}
\bvolume{712}(\bissue{2}),
\bfpage{833}--\blpage{857}
(\byear{2010})
\doiurl{10.1088/0004-637X/712/2/833}
{\href{https://arxiv.org/abs/0911.3151}{{arXiv:0911.3151}}}
{[astro-ph.CO]}
\end{barticle}
\endbibitem

\bibitem[\protect\citeauthoryear{{Kroupa}}{2001}]{Kroupa2001}
\begin{barticle}
\bauthor{\bsnm{{Kroupa}}, \binits{P.}}:
\batitle{{On the variation of the initial mass function}}.
\bjtitle{\mnras}
\bvolume{322}(\bissue{2}),
\bfpage{231}--\blpage{246}
(\byear{2001})
\doiurl{10.1046/j.1365-8711.2001.04022.x}
{\href{https://arxiv.org/abs/astro-ph/0009005}{{arXiv:astro-ph/0009005}}}
{[astro-ph]}
\end{barticle}
\endbibitem

\bibitem[\protect\citeauthoryear{{Chabrier}}{2003}]{Chabrier2003}
\begin{barticle}
\bauthor{\bsnm{{Chabrier}}, \binits{G.}}:
\batitle{{Galactic Stellar and Substellar Initial Mass Function}}.
\bjtitle{\pasp}
\bvolume{115}(\bissue{809}),
\bfpage{763}--\blpage{795}
(\byear{2003})
\doiurl{10.1086/376392}
{\href{https://arxiv.org/abs/astro-ph/0304382}{{arXiv:astro-ph/0304382}}}
{[astro-ph]}
\end{barticle}
\endbibitem

\bibitem[\protect\citeauthoryear{{Marley} and {Robinson}}{2015}]{Marley2015}
\begin{barticle}
\bauthor{\bsnm{{Marley}}, \binits{M.S.}},
\bauthor{\bsnm{{Robinson}}, \binits{T.D.}}:
\batitle{{On the Cool Side: Modeling the Atmospheres of Brown Dwarfs and Giant
  Planets}}.
\bjtitle{\araa}
\bvolume{53},
\bfpage{279}--\blpage{323}
(\byear{2015})
\doiurl{10.1146/annurev-astro-082214-122522}
{\href{https://arxiv.org/abs/1410.6512}{{arXiv:1410.6512}}}
{[astro-ph.EP]}
\end{barticle}
\endbibitem

\bibitem[\protect\citeauthoryear{{Allard} et~al.}{2012}]{Allard2012}
\begin{barticle}
\bauthor{\bsnm{{Allard}}, \binits{F.}},
\bauthor{\bsnm{{Homeier}}, \binits{D.}},
\bauthor{\bsnm{{Freytag}}, \binits{B.}}:
\batitle{{Models of very-low-mass stars, brown dwarfs and exoplanets}}.
\bjtitle{Philosophical Transactions of the Royal Society of London Series A}
\bvolume{370}(\bissue{1968}),
\bfpage{2765}--\blpage{2777}
(\byear{2012})
\doiurl{10.1098/rsta.2011.0269}
{\href{https://arxiv.org/abs/1112.3591}{{arXiv:1112.3591}}}
{[astro-ph.SR]}
\end{barticle}
\endbibitem

\bibitem[\protect\citeauthoryear{{Speagle}}{2020}]{Speagle2020}
\begin{barticle}
\bauthor{\bsnm{{Speagle}}, \binits{J.S.}}:
\batitle{{DYNESTY: a dynamic nested sampling package for estimating Bayesian
  posteriors and evidences}}.
\bjtitle{\mnras}
\bvolume{493}(\bissue{3}),
\bfpage{3132}--\blpage{3158}
(\byear{2020})
\doiurl{10.1093/mnras/staa278}
{\href{https://arxiv.org/abs/1904.02180}{{arXiv:1904.02180}}}
{[astro-ph.IM]}
\end{barticle}
\endbibitem

\bibitem[\protect\citeauthoryear{{Koposov} et~al.}{2024}]{Koposov2024}
\begin{botherref}
\oauthor{\bsnm{{Koposov}}, \binits{S.}},
\oauthor{\bsnm{{Speagle}}, \binits{J.}},
\oauthor{\bsnm{{Barbary}}, \binits{K.}},
\oauthor{\bsnm{{Ashton}}, \binits{G.}},
\oauthor{\bsnm{{Bennett}}, \binits{E.}},
\oauthor{\bsnm{{Buchner}}, \binits{J.}},
\oauthor{\bsnm{{Scheffler}}, \binits{C.}},
\oauthor{\bsnm{{Cook}}, \binits{B.}},
\oauthor{\bsnm{{Talbot}}, \binits{C.}},
\oauthor{\bsnm{{Guillochon}}, \binits{J.}},
\oauthor{\bsnm{{Cubillos}}, \binits{P.}},
\oauthor{\bsnm{{Asensio Ramos}}, \binits{A.}},
\oauthor{\bsnm{{Dartiailh}}, \binits{M.}},
\oauthor{\bsnm{{Ilya}}},
\oauthor{\bsnm{{Tollerud}}, \binits{E.}},
\oauthor{\bsnm{{Lang}}, \binits{D.}},
\oauthor{\bsnm{{Johnson}}, \binits{B.}},
\oauthor{\bsnm{{jtmendel}}},
\oauthor{\bsnm{{Higson}}, \binits{E.}},
\oauthor{\bsnm{{Vandal}}, \binits{T.}},
\oauthor{\bsnm{{Daylan}}, \binits{T.}},
\oauthor{\bsnm{{Angus}}, \binits{R.}},
\oauthor{\bsnm{{patelR}}},
\oauthor{\bsnm{{Cargile}}, \binits{P.}},
\oauthor{\bsnm{{Sheehan}}, \binits{P.}},
\oauthor{\bsnm{{Pitkin}}, \binits{M.}},
\oauthor{\bsnm{{Kirk}}, \binits{M.}},
\oauthor{\bsnm{{Leja}}, \binits{J.}},
\oauthor{\bsnm{{joezuntz}}},
\oauthor{\bsnm{{Goldstein}}, \binits{D.}}:
{joshspeagle/dynesty: V2.1.4}.
\doiurl{10.5281/zenodo.12537467}
\end{botherref}
\endbibitem

\bibitem[\protect\citeauthoryear{{Sun} et~al.}{2026}]{Sun2026}
\begin{botherref}
\oauthor{\bsnm{{Sun}}, \binits{W.Q.}},
\oauthor{\bsnm{{Naidu}}, \binits{R.P.}},
\oauthor{\bsnm{{Matthee}}, \binits{J.}},
\oauthor{\bsnm{{de Graaff}}, \binits{A.}},
\oauthor{\bsnm{{Chisholm}}, \binits{J.}},
\oauthor{\bsnm{{Greene}}, \binits{J.E.}},
\oauthor{\bsnm{{Oesch}}, \binits{P.A.}},
\oauthor{\bsnm{{Torralba}}, \binits{A.}},
\oauthor{\bsnm{{Hviding}}, \binits{R.E.}},
\oauthor{\bsnm{{Brammer}}, \binits{G.}},
\oauthor{\bsnm{{Simcoe}}, \binits{R.A.}},
\oauthor{\bsnm{{Bose}}, \binits{S.}},
\oauthor{\bsnm{{Bouwens}}, \binits{R.}},
\oauthor{\bsnm{{Dayal}}, \binits{P.}},
\oauthor{\bsnm{{Eilers}}, \binits{A.-C.}},
\oauthor{\bsnm{{Fei}}, \binits{Q.}},
\oauthor{\bsnm{{Furtak}}, \binits{L.J.}},
\oauthor{\bsnm{{Gottumukkala}}, \binits{R.}},
\oauthor{\bsnm{{Goulding}}, \binits{A.}},
\oauthor{\bsnm{{Heintz}}, \binits{K.E.}},
\oauthor{\bsnm{{Hirschmann}}, \binits{M.}},
\oauthor{\bsnm{{Kokorev}}, \binits{V.}},
\oauthor{\bsnm{{Leja}}, \binits{J.}},
\oauthor{\bsnm{{Liu}}, \binits{Z.}},
\oauthor{\bsnm{{Natarajan}}, \binits{P.}},
\oauthor{\bsnm{{Santarelli}}, \binits{A.D.}},
\oauthor{\bsnm{{Setton}}, \binits{D.J.}},
\oauthor{\bsnm{{Smith}}, \binits{A.}},
\oauthor{\bsnm{{Tacchella}}, \binits{S.}},
\oauthor{\bsnm{{Volonteri}}, \binits{M.}},
\oauthor{\bsnm{{Walter}}, \binits{F.}},
\oauthor{\bsnm{{Weibel}}, \binits{A.}},
\oauthor{\bsnm{{Williams}}, \binits{C.C.}}:
{Little Red Dot $-$ Host Galaxy $=$ Black Hole Star: A Gas-Enshrouded Heart at
  the Center of Every Little Red Dot}.
arXiv e-prints,
2601--20929
(2026)
\doiurl{10.48550/arXiv.2601.20929}
{\href{https://arxiv.org/abs/2601.20929}{{arXiv:2601.20929}}}
{[astro-ph.GA]}
\end{botherref}
\endbibitem

\bibitem[\protect\citeauthoryear{{Hubeny}}{1988}]{Hubeny1988}
\begin{barticle}
\bauthor{\bsnm{{Hubeny}}, \binits{I.}}:
\batitle{{A computer program for calculating non-LTE model stellar
  atmospheres}}.
\bjtitle{Computer Physics Communications}
\bvolume{52}(\bissue{1}),
\bfpage{103}--\blpage{132}
(\byear{1988})
\doiurl{10.1016/0010-4655(88)90177-4}
\end{barticle}
\endbibitem

\bibitem[\protect\citeauthoryear{{Hubeny} et~al.}{2021}]{Hubeny2021}
\begin{botherref}
\oauthor{\bsnm{{Hubeny}}, \binits{I.}},
\oauthor{\bsnm{{Allende Prieto}}, \binits{C.}},
\oauthor{\bsnm{{Osorio}}, \binits{Y.}},
\oauthor{\bsnm{{Lanz}}, \binits{T.}}:
{TLUSTY and SYNSPEC Users's Guide IV: Upgraded Versions 208 and 54}.
arXiv e-prints,
2104--02829
(2021)
\doiurl{10.48550/arXiv.2104.02829}
{\href{https://arxiv.org/abs/2104.02829}{{arXiv:2104.02829}}}
{[astro-ph.SR]}
\end{botherref}
\endbibitem

\bibitem[\protect\citeauthoryear{{Lanz} and {Hubeny}}{2007}]{Lanz2007}
\begin{barticle}
\bauthor{\bsnm{{Lanz}}, \binits{T.}},
\bauthor{\bsnm{{Hubeny}}, \binits{I.}}:
\batitle{{A Grid of NLTE Line-blanketed Model Atmospheres of Early B-Type
  Stars}}.
\bjtitle{\apjs}
\bvolume{169}(\bissue{1}),
\bfpage{83}--\blpage{104}
(\byear{2007})
\doiurl{10.1086/511270}
{\href{https://arxiv.org/abs/astro-ph/0611891}{{arXiv:astro-ph/0611891}}}
{[astro-ph]}
\end{barticle}
\endbibitem

\bibitem[\protect\citeauthoryear{{Osorio} et~al.}{2020}]{Osorio2020}
\begin{barticle}
\bauthor{\bsnm{{Osorio}}, \binits{Y.}},
\bauthor{\bsnm{{Allende Prieto}}, \binits{C.}},
\bauthor{\bsnm{{Hubeny}}, \binits{I.}},
\bauthor{\bsnm{{M{\'e}sz{\'a}ros}}, \binits{S.}},
\bauthor{\bsnm{{Shetrone}}, \binits{M.}}:
\batitle{{NLTE for APOGEE: simultaneous multi-element NLTE radiative
  transfer}}.
\bjtitle{\aap}
\bvolume{637},
\bfpage{80}
(\byear{2020})
\doiurl{10.1051/0004-6361/201937054}
{\href{https://arxiv.org/abs/2003.13353}{{arXiv:2003.13353}}}
{[astro-ph.SR]}
\end{barticle}
\endbibitem

\bibitem[\protect\citeauthoryear{{Hubeny} and {Hubeny}}{1998}]{Hubeny1998}
\begin{barticle}
\bauthor{\bsnm{{Hubeny}}, \binits{I.}},
\bauthor{\bsnm{{Hubeny}}, \binits{V.}}:
\batitle{{Non-LTE Models and Theoretical Spectra of Accretion Disks in Active
  Galactic Nuclei. II. Vertical Structure of the Disk}}.
\bjtitle{\apj}
\bvolume{505}(\bissue{2}),
\bfpage{558}--\blpage{576}
(\byear{1998})
\doiurl{10.1086/306207}
{\href{https://arxiv.org/abs/astro-ph/9804288}{{arXiv:astro-ph/9804288}}}
{[astro-ph]}
\end{barticle}
\endbibitem

\bibitem[\protect\citeauthoryear{{Hubeny} et~al.}{2001}]{Hubeny2001}
\begin{barticle}
\bauthor{\bsnm{{Hubeny}}, \binits{I.}},
\bauthor{\bsnm{{Blaes}}, \binits{O.}},
\bauthor{\bsnm{{Krolik}}, \binits{J.H.}},
\bauthor{\bsnm{{Agol}}, \binits{E.}}:
\batitle{{Non-LTE Models and Theoretical Spectra of Accretion Disks in Active
  Galactic Nuclei. IV. Effects of Compton Scattering and Metal Opacities}}.
\bjtitle{\apj}
\bvolume{559}(\bissue{2}),
\bfpage{680}--\blpage{702}
(\byear{2001})
\doiurl{10.1086/322344}
{\href{https://arxiv.org/abs/astro-ph/0105507}{{arXiv:astro-ph/0105507}}}
{[astro-ph]}
\end{barticle}
\endbibitem

\bibitem[\protect\citeauthoryear{{Hui} et~al.}{2005}]{Hui2005}
\begin{barticle}
\bauthor{\bsnm{{Hui}}, \binits{Y.}},
\bauthor{\bsnm{{Krolik}}, \binits{J.H.}},
\bauthor{\bsnm{{Hubeny}}, \binits{I.}}:
\batitle{{Non-LTE Spectra of Accretion Disks around Intermediate-Mass Black
  Holes}}.
\bjtitle{\apj}
\bvolume{625}(\bissue{2}),
\bfpage{913}--\blpage{922}
(\byear{2005})
\doiurl{10.1086/429884}
{\href{https://arxiv.org/abs/astro-ph/0502355}{{arXiv:astro-ph/0502355}}}
{[astro-ph]}
\end{barticle}
\endbibitem

\bibitem[\protect\citeauthoryear{{Asplund} et~al.}{2009}]{Asplund2009}
\begin{barticle}
\bauthor{\bsnm{{Asplund}}, \binits{M.}},
\bauthor{\bsnm{{Grevesse}}, \binits{N.}},
\bauthor{\bsnm{{Sauval}}, \binits{A.J.}},
\bauthor{\bsnm{{Scott}}, \binits{P.}}:
\batitle{{The Chemical Composition of the Sun}}.
\bjtitle{\araa}
\bvolume{47}(\bissue{1}),
\bfpage{481}--\blpage{522}
(\byear{2009})
\doiurl{10.1146/annurev.astro.46.060407.145222}
{\href{https://arxiv.org/abs/0909.0948}{{arXiv:0909.0948}}}
{[astro-ph.SR]}
\end{barticle}
\endbibitem

\bibitem[\protect\citeauthoryear{{Polyansky} et~al.}{2018}]{Polyansky2018}
\begin{barticle}
\bauthor{\bsnm{{Polyansky}}, \binits{O.L.}},
\bauthor{\bsnm{{Kyuberis}}, \binits{A.A.}},
\bauthor{\bsnm{{Zobov}}, \binits{N.F.}},
\bauthor{\bsnm{{Tennyson}}, \binits{J.}},
\bauthor{\bsnm{{Yurchenko}}, \binits{S.N.}},
\bauthor{\bsnm{{Lodi}}, \binits{L.}}:
\batitle{{ExoMol molecular line lists XXX: a complete high-accuracy line list
  for water}}.
\bjtitle{\mnras}
\bvolume{480}(\bissue{2}),
\bfpage{2597}--\blpage{2608}
(\byear{2018})
\doiurl{10.1093/mnras/sty1877}
{\href{https://arxiv.org/abs/1807.04529}{{arXiv:1807.04529}}}
{[astro-ph.EP]}
\end{barticle}
\endbibitem

\bibitem[\protect\citeauthoryear{{Liu} et~al.}{2025}]{Liu:inprep}
\begin{botherref}
\oauthor{\bsnm{{Liu}}, \binits{H.}}, et al.
in prep
(2025)
\end{botherref}
\endbibitem

\bibitem[\protect\citeauthoryear{{Ji} et~al.}{2026}]{Ji2025:lord}
\begin{barticle}
\bauthor{\bsnm{{Ji}}, \binits{X.}},
\bauthor{\bsnm{{D'Eugenio}}, \binits{F.}},
\bauthor{\bsnm{{Juod{\v{z}}balis}}, \binits{I.}},
\bauthor{\bsnm{{Walton}}, \binits{D.J.}},
\bauthor{\bsnm{{Fabian}}, \binits{A.C.}},
\bauthor{\bsnm{{Maiolino}}, \binits{R.}},
\bauthor{\bsnm{{Ramos Almeida}}, \binits{C.}},
\bauthor{\bsnm{{Acosta Pulido}}, \binits{J.A.}},
\bauthor{\bsnm{{Belokurov}}, \binits{V.A.}},
\bauthor{\bsnm{{Isobe}}, \binits{Y.}},
\bauthor{\bsnm{{Jones}}, \binits{G.}},
\bauthor{\bsnm{{Maraston}}, \binits{C.}},
\bauthor{\bsnm{{Scholtz}}, \binits{J.}},
\bauthor{\bsnm{{Simmonds}}, \binits{C.}},
\bauthor{\bsnm{{Tacchella}}, \binits{S.}},
\bauthor{\bsnm{{Terlevich}}, \binits{E.}},
\bauthor{\bsnm{{Terlevich}}, \binits{R.}}:
\batitle{{Lord of LRDs: insights into a 'Little Red Dot' with a low-ionization
  spectrum at z = 0.1}}.
\bjtitle{\mnras}
\bvolume{545}(\bissue{3}),
\bfpage{2235}
(\byear{2026})
\doiurl{10.1093/mnras/staf2235}
{\href{https://arxiv.org/abs/2507.23774}{{arXiv:2507.23774}}}
{[astro-ph.GA]}
\end{barticle}
\endbibitem

\bibitem[\protect\citeauthoryear{{Labb{\'e}} et~al.}{2023}]{Labbe2023}
\begin{barticle}
\bauthor{\bsnm{{Labb{\'e}}}, \binits{I.}},
\bauthor{\bsnm{{van Dokkum}}, \binits{P.}},
\bauthor{\bsnm{{Nelson}}, \binits{E.}},
\bauthor{\bsnm{{Bezanson}}, \binits{R.}},
\bauthor{\bsnm{{Suess}}, \binits{K.A.}},
\bauthor{\bsnm{{Leja}}, \binits{J.}},
\bauthor{\bsnm{{Brammer}}, \binits{G.}},
\bauthor{\bsnm{{Whitaker}}, \binits{K.}},
\bauthor{\bsnm{{Mathews}}, \binits{E.}},
\bauthor{\bsnm{{Stefanon}}, \binits{M.}},
\bauthor{\bsnm{{Wang}}, \binits{B.}}:
\batitle{{A population of red candidate massive galaxies 600 Myr after the Big
  Bang}}.
\bjtitle{\nat}
\bvolume{616}(\bissue{7956}),
\bfpage{266}--\blpage{269}
(\byear{2023})
\doiurl{10.1038/s41586-023-05786-2}
{\href{https://arxiv.org/abs/2207.12446}{{arXiv:2207.12446}}}
{[astro-ph.GA]}
\end{barticle}
\endbibitem

\bibitem[\protect\citeauthoryear{{Maiolino}
  et~al.}{2025}]{Maiolino2025:metpoor}
\begin{botherref}
\oauthor{\bsnm{{Maiolino}}, \binits{R.}},
\oauthor{\bsnm{{Uebler}}, \binits{H.}},
\oauthor{\bsnm{{D'Eugenio}}, \binits{F.}},
\oauthor{\bsnm{{Scholtz}}, \binits{J.}},
\oauthor{\bsnm{{Juodzbalis}}, \binits{I.}},
\oauthor{\bsnm{{Ji}}, \binits{X.}},
\oauthor{\bsnm{{Perna}}, \binits{M.}},
\oauthor{\bsnm{{Bromm}}, \binits{V.}},
\oauthor{\bsnm{{Dayal}}, \binits{P.}},
\oauthor{\bsnm{{Koudmani}}, \binits{S.}},
\oauthor{\bsnm{{Liu}}, \binits{B.}},
\oauthor{\bsnm{{Schneider}}, \binits{R.}},
\oauthor{\bsnm{{Sijacki}}, \binits{D.}},
\oauthor{\bsnm{{Valiante}}, \binits{R.}},
\oauthor{\bsnm{{Trinca}}, \binits{A.}},
\oauthor{\bsnm{{Zhang}}, \binits{S.}},
\oauthor{\bsnm{{Volonteri}}, \binits{M.}},
\oauthor{\bsnm{{Inayoshi}}, \binits{K.}},
\oauthor{\bsnm{{Carniani}}, \binits{S.}},
\oauthor{\bsnm{{Nakajima}}, \binits{K.}},
\oauthor{\bsnm{{Isobe}}, \binits{Y.}},
\oauthor{\bsnm{{Witstok}}, \binits{J.}},
\oauthor{\bsnm{{Jones}}, \binits{G.C.}},
\oauthor{\bsnm{{Tacchella}}, \binits{S.}},
\oauthor{\bsnm{{Arribas}}, \binits{S.}},
\oauthor{\bsnm{{Bunker}}, \binits{A.}},
\oauthor{\bsnm{{Cataldi}}, \binits{E.}},
\oauthor{\bsnm{{Charlot}}, \binits{S.}},
\oauthor{\bsnm{{Cresci}}, \binits{G.}},
\oauthor{\bsnm{{Curti}}, \binits{M.}},
\oauthor{\bsnm{{Fabian}}, \binits{A.C.}},
\oauthor{\bsnm{{Katz}}, \binits{H.}},
\oauthor{\bsnm{{Kumari}}, \binits{N.}},
\oauthor{\bsnm{{Laporte}}, \binits{N.}},
\oauthor{\bsnm{{Mazzolari}}, \binits{G.}},
\oauthor{\bsnm{{Robertson}}, \binits{B.}},
\oauthor{\bsnm{{Sun}}, \binits{F.}},
\oauthor{\bsnm{{Rodriguez Del Pino}}, \binits{B.}},
\oauthor{\bsnm{{Venturi}}, \binits{G.}}:
{A black hole in a near-pristine galaxy 700 million years after the Big Bang}.
arXiv e-prints,
2505--22567
(2025)
\doiurl{10.48550/arXiv.2505.22567}
{\href{https://arxiv.org/abs/2505.22567}{{arXiv:2505.22567}}}
{[astro-ph.GA]}
\end{botherref}
\endbibitem

\bibitem[\protect\citeauthoryear{{Lambrides} et~al.}{2025}]{Lambrides2025}
\begin{botherref}
\oauthor{\bsnm{{Lambrides}}, \binits{E.}},
\oauthor{\bsnm{{Larson}}, \binits{R.}},
\oauthor{\bsnm{{Hutchison}}, \binits{T.}},
\oauthor{\bsnm{{Arrabal Haro}}, \binits{P.}},
\oauthor{\bsnm{{Wang}}, \binits{B.}},
\oauthor{\bsnm{{Welch}}, \binits{B.}},
\oauthor{\bsnm{{Kocevski}}, \binits{D.D.}},
\oauthor{\bsnm{{Richardson}}, \binits{C.T.}},
\oauthor{\bsnm{{Papovich}}, \binits{C.}},
\oauthor{\bsnm{{Trump}}, \binits{J.R.}},
\oauthor{\bsnm{{Bosman}}, \binits{S.E.I.}},
\oauthor{\bsnm{{Rigby}}, \binits{J.R.}},
\oauthor{\bsnm{{Finkelstein}}, \binits{S.L.}},
\oauthor{\bsnm{{Barro}}, \binits{G.}},
\oauthor{\bsnm{{Antwi-Danso}}, \binits{J.}},
\oauthor{\bsnm{{Long}}, \binits{A.}},
\oauthor{\bsnm{{Taylor}}, \binits{A.J.}},
\oauthor{\bsnm{{Cann}}, \binits{J.}},
\oauthor{\bsnm{{McKaig}}, \binits{J.}},
\oauthor{\bsnm{{Koekemoer}}, \binits{A.M.}},
\oauthor{\bsnm{{Cleri}}, \binits{N.J.}},
\oauthor{\bsnm{{Akins}}, \binits{H.B.}},
\oauthor{\bsnm{{Bagley}}, \binits{M.B.}},
\oauthor{\bsnm{{Berg}}, \binits{D.A.}},
\oauthor{\bsnm{{Bromm}}, \binits{V.}},
\oauthor{\bsnm{{Chisholm}}, \binits{J.}},
\oauthor{\bsnm{{Chworowsky}}, \binits{K.}},
\oauthor{\bsnm{{Coffin}}, \binits{S.}},
\oauthor{\bsnm{{Cooper}}, \binits{M.C.}},
\oauthor{\bsnm{{Cooper}}, \binits{O.}},
\oauthor{\bsnm{{Cox}}, \binits{I.}},
\oauthor{\bsnm{{Dickinson}}, \binits{M.}},
\oauthor{\bsnm{{Ferguson}}, \binits{H.C.}},
\oauthor{\bsnm{{Franco}}, \binits{M.}},
\oauthor{\bsnm{{Gardner}}, \binits{J.P.}},
\oauthor{\bsnm{{Grogin}}, \binits{N.A.}},
\oauthor{\bsnm{{Hirschmann}}, \binits{M.}},
\oauthor{\bsnm{{Huertas-Company}}, \binits{M.}},
\oauthor{\bsnm{{Jung}}, \binits{I.}},
\oauthor{\bsnm{{Kartaltepe}}, \binits{J.S.}},
\oauthor{\bsnm{{Khullar}}, \binits{G.P.}},
\oauthor{\bsnm{{Lucas}}, \binits{R.A.}},
\oauthor{\bsnm{{McGrath}}, \binits{E.J.}},
\oauthor{\bsnm{{Morales}}, \binits{A.M.}},
\oauthor{\bsnm{{Olivier}}, \binits{G.M.}},
\oauthor{\bsnm{{Ch{\'a}vez Ortiz}}, \binits{{\'O}.A.}},
\oauthor{\bsnm{{P{\'e}rez-Gonz{\'a}lez}}, \binits{P.G.}},
\oauthor{\bsnm{{Pirzkal}}, \binits{N.}},
\oauthor{\bsnm{{Somerville}}, \binits{R.S.}},
\oauthor{\bsnm{{Vanderhoof}}, \binits{B.}},
\oauthor{\bsnm{{Weiner}}, \binits{B.J.}},
\oauthor{\bsnm{{Yung}}, \binits{L.Y.A.}},
\oauthor{\bsnm{{Zavala}}, \binits{J.A.}}:
{Discovery of Multiply Ionized Iron Emission Powered by an Active Galactic
  Nucleus in a z\raisebox{-0.5ex}\textasciitilde7 Little Red Dot}.
arXiv e-prints,
2509--09607
(2025)
\doiurl{10.48550/arXiv.2509.09607}
{\href{https://arxiv.org/abs/2509.09607}{{arXiv:2509.09607}}}
{[astro-ph.GA]}
\end{botherref}
\endbibitem

\bibitem[\protect\citeauthoryear{{Armandroff} and
  {Zinn}}{1988}]{Armandroff1988}
\begin{barticle}
\bauthor{\bsnm{{Armandroff}}, \binits{T.E.}},
\bauthor{\bsnm{{Zinn}}, \binits{R.}}:
\batitle{{Integrated-Light Spectroscopy of Globular Clusters at the Infrared CA
  II Lines}}.
\bjtitle{\aj}
\bvolume{96},
\bfpage{92}
(\byear{1988})
\doiurl{10.1086/114792}
\end{barticle}
\endbibitem

\bibitem[\protect\citeauthoryear{{Carrera} et~al.}{2013}]{Carrera2013}
\begin{barticle}
\bauthor{\bsnm{{Carrera}}, \binits{R.}},
\bauthor{\bsnm{{Pancino}}, \binits{E.}},
\bauthor{\bsnm{{Gallart}}, \binits{C.}},
\bauthor{\bsnm{{del Pino}}, \binits{A.}}:
\batitle{{The near-infrared Ca II triplet as a metallicity indicator - II.
  Extension to extremely metal-poor metallicity regimes}}.
\bjtitle{\mnras}
\bvolume{434}(\bissue{2}),
\bfpage{1681}--\blpage{1691}
(\byear{2013})
\doiurl{10.1093/mnras/stt1126}
{\href{https://arxiv.org/abs/1306.3883}{{arXiv:1306.3883}}}
{[astro-ph.GA]}
\end{barticle}
\endbibitem

\bibitem[\protect\citeauthoryear{{Inayoshi} et~al.}{2022}]{Inayoshi2022}
\begin{barticle}
\bauthor{\bsnm{{Inayoshi}}, \binits{K.}},
\bauthor{\bsnm{{Onoue}}, \binits{M.}},
\bauthor{\bsnm{{Sugahara}}, \binits{Y.}},
\bauthor{\bsnm{{Inoue}}, \binits{A.K.}},
\bauthor{\bsnm{{Ho}}, \binits{L.C.}}:
\batitle{{The Age of Discovery with the James Webb Space Telescope: Excavating
  the Spectral Signatures of the First Massive Black Holes}}.
\bjtitle{\apjl}
\bvolume{931}(\bissue{2}),
\bfpage{25}
(\byear{2022})
\doiurl{10.3847/2041-8213/ac6f01}
{\href{https://arxiv.org/abs/2204.09692}{{arXiv:2204.09692}}}
{[astro-ph.GA]}
\end{barticle}
\endbibitem

\bibitem[\protect\citeauthoryear{{Tripodi} et~al.}{2025}]{Tripodi2025}
\begin{barticle}
\bauthor{\bsnm{{Tripodi}}, \binits{R.}},
\bauthor{\bsnm{{Brada{\v{c}}}}, \binits{M.}},
\bauthor{\bsnm{{D'Eugenio}}, \binits{F.}},
\bauthor{\bsnm{{Martis}}, \binits{N.}},
\bauthor{\bsnm{{Rihtar{\v{s}}i{\v{c}}}}, \binits{G.}},
\bauthor{\bsnm{{Willott}}, \binits{C.}},
\bauthor{\bsnm{{Pentericci}}, \binits{L.}},
\bauthor{\bsnm{{Moreschini}}, \binits{B.}},
\bauthor{\bsnm{{Markevitch}}, \binits{M.}},
\bauthor{\bsnm{{Asada}}, \binits{Y.}},
\bauthor{\bsnm{{Calabr{\'o}}}, \binits{A.}},
\bauthor{\bsnm{{Desprez}}, \binits{G.}},
\bauthor{\bsnm{{Felicioni}}, \binits{G.}},
\bauthor{\bsnm{{Gaspar}}, \binits{G.}},
\bauthor{\bsnm{{Gonzalez}}, \binits{A.H.}},
\bauthor{\bsnm{{Harshan}}, \binits{A.}},
\bauthor{\bsnm{{Ji}}, \binits{X.}},
\bauthor{\bsnm{{Jude{\v{z}}}}, \binits{J.}},
\bauthor{\bsnm{{Lemaux}}, \binits{B.C.}},
\bauthor{\bsnm{{Marconi}}, \binits{A.}},
\bauthor{\bsnm{{Markov}}, \binits{V.}},
\bauthor{\bsnm{{Merida}}, \binits{R.M.}},
\bauthor{\bsnm{{Napolitano}}, \binits{L.}},
\bauthor{\bsnm{{Noirot}}, \binits{G.}},
\bauthor{\bsnm{{Parente}}, \binits{M.}},
\bauthor{\bsnm{{Peter}}, \binits{A.H.G.}},
\bauthor{\bsnm{{Robbins}}, \binits{L.}},
\bauthor{\bsnm{{Robertson}}, \binits{A.}},
\bauthor{\bsnm{{Sarrouh}}, \binits{G.T.E.}},
\bauthor{\bsnm{{Sawicki}}, \binits{M.}}:
\batitle{{A Deep Dive down the Broad-line Region: Permitted O I, Ca II, and Fe
  II Emission in an Active Galactic Nucleus Little Red Dot at z = 5.3}}.
\bjtitle{\apjl}
\bvolume{994}(\bissue{1}),
\bfpage{6}
(\byear{2025})
\doiurl{10.3847/2041-8213/ae13a9}
{\href{https://arxiv.org/abs/2507.20684}{{arXiv:2507.20684}}}
{[astro-ph.GA]}
\end{barticle}
\endbibitem

\bibitem[\protect\citeauthoryear{{Kirkpatrick} et~al.}{1991}]{Kirkpatrick1991}
\begin{barticle}
\bauthor{\bsnm{{Kirkpatrick}}, \binits{J.D.}},
\bauthor{\bsnm{{Henry}}, \binits{T.J.}},
\bauthor{\bsnm{{McCarthy}}, \binits{D.W.} \bsuffix{Jr.}}:
\batitle{{A Standard Stellar Spectral Sequence in the Red/Near-Infrared:
  Classes K5 to M9}}.
\bjtitle{\apjs}
\bvolume{77},
\bfpage{417}
(\byear{1991})
\doiurl{10.1086/191611}
\end{barticle}
\endbibitem

\bibitem[\protect\citeauthoryear{{Ji} et~al.}{2025}]{Ji2025:qso}
\begin{barticle}
\bauthor{\bsnm{{Ji}}, \binits{X.}},
\bauthor{\bsnm{{Maiolino}}, \binits{R.}},
\bauthor{\bsnm{{{\"U}bler}}, \binits{H.}},
\bauthor{\bsnm{{Scholtz}}, \binits{J.}},
\bauthor{\bsnm{{D'Eugenio}}, \binits{F.}},
\bauthor{\bsnm{{Sun}}, \binits{F.}},
\bauthor{\bsnm{{Perna}}, \binits{M.}},
\bauthor{\bsnm{{Turner}}, \binits{H.}},
\bauthor{\bsnm{{Carniani}}, \binits{S.}},
\bauthor{\bsnm{{Arribas}}, \binits{S.}},
\bauthor{\bsnm{{Bennett}}, \binits{J.S.}},
\bauthor{\bsnm{{Bunker}}, \binits{A.}},
\bauthor{\bsnm{{Charlot}}, \binits{S.}},
\bauthor{\bsnm{{Cresci}}, \binits{G.}},
\bauthor{\bsnm{{Curti}}, \binits{M.}},
\bauthor{\bsnm{{Egami}}, \binits{E.}},
\bauthor{\bsnm{{Fabian}}, \binits{A.}},
\bauthor{\bsnm{{Inayoshi}}, \binits{K.}},
\bauthor{\bsnm{{Isobe}}, \binits{Y.}},
\bauthor{\bsnm{{Jones}}, \binits{G.}},
\bauthor{\bsnm{{Juod{\v{z}}balis}}, \binits{I.}},
\bauthor{\bsnm{{Kumari}}, \binits{N.}},
\bauthor{\bsnm{{Lyu}}, \binits{J.}},
\bauthor{\bsnm{{Mazzolari}}, \binits{G.}},
\bauthor{\bsnm{{Parlanti}}, \binits{E.}},
\bauthor{\bsnm{{Robertson}}, \binits{B.}},
\bauthor{\bsnm{{Rodr{\'\i}guez Del Pino}}, \binits{B.}},
\bauthor{\bsnm{{Schneider}}, \binits{R.}},
\bauthor{\bsnm{{Sijacki}}, \binits{D.}},
\bauthor{\bsnm{{Tacchella}}, \binits{S.}},
\bauthor{\bsnm{{Trinca}}, \binits{A.}},
\bauthor{\bsnm{{Valiante}}, \binits{R.}},
\bauthor{\bsnm{{Venturi}}, \binits{G.}},
\bauthor{\bsnm{{Volonteri}}, \binits{M.}},
\bauthor{\bsnm{{Willott}}, \binits{C.}},
\bauthor{\bsnm{{Witten}}, \binits{C.}},
\bauthor{\bsnm{{Witstok}}, \binits{J.}}:
\batitle{{BlackTHUNDER -- A non-stellar Balmer break in a black hole-dominated
  little red dot at z = 7.04}}.
\bjtitle{\mnras}
\bvolume{544}(\bissue{4}),
\bfpage{3900}--\blpage{3935}
(\byear{2025})
\doiurl{10.1093/mnras/staf1867}
{\href{https://arxiv.org/abs/2501.13082}{{arXiv:2501.13082}}}
{[astro-ph.GA]}
\end{barticle}
\endbibitem

\bibitem[\protect\citeauthoryear{{Freytag} et~al.}{2002}]{Freytag2002}
\begin{barticle}
\bauthor{\bsnm{{Freytag}}, \binits{B.}},
\bauthor{\bsnm{{Steffen}}, \binits{M.}},
\bauthor{\bsnm{{Dorch}}, \binits{B.}}:
\batitle{{Spots on the surface of Betelgeuse -- Results from new 3D stellar
  convection models}}.
\bjtitle{Astronomische Nachrichten}
\bvolume{323},
\bfpage{213}--\blpage{219}
(\byear{2002})
\doiurl{10.1002/1521-3994(200208)323:3/4<213::AID-ASNA213>3.0.CO;2-H}
\end{barticle}
\endbibitem

\bibitem[\protect\citeauthoryear{{Chiavassa} et~al.}{2010}]{Chiavassa2010}
\begin{barticle}
\bauthor{\bsnm{{Chiavassa}}, \binits{A.}},
\bauthor{\bsnm{{Haubois}}, \binits{X.}},
\bauthor{\bsnm{{Young}}, \binits{J.S.}},
\bauthor{\bsnm{{Plez}}, \binits{B.}},
\bauthor{\bsnm{{Josselin}}, \binits{E.}},
\bauthor{\bsnm{{Perrin}}, \binits{G.}},
\bauthor{\bsnm{{Freytag}}, \binits{B.}}:
\batitle{{Radiative hydrodynamics simulations of red supergiant stars. II.
  Simulations of convection on Betelgeuse match interferometric observations}}.
\bjtitle{\aap}
\bvolume{515},
\bfpage{12}
(\byear{2010})
\doiurl{10.1051/0004-6361/200913907}
{\href{https://arxiv.org/abs/1003.1407}{{arXiv:1003.1407}}}
{[astro-ph.SR]}
\end{barticle}
\endbibitem

\bibitem[\protect\citeauthoryear{{Montarg{\`e}s} et~al.}{2021}]{Montarges2021}
\begin{barticle}
\bauthor{\bsnm{{Montarg{\`e}s}}, \binits{M.}},
\bauthor{\bsnm{{Cannon}}, \binits{E.}},
\bauthor{\bsnm{{Lagadec}}, \binits{E.}},
\bauthor{\bsnm{{de Koter}}, \binits{A.}},
\bauthor{\bsnm{{Kervella}}, \binits{P.}},
\bauthor{\bsnm{{Sanchez-Bermudez}}, \binits{J.}},
\bauthor{\bsnm{{Paladini}}, \binits{C.}},
\bauthor{\bsnm{{Cantalloube}}, \binits{F.}},
\bauthor{\bsnm{{Decin}}, \binits{L.}},
\bauthor{\bsnm{{Scicluna}}, \binits{P.}},
\bauthor{\bsnm{{Kravchenko}}, \binits{K.}},
\bauthor{\bsnm{{Dupree}}, \binits{A.K.}},
\bauthor{\bsnm{{Ridgway}}, \binits{S.}},
\bauthor{\bsnm{{Wittkowski}}, \binits{M.}},
\bauthor{\bsnm{{Anugu}}, \binits{N.}},
\bauthor{\bsnm{{Norris}}, \binits{R.}},
\bauthor{\bsnm{{Rau}}, \binits{G.}},
\bauthor{\bsnm{{Perrin}}, \binits{G.}},
\bauthor{\bsnm{{Chiavassa}}, \binits{A.}},
\bauthor{\bsnm{{Kraus}}, \binits{S.}},
\bauthor{\bsnm{{Monnier}}, \binits{J.D.}},
\bauthor{\bsnm{{Millour}}, \binits{F.}},
\bauthor{\bsnm{{Le Bouquin}}, \binits{J.-B.}},
\bauthor{\bsnm{{Haubois}}, \binits{X.}},
\bauthor{\bsnm{{Lopez}}, \binits{B.}},
\bauthor{\bsnm{{Stee}}, \binits{P.}},
\bauthor{\bsnm{{Danchi}}, \binits{W.}}:
\batitle{{A dusty veil shading Betelgeuse during its Great Dimming}}.
\bjtitle{\nat}
\bvolume{594}(\bissue{7863}),
\bfpage{365}--\blpage{368}
(\byear{2021})
\doiurl{10.1038/s41586-021-03546-8}
{\href{https://arxiv.org/abs/2201.10551}{{arXiv:2201.10551}}}
{[astro-ph.SR]}
\end{barticle}
\endbibitem

\bibitem[\protect\citeauthoryear{{Hall}}{1970}]{Hall1970}
\begin{botherref}
\oauthor{\bsnm{{Hall}}, \binits{D.N.B.}}:
{Observations of the Infrared Sunspot Spectrum Between 11340 and 24778
  Angstroms.}
PhD thesis,
Harvard University, Massachusetts
(January 1970)
\end{botherref}
\endbibitem

\bibitem[\protect\citeauthoryear{{Polyansky} et~al.}{1997}]{Polyansky1997}
\begin{barticle}
\bauthor{\bsnm{{Polyansky}}, \binits{O.L.}},
\bauthor{\bsnm{{Zobov}}, \binits{N.F.}},
\bauthor{\bsnm{{Viti}}, \binits{S.}},
\bauthor{\bsnm{{Tennyson}}, \binits{J.}},
\bauthor{\bsnm{{Bernath}}, \binits{P.F.}},
\bauthor{\bsnm{{Wallace}}, \binits{L.}}:
\batitle{{Water on the Sun: line assignments based on variational
  calculations.}}
\bjtitle{Science}
\bvolume{277}(\bissue{5324}),
\bfpage{346}--\blpage{348}
(\byear{1997})
\doiurl{10.1126/science.277.5324.346}
\end{barticle}
\endbibitem

\bibitem[\protect\citeauthoryear{{Inayoshi} et~al.}{2025}]{Inayoshi2025:binary}
\begin{botherref}
\oauthor{\bsnm{{Inayoshi}}, \binits{K.}},
\oauthor{\bsnm{{Shangguan}}, \binits{J.}},
\oauthor{\bsnm{{Chen}}, \binits{X.}},
\oauthor{\bsnm{{Ho}}, \binits{L.C.}},
\oauthor{\bsnm{{Haiman}}, \binits{Z.}}:
{The Emergence of Little Red Dots from Binary Massive Black Holes}.
arXiv e-prints,
2505--05322
(2025)
\doiurl{10.48550/arXiv.2505.05322}
{\href{https://arxiv.org/abs/2505.05322}{{arXiv:2505.05322}}}
{[astro-ph.HE]}
\end{botherref}
\endbibitem

\bibitem[\protect\citeauthoryear{{Zhang} et~al.}{2025}]{Zhang2025}
\begin{botherref}
\oauthor{\bsnm{{Zhang}}, \binits{C.}},
\oauthor{\bsnm{{Wu}}, \binits{Q.}},
\oauthor{\bsnm{{Fan}}, \binits{X.}},
\oauthor{\bsnm{{Ho}}, \binits{L.C.}},
\oauthor{\bsnm{{Wu}}, \binits{J.}},
\oauthor{\bsnm{{Zhang}}, \binits{H.}},
\oauthor{\bsnm{{Lyu}}, \binits{B.}},
\oauthor{\bsnm{{Cao}}, \binits{X.}},
\oauthor{\bsnm{{Wang}}, \binits{J.}}:
{The Composite Spectrum of the Little Red Dots from an Inner Standard Disk and
  an Outer Gravitationally Unstable Disk}.
arXiv e-prints,
2505--12719
(2025)
\doiurl{10.48550/arXiv.2505.12719}
{\href{https://arxiv.org/abs/2505.12719}{{arXiv:2505.12719}}}
{[astro-ph.HE]}
\end{botherref}
\endbibitem

\bibitem[\protect\citeauthoryear{{Zwick} et~al.}{2025}]{Zwick2025}
\begin{botherref}
\oauthor{\bsnm{{Zwick}}, \binits{L.}},
\oauthor{\bsnm{{Tiede}}, \binits{C.}},
\oauthor{\bsnm{{Mayer}}, \binits{L.}}:
{Little Red Dots as self-gravitating discs accreting on supermassive stars:
  Spectral appearance and formation pathway of the progenitors to direct
  collapse black holes}.
arXiv e-prints,
2507--22014
(2025)
\doiurl{10.48550/arXiv.2507.22014}
{\href{https://arxiv.org/abs/2507.22014}{{arXiv:2507.22014}}}
{[astro-ph.GA]}
\end{botherref}
\endbibitem

\bibitem[\protect\citeauthoryear{{Hillenbrand} et~al.}{1992}]{Hillenbrand1992}
\begin{barticle}
\bauthor{\bsnm{{Hillenbrand}}, \binits{L.A.}},
\bauthor{\bsnm{{Strom}}, \binits{S.E.}},
\bauthor{\bsnm{{Vrba}}, \binits{F.J.}},
\bauthor{\bsnm{{Keene}}, \binits{J.}}:
\batitle{{Herbig Ae/Be Stars: Intermediate-Mass Stars Surrounded by Massive
  Circumstellar Accretion Disks}}.
\bjtitle{\apj}
\bvolume{397},
\bfpage{613}
(\byear{1992})
\doiurl{10.1086/171819}
\end{barticle}
\endbibitem

\bibitem[\protect\citeauthoryear{{Rivinius} et~al.}{2013}]{Rivinius2013}
\begin{barticle}
\bauthor{\bsnm{{Rivinius}}, \binits{T.}},
\bauthor{\bsnm{{Carciofi}}, \binits{A.C.}},
\bauthor{\bsnm{{Martayan}}, \binits{C.}}:
\batitle{{Classical Be stars. Rapidly rotating B stars with viscous Keplerian
  decretion disks}}.
\bjtitle{\aapr}
\bvolume{21},
\bfpage{69}
(\byear{2013})
\doiurl{10.1007/s00159-013-0069-0}
{\href{https://arxiv.org/abs/1310.3962}{{arXiv:1310.3962}}}
{[astro-ph.SR]}
\end{barticle}
\endbibitem

\bibitem[\protect\citeauthoryear{{Matsuura} et~al.}{1999}]{Matsuura1999}
\begin{barticle}
\bauthor{\bsnm{{Matsuura}}, \binits{M.}},
\bauthor{\bsnm{{Yamamura}}, \binits{I.}},
\bauthor{\bsnm{{Murakami}}, \binits{H.}},
\bauthor{\bsnm{{Freund}}, \binits{M.M.}},
\bauthor{\bsnm{{Tanaka}}, \binits{M.}}:
\batitle{{Water vapor absorption in early M-type stars}}.
\bjtitle{\aap}
\bvolume{348},
\bfpage{579}--\blpage{583}
(\byear{1999})
\doiurl{10.48550/arXiv.astro-ph/9906264}
{\href{https://arxiv.org/abs/astro-ph/9906264}{{arXiv:astro-ph/9906264}}}
{[astro-ph]}
\end{barticle}
\endbibitem

\bibitem[\protect\citeauthoryear{{Tsuji}}{2000}]{Tsuji2000}
\begin{barticle}
\bauthor{\bsnm{{Tsuji}}, \binits{T.}}:
\batitle{{Water on the Early M Supergiant Stars {\ensuremath{\alpha}} Orionis
  and {\ensuremath{\mu}} Cephei}}.
\bjtitle{\apj}
\bvolume{538}(\bissue{2}),
\bfpage{801}--\blpage{807}
(\byear{2000})
\doiurl{10.1086/309185}
\end{barticle}
\endbibitem

\bibitem[\protect\citeauthoryear{{Ryde} et~al.}{2006}]{Ryde2006}
\begin{barticle}
\bauthor{\bsnm{{Ryde}}, \binits{N.}},
\bauthor{\bsnm{{Richter}}, \binits{M.J.}},
\bauthor{\bsnm{{Harper}}, \binits{G.M.}},
\bauthor{\bsnm{{Eriksson}}, \binits{K.}},
\bauthor{\bsnm{{Lambert}}, \binits{D.L.}}:
\batitle{{Water Vapor on Supergiants: The 12 {\ensuremath{\mu}}m TEXES Spectra
  of {\ensuremath{\mu}} Cephei}}.
\bjtitle{\apj}
\bvolume{645}(\bissue{1}),
\bfpage{652}--\blpage{658}
(\byear{2006})
\doiurl{10.1086/504287}
{\href{https://arxiv.org/abs/astro-ph/0603384}{{arXiv:astro-ph/0603384}}}
{[astro-ph]}
\end{barticle}
\endbibitem

\bibitem[\protect\citeauthoryear{{Kritos} et~al.}{2025}]{Kritos2025}
\begin{barticle}
\bauthor{\bsnm{{Kritos}}, \binits{K.}},
\bauthor{\bsnm{{Beckmann}}, \binits{R.S.}},
\bauthor{\bsnm{{Silk}}, \binits{J.}},
\bauthor{\bsnm{{Berti}}, \binits{E.}},
\bauthor{\bsnm{{Yi}}, \binits{S.}},
\bauthor{\bsnm{{Volonteri}}, \binits{M.}},
\bauthor{\bsnm{{Dubois}}, \binits{Y.}},
\bauthor{\bsnm{{Devriendt}}, \binits{J.}}:
\batitle{{Supermassive Black Hole Growth in Hierarchically Merging Nuclear Star
  Clusters}}.
\bjtitle{\apj}
\bvolume{991}(\bissue{1}),
\bfpage{58}
(\byear{2025})
\doiurl{10.3847/1538-4357/adeb44}
{\href{https://arxiv.org/abs/2412.15334}{{arXiv:2412.15334}}}
{[astro-ph.GA]}
\end{barticle}
\endbibitem

\bibitem[\protect\citeauthoryear{{Inayoshi} et~al.}{2025}]{Inayoshi2025:stars}
\begin{botherref}
\oauthor{\bsnm{{Inayoshi}}, \binits{K.}},
\oauthor{\bsnm{{Murase}}, \binits{K.}},
\oauthor{\bsnm{{Kashiyama}}, \binits{K.}}:
{Spectral Uniformity of Little Red Dots: A Natural Outcome of Coevolving Seed
  Black Holes and Nascent Starbursts}.
arXiv e-prints,
2509--19422
(2025)
\doiurl{10.48550/arXiv.2509.19422}
{\href{https://arxiv.org/abs/2509.19422}{{arXiv:2509.19422}}}
{[astro-ph.GA]}
\end{botherref}
\endbibitem

\bibitem[\protect\citeauthoryear{{Neumayer} et~al.}{2020}]{Neumayer2020}
\begin{barticle}
\bauthor{\bsnm{{Neumayer}}, \binits{N.}},
\bauthor{\bsnm{{Seth}}, \binits{A.}},
\bauthor{\bsnm{{B{\"o}ker}}, \binits{T.}}:
\batitle{{Nuclear star clusters}}.
\bjtitle{\aapr}
\bvolume{28}(\bissue{1}),
\bfpage{4}
(\byear{2020})
\doiurl{10.1007/s00159-020-00125-0}
{\href{https://arxiv.org/abs/2001.03626}{{arXiv:2001.03626}}}
{[astro-ph.GA]}
\end{barticle}
\endbibitem

\bibitem[\protect\citeauthoryear{{Bartko} et~al.}{2010}]{Bartko2010}
\begin{barticle}
\bauthor{\bsnm{{Bartko}}, \binits{H.}},
\bauthor{\bsnm{{Martins}}, \binits{F.}},
\bauthor{\bsnm{{Trippe}}, \binits{S.}},
\bauthor{\bsnm{{Fritz}}, \binits{T.K.}},
\bauthor{\bsnm{{Genzel}}, \binits{R.}},
\bauthor{\bsnm{{Ott}}, \binits{T.}},
\bauthor{\bsnm{{Eisenhauer}}, \binits{F.}},
\bauthor{\bsnm{{Gillessen}}, \binits{S.}},
\bauthor{\bsnm{{Paumard}}, \binits{T.}},
\bauthor{\bsnm{{Alexander}}, \binits{T.}},
\bauthor{\bsnm{{Dodds-Eden}}, \binits{K.}},
\bauthor{\bsnm{{Gerhard}}, \binits{O.}},
\bauthor{\bsnm{{Levin}}, \binits{Y.}},
\bauthor{\bsnm{{Mascetti}}, \binits{L.}},
\bauthor{\bsnm{{Nayakshin}}, \binits{S.}},
\bauthor{\bsnm{{Perets}}, \binits{H.B.}},
\bauthor{\bsnm{{Perrin}}, \binits{G.}},
\bauthor{\bsnm{{Pfuhl}}, \binits{O.}},
\bauthor{\bsnm{{Reid}}, \binits{M.J.}},
\bauthor{\bsnm{{Rouan}}, \binits{D.}},
\bauthor{\bsnm{{Zilka}}, \binits{M.}},
\bauthor{\bsnm{{Sternberg}}, \binits{A.}}:
\batitle{{An Extremely Top-Heavy Initial Mass Function in the Galactic Center
  Stellar Disks}}.
\bjtitle{\apj}
\bvolume{708}(\bissue{1}),
\bfpage{834}--\blpage{840}
(\byear{2010})
\doiurl{10.1088/0004-637X/708/1/834}
{\href{https://arxiv.org/abs/0908.2177}{{arXiv:0908.2177}}}
{[astro-ph.GA]}
\end{barticle}
\endbibitem

\bibitem[\protect\citeauthoryear{{Genzel} et~al.}{2010}]{Genzel2010}
\begin{barticle}
\bauthor{\bsnm{{Genzel}}, \binits{R.}},
\bauthor{\bsnm{{Eisenhauer}}, \binits{F.}},
\bauthor{\bsnm{{Gillessen}}, \binits{S.}}:
\batitle{{The Galactic Center massive black hole and nuclear star cluster}}.
\bjtitle{Reviews of Modern Physics}
\bvolume{82}(\bissue{4}),
\bfpage{3121}--\blpage{3195}
(\byear{2010})
\doiurl{10.1103/RevModPhys.82.3121}
{\href{https://arxiv.org/abs/1006.0064}{{arXiv:1006.0064}}}
{[astro-ph.GA]}
\end{barticle}
\endbibitem

\bibitem[\protect\citeauthoryear{{Goodman}}{2003}]{Goodman2003}
\begin{barticle}
\bauthor{\bsnm{{Goodman}}, \binits{J.}}:
\batitle{{Self-gravity and quasi-stellar object discs}}.
\bjtitle{\mnras}
\bvolume{339}(\bissue{4}),
\bfpage{937}--\blpage{948}
(\byear{2003})
\doiurl{10.1046/j.1365-8711.2003.06241.x}
{\href{https://arxiv.org/abs/astro-ph/0201001}{{arXiv:astro-ph/0201001}}}
{[astro-ph]}
\end{barticle}
\endbibitem

\bibitem[\protect\citeauthoryear{{Wang} et~al.}{2025}]{Wang2025:qion}
\begin{botherref}
\oauthor{\bsnm{{Wang}}, \binits{B.}},
\oauthor{\bsnm{{Leja}}, \binits{J.}},
\oauthor{\bsnm{{Katz}}, \binits{H.}},
\oauthor{\bsnm{{Inayoshi}}, \binits{K.}},
\oauthor{\bsnm{{Cleri}}, \binits{N.J.}},
\oauthor{\bsnm{{de Graaff}}, \binits{A.}},
\oauthor{\bsnm{{Hviding}}, \binits{R.E.}},
\oauthor{\bsnm{{van Dokkum}}, \binits{P.}},
\oauthor{\bsnm{{Greene}}, \binits{J.E.}},
\oauthor{\bsnm{{Labb{\'e}}}, \binits{I.}},
\oauthor{\bsnm{{Matthee}}, \binits{J.}},
\oauthor{\bsnm{{McConachie}}, \binits{I.}},
\oauthor{\bsnm{{Naidu}}, \binits{R.P.}},
\oauthor{\bsnm{{Nelson}}, \binits{E.J.}}:
{The Missing Hard Photons of Little Red Dots: Their Incident Ionizing Spectra
  Resemble Massive Stars}.
arXiv e-prints,
2508--18358
(2025)
\doiurl{10.48550/arXiv.2508.18358}
{\href{https://arxiv.org/abs/2508.18358}{{arXiv:2508.18358}}}
{[astro-ph.GA]}
\end{botherref}
\endbibitem

\bibitem[\protect\citeauthoryear{{Tanaka} et~al.}{2024}]{Tanaka2024}
\begin{botherref}
\oauthor{\bsnm{{Tanaka}}, \binits{T.S.}},
\oauthor{\bsnm{{Silverman}}, \binits{J.D.}},
\oauthor{\bsnm{{Shimasaku}}, \binits{K.}},
\oauthor{\bsnm{{Arita}}, \binits{J.}},
\oauthor{\bsnm{{Akins}}, \binits{H.B.}},
\oauthor{\bsnm{{Inayoshi}}, \binits{K.}},
\oauthor{\bsnm{{Ding}}, \binits{X.}},
\oauthor{\bsnm{{Onoue}}, \binits{M.}},
\oauthor{\bsnm{{Liu}}, \binits{Z.}},
\oauthor{\bsnm{{Casey}}, \binits{C.M.}},
\oauthor{\bsnm{{Lambrides}}, \binits{E.}},
\oauthor{\bsnm{{Kokorev}}, \binits{V.}},
\oauthor{\bsnm{{Jin}}, \binits{S.}},
\oauthor{\bsnm{{Faisst}}, \binits{A.L.}},
\oauthor{\bsnm{{Drakos}}, \binits{N.}},
\oauthor{\bsnm{{Shen}}, \binits{Y.}},
\oauthor{\bsnm{{Li}}, \binits{J.}},
\oauthor{\bsnm{{Zhuang}}, \binits{M.}},
\oauthor{\bsnm{{Fei}}, \binits{Q.}},
\oauthor{\bsnm{{Ito}}, \binits{K.}},
\oauthor{\bsnm{{Ren}}, \binits{W.}},
\oauthor{\bsnm{{Matsui}}, \binits{S.}},
\oauthor{\bsnm{{Ando}}, \binits{M.}},
\oauthor{\bsnm{{Hatano}}, \binits{S.}},
\oauthor{\bsnm{{Fujii}}, \binits{M.S.}},
\oauthor{\bsnm{{Kartaltepe}}, \binits{J.S.}},
\oauthor{\bsnm{{Koekemoer}}, \binits{A.M.}},
\oauthor{\bsnm{{Liu}}, \binits{D.}},
\oauthor{\bsnm{{McCracken}}, \binits{H.J.}},
\oauthor{\bsnm{{Rhodes}}, \binits{J.}},
\oauthor{\bsnm{{Robertson}}, \binits{B.E.}},
\oauthor{\bsnm{{Franco}}, \binits{M.}},
\oauthor{\bsnm{{Andika}}, \binits{I.T.}},
\oauthor{\bsnm{{Cloonan}}, \binits{A.P.}},
\oauthor{\bsnm{{Fan}}, \binits{X.}},
\oauthor{\bsnm{{Gozaliasl}}, \binits{G.}},
\oauthor{\bsnm{{Harish}}, \binits{S.}},
\oauthor{\bsnm{{Hayward}}, \binits{C.C.}},
\oauthor{\bsnm{{Huertas-Company}}, \binits{M.}},
\oauthor{\bsnm{{Kakkad}}, \binits{D.}},
\oauthor{\bsnm{{Kinugawa}}, \binits{T.}},
\oauthor{\bsnm{{Roy}}, \binits{N.}},
\oauthor{\bsnm{{Shuntov}}, \binits{M.}},
\oauthor{\bsnm{{Talia}}, \binits{M.}},
\oauthor{\bsnm{{Toft}}, \binits{S.}},
\oauthor{\bsnm{{Vijayan}}, \binits{A.P.}},
\oauthor{\bsnm{{Zhang}}, \binits{Y.}}:
{Discovery of dual ``little red dots'' indicates excess clustering on
  kilo-parsec scales}.
arXiv e-prints,
2412--14246
(2024)
\doiurl{10.48550/arXiv.2412.14246}
{\href{https://arxiv.org/abs/2412.14246}{{arXiv:2412.14246}}}
{[astro-ph.GA]}
\end{botherref}
\endbibitem

\bibitem[\protect\citeauthoryear{{Yanagisawa} et~al.}{2026}]{Yanagisawa2026}
\begin{botherref}
\oauthor{\bsnm{{Yanagisawa}}, \binits{H.}},
\oauthor{\bsnm{{Ouchi}}, \binits{M.}},
\oauthor{\bsnm{{Golubchik}}, \binits{M.}},
\oauthor{\bsnm{{Oguri}}, \binits{M.}},
\oauthor{\bsnm{{Fujimoto}}, \binits{S.}},
\oauthor{\bsnm{{Kokorev}}, \binits{V.}},
\oauthor{\bsnm{{Brammer}}, \binits{G.}},
\oauthor{\bsnm{{Sun}}, \binits{F.}},
\oauthor{\bsnm{{Nakane}}, \binits{M.}},
\oauthor{\bsnm{{Harikane}}, \binits{Y.}},
\oauthor{\bsnm{{Umeda}}, \binits{H.}},
\oauthor{\bsnm{{Akins}}, \binits{H.B.}},
\oauthor{\bsnm{{Atek}}, \binits{H.}},
\oauthor{\bsnm{{Bauer}}, \binits{F.E.}},
\oauthor{\bsnm{{Brada{\v{c}}}}, \binits{M.}},
\oauthor{\bsnm{{Chisholm}}, \binits{J.}},
\oauthor{\bsnm{{Coe}}, \binits{D.}},
\oauthor{\bsnm{{Diego}}, \binits{J.M.}},
\oauthor{\bsnm{{Ferguson}}, \binits{H.C.}},
\oauthor{\bsnm{{Finkelstein}}, \binits{S.L.}},
\oauthor{\bsnm{{Furtak}}, \binits{L.J.}},
\oauthor{\bsnm{{Inayoshi}}, \binits{K.}},
\oauthor{\bsnm{{Koekemoer}}, \binits{A.M.}},
\oauthor{\bsnm{{Matthee}}, \binits{J.}},
\oauthor{\bsnm{{Naidu}}, \binits{R.P.}},
\oauthor{\bsnm{{Ono}}, \binits{Y.}},
\oauthor{\bsnm{{Pan}}, \binits{R.}},
\oauthor{\bsnm{{Richard}}, \binits{J.}},
\oauthor{\bsnm{{Robbins}}, \binits{L.}},
\oauthor{\bsnm{{Willott}}, \binits{C.}},
\oauthor{\bsnm{{Zitrin}}, \binits{A.}},
\oauthor{\bsnm{{Amor{\'\i}n}}, \binits{R.O.}},
\oauthor{\bsnm{{Bradley}}, \binits{L.D.}},
\oauthor{\bsnm{{Bromm}}, \binits{V.}},
\oauthor{\bsnm{{Conselice}}, \binits{C.J.}},
\oauthor{\bsnm{{Dayal}}, \binits{P.}},
\oauthor{\bsnm{{Kartaltepe}}, \binits{J.S.}},
\oauthor{\bsnm{{Lopes}}, \binits{P.A.A.}},
\oauthor{\bsnm{{Lucas}}, \binits{R.A.}},
\oauthor{\bsnm{{Magdis}}, \binits{G.E.}},
\oauthor{\bsnm{{Martis}}, \binits{N.S.}},
\oauthor{\bsnm{{Papovich}}, \binits{C.}},
\oauthor{\bsnm{{Schaerer}}, \binits{D.}},
\oauthor{\bsnm{{Valentino}}, \binits{F.}},
\oauthor{\bsnm{{Vanzella}}, \binits{E.}},
\oauthor{\bsnm{{Allingham}}, \binits{J.F.V.}},
\oauthor{\bsnm{{Grogin}}, \binits{N.A.}},
\oauthor{\bsnm{{Gonz{\'a}lez-Otero}}, \binits{M.}},
\oauthor{\bsnm{{Ricotti}}, \binits{M.}},
\oauthor{\bsnm{{Windhorst}}, \binits{R.A.}}:
{VENUS: Two Faint Little Red Dots Separated by $\sim70\,\mathrm{pc}$ Hidden in
  a Single Lensed Galaxy at $z\sim7$}.
arXiv e-prints,
2601--06015
(2026)
{\href{https://arxiv.org/abs/2601.06015}{{arXiv:2601.06015}}}
{[astro-ph.GA]}
\end{botherref}
\endbibitem

\bibitem[\protect\citeauthoryear{{Weaver} et~al.}{2024}]{Weaver2024}
\begin{barticle}
\bauthor{\bsnm{{Weaver}}, \binits{J.R.}},
\bauthor{\bsnm{{Cutler}}, \binits{S.E.}},
\bauthor{\bsnm{{Pan}}, \binits{R.}},
\bauthor{\bsnm{{Whitaker}}, \binits{K.E.}},
\bauthor{\bsnm{{Labb{\'e}}}, \binits{I.}},
\bauthor{\bsnm{{Price}}, \binits{S.H.}},
\bauthor{\bsnm{{Bezanson}}, \binits{R.}},
\bauthor{\bsnm{{Brammer}}, \binits{G.}},
\bauthor{\bsnm{{Marchesini}}, \binits{D.}},
\bauthor{\bsnm{{Leja}}, \binits{J.}},
\bauthor{\bsnm{{Wang}}, \binits{B.}},
\bauthor{\bsnm{{Furtak}}, \binits{L.J.}},
\bauthor{\bsnm{{Zitrin}}, \binits{A.}},
\bauthor{\bsnm{{Atek}}, \binits{H.}},
\bauthor{\bsnm{{Chemerynska}}, \binits{I.}},
\bauthor{\bsnm{{Coe}}, \binits{D.}},
\bauthor{\bsnm{{Dayal}}, \binits{P.}},
\bauthor{\bsnm{{van Dokkum}}, \binits{P.}},
\bauthor{\bsnm{{Feldmann}}, \binits{R.}},
\bauthor{\bsnm{{F{\"o}rster Schreiber}}, \binits{N.M.}},
\bauthor{\bsnm{{Franx}}, \binits{M.}},
\bauthor{\bsnm{{Fujimoto}}, \binits{S.}},
\bauthor{\bsnm{{Fudamoto}}, \binits{Y.}},
\bauthor{\bsnm{{Glazebrook}}, \binits{K.}},
\bauthor{\bsnm{{de Graaff}}, \binits{A.}},
\bauthor{\bsnm{{Greene}}, \binits{J.E.}},
\bauthor{\bsnm{{Juneau}}, \binits{S.}},
\bauthor{\bsnm{{Kassin}}, \binits{S.}},
\bauthor{\bsnm{{Kriek}}, \binits{M.}},
\bauthor{\bsnm{{Khullar}}, \binits{G.}},
\bauthor{\bsnm{{Maseda}}, \binits{M.V.}},
\bauthor{\bsnm{{Mowla}}, \binits{L.A.}},
\bauthor{\bsnm{{Muzzin}}, \binits{A.}},
\bauthor{\bsnm{{Nanayakkara}}, \binits{T.}},
\bauthor{\bsnm{{Nelson}}, \binits{E.J.}},
\bauthor{\bsnm{{Oesch}}, \binits{P.A.}},
\bauthor{\bsnm{{Pacifici}}, \binits{C.}},
\bauthor{\bsnm{{Papovich}}, \binits{C.}},
\bauthor{\bsnm{{Setton}}, \binits{D.J.}},
\bauthor{\bsnm{{Shapley}}, \binits{A.E.}},
\bauthor{\bsnm{{Shipley}}, \binits{H.V.}},
\bauthor{\bsnm{{Smit}}, \binits{R.}},
\bauthor{\bsnm{{Stefanon}}, \binits{M.}},
\bauthor{\bsnm{{Taylor}}, \binits{E.N.}},
\bauthor{\bsnm{{Weibel}}, \binits{A.}},
\bauthor{\bsnm{{Williams}}, \binits{C.C.}}:
\batitle{{The UNCOVER Survey: A First-look HST + JWST Catalog of 60,000
  Galaxies near A2744 and beyond}}.
\bjtitle{\apjs}
\bvolume{270}(\bissue{1}),
\bfpage{7}
(\byear{2024})
\doiurl{10.3847/1538-4365/ad07e0}
{\href{https://arxiv.org/abs/2301.02671}{{arXiv:2301.02671}}}
{[astro-ph.GA]}
\end{barticle}
\endbibitem

\bibitem[\protect\citeauthoryear{{Wang} et~al.}{2024}]{Wang2024:sps}
\begin{barticle}
\bauthor{\bsnm{{Wang}}, \binits{B.}},
\bauthor{\bsnm{{Leja}}, \binits{J.}},
\bauthor{\bsnm{{Labb{\'e}}}, \binits{I.}},
\bauthor{\bsnm{{Bezanson}}, \binits{R.}},
\bauthor{\bsnm{{Whitaker}}, \binits{K.E.}},
\bauthor{\bsnm{{Brammer}}, \binits{G.}},
\bauthor{\bsnm{{Furtak}}, \binits{L.J.}},
\bauthor{\bsnm{{Weaver}}, \binits{J.R.}},
\bauthor{\bsnm{{Price}}, \binits{S.H.}},
\bauthor{\bsnm{{Zitrin}}, \binits{A.}},
\bauthor{\bsnm{{Atek}}, \binits{H.}},
\bauthor{\bsnm{{Coe}}, \binits{D.}},
\bauthor{\bsnm{{Cutler}}, \binits{S.E.}},
\bauthor{\bsnm{{Dayal}}, \binits{P.}},
\bauthor{\bsnm{{van Dokkum}}, \binits{P.}},
\bauthor{\bsnm{{Feldmann}}, \binits{R.}},
\bauthor{\bsnm{{Marchesini}}, \binits{D.}},
\bauthor{\bsnm{{Franx}}, \binits{M.}},
\bauthor{\bsnm{{F{\"o}rster Schreiber}}, \binits{N.}},
\bauthor{\bsnm{{Fujimoto}}, \binits{S.}},
\bauthor{\bsnm{{Geha}}, \binits{M.}},
\bauthor{\bsnm{{Glazebrook}}, \binits{K.}},
\bauthor{\bsnm{{de Graaff}}, \binits{A.}},
\bauthor{\bsnm{{Greene}}, \binits{J.E.}},
\bauthor{\bsnm{{Juneau}}, \binits{S.}},
\bauthor{\bsnm{{Kassin}}, \binits{S.}},
\bauthor{\bsnm{{Kriek}}, \binits{M.}},
\bauthor{\bsnm{{Khullar}}, \binits{G.}},
\bauthor{\bsnm{{Maseda}}, \binits{M.}},
\bauthor{\bsnm{{Mowla}}, \binits{L.A.}},
\bauthor{\bsnm{{Muzzin}}, \binits{A.}},
\bauthor{\bsnm{{Nanayakkara}}, \binits{T.}},
\bauthor{\bsnm{{Nelson}}, \binits{E.J.}},
\bauthor{\bsnm{{Oesch}}, \binits{P.A.}},
\bauthor{\bsnm{{Pacifici}}, \binits{C.}},
\bauthor{\bsnm{{Pan}}, \binits{R.}},
\bauthor{\bsnm{{Papovich}}, \binits{C.}},
\bauthor{\bsnm{{Setton}}, \binits{D.J.}},
\bauthor{\bsnm{{Shapley}}, \binits{A.E.}},
\bauthor{\bsnm{{Smit}}, \binits{R.}},
\bauthor{\bsnm{{Stefanon}}, \binits{M.}},
\bauthor{\bsnm{{Suess}}, \binits{K.A.}},
\bauthor{\bsnm{{Taylor}}, \binits{E.N.}},
\bauthor{\bsnm{{Williams}}, \binits{C.C.}}:
\batitle{{The UNCOVER Survey: A First-look HST+JWST Catalog of Galaxy Redshifts
  and Stellar Population Properties Spanning 0.2 {\ensuremath{\lesssim}} z
  {\ensuremath{\lesssim}} 15}}.
\bjtitle{\apjs}
\bvolume{270}(\bissue{1}),
\bfpage{12}
(\byear{2024})
\doiurl{10.3847/1538-4365/ad0846}
{\href{https://arxiv.org/abs/2310.01276}{{arXiv:2310.01276}}}
{[astro-ph.GA]}
\end{barticle}
\endbibitem

\bibitem[\protect\citeauthoryear{{Brammer}}{2023}]{Brammer2022}
\begin{botherref}
\oauthor{\bsnm{{Brammer}}, \binits{G.}}:
{msaexp: NIRSpec analyis tools}.
Zenodo
(2023).
\doiurl{10.5281/zenodo.7299500}
\end{botherref}
\endbibitem

\end{thebibliography}

\end{document}